\documentclass[useAMS,usenatbib,usegraphicx]{mn2e}
\usepackage{times}
\usepackage{amsmath}
\usepackage{amssymb}
\usepackage{flushend}

\newcommand{\aj}{AJ}                   
\newcommand{\araa}{ARA\&A}             
\newcommand{\apj}{ApJ}                 
\newcommand{\apjl}{ApJ}                

\newcommand{\apjs}{ApJS}               
     
\newcommand{\apss}{Ap\&SS}             
\newcommand{\aap}{A\&A}                
  
\newcommand{\aapr}{A\&A~Rev.}          
\newcommand{\aaps}{A\&AS}              
\newcommand{\mnras}{MNRAS}             
\newcommand{\pasp}{PASP}               

\newcommand{\echa}{$\eta$~Cha}
\newcommand{\epscha}{$\epsilon$~Cha}
\newcommand{\masyr}{mas~yr$^{-1}$}
\newcommand{\kms}{km~s$^{-1}$}

\newcommand{\msun}{$M_{\odot}$}
\newcommand{\rxjeleventhirtyseven}{RX J1137.4$-$7648}
\newcommand{\rxjelevenfortyseven}{RX J1147.7$-$7842}
\newcommand{\rxjelevenfiftyfour}{RX J1150.4$-$7704}
\newcommand{\rxjelevenfiftynine}{RX J1150.9$-$7411}
\newcommand{\rxjelevenfiftyeight}{RX J1158.5$-$7754A}
\newcommand{\rxjtwelvetwo}{RX J1202.8$-$7718}
\newcommand{\rxjtwelveseven}{RX J1207.7$-$7953}
\newcommand{\rxjtwelvefortythree}{RX J1243.1$-$7458}

\begin{document}
\title[Re-examining the $\epsilon$~Cha association]{Re-examining the membership and origin of the $\epsilon$~Cha association}
\author[S. J. Murphy, W. A. Lawson and M. S. Bessell]{Simon~J.~Murphy$^{1,2}$\thanks{Corresponding author: murphy@ari.uni-heidelberg.de}, Warrick~A.~Lawson$^3$ and Michael~S.~Bessell$^2$ \\
$^1$ Gliese Fellow, Astronomisches Rechen-Institut, Zentrum f\"{u}r Astronomie der Universit\"{a}t Heidelberg, Germany 69120\\
$^2$ Research School of Astronomy and Astrophysics, Australian National University, Canberra, ACT 2611, Australia \\
$^3$  School of Physical, Environmental and Mathematical Sciences, University of New South Wales, Canberra, ACT 2600, Australia}

\maketitle
\begin{abstract}
We present a comprehensive investigation of the $\epsilon$ Chamaeleontis association (\epscha), one of several young moving groups spread across the southern sky. We re-assess the putative membership of \epscha\ using the best-available proper motion and spectroscopic measurements, including new ANU 2.3-m/WiFeS observations. After applying a kinematic analysis our final membership comprises 35--41 stars from B9 to mid-M spectral types, with a mean distance of $110\pm7$~pc and a mean space motion of $(U,V,W)=(-10.9\pm0.8,-20.4\pm1.3,-9.9\pm1.4)$~\kms.  Theoretical evolutionary models suggest \epscha\ is  3--5~Myr old, distinguishing it as the youngest moving group in the solar neighbourhood. Fifteen members show 3--22~\micron\ spectral energy distributions attributable to circumstellar discs, including 11 stars which appear to be actively accreting. \epscha's disc and accretion fractions ($29^{+8}_{-6}$ and $32^{+9}_{-7}$ per cent, respectively) are both consistent with a typical 3--5 Myr-old population. Multi-epoch spectroscopy reveals three M-type members with broad and highly-variable H$\alpha$ emission as well as several new spectroscopic binaries. We reject 11 stars proposed as members in the literature and suggest they may belong to the background Cha I and II clouds or other nearby young groups. Our analysis underscores the importance of a holistic and conservative approach to assigning young stars to kinematic groups, many of which have only subtly different properties and ill-defined memberships. We conclude with a brief discussion of \epscha's connection to the young open cluster $\eta$ Cha and the Scorpius-Centaurus OB association (Sco-Cen). Contrary to earlier  studies which assumed $\eta$ and $\epsilon$ Cha are coeval and were born in the same location,  we find the groups were separated by $\sim$30~pc when $\eta$ Cha formed 4--8 Myr ago in the outskirts of Sco-Cen, 1--3 Myr before the majority of \epscha\ members. 
\end{abstract}
\begin{keywords}
open clusters and associations: individual: $\epsilon$ Chamaeleontis -- stars: pre-main sequence -- stars: kinematics and dynamics -- stars: formation -- stars: low-mass
\end{keywords}

\section{Introduction}\label{epschaintro}

Young stars in the solar neighbourhood are ideal laboratories for studying circumstellar discs and nascent planetary systems at high resolution and sensitivity. Well-characterised samples of stars with ages $\lesssim$10~Myr are particularly important, as it is during these epochs that discs rapidly evolve from massive gas-rich systems capable of supporting accretion and the formation of giant planets, to more-quiescent dusty debris discs which are the precursors of terrestrial planets. \citep[e.g.][]{Haisch01,Williams11}.

Members of nearby ($\lesssim100$~pc) kinematic associations, like those around the young stars TW Hydrae (age 8--10~Myr) and $\beta$ Pictoris ($\sim$12 Myr), have proven fruitful targets for such work \citep{Zuckerman04a,Torres08}. However, the majority of stars in these groups show signs their primordial discs are already evolved \citep{Schneider12b,Simon12}. With an age of 3--5~Myr (see \S\ref{sec:ages}), the $\epsilon$ Chamaeleontis association \citep{Mamajek00,Feigelson03} is therefore ideally suited to studies of the rapid disc evolution taking place at these intermediate ages  \citep{Fang13,Sicilia-Aguilar09}.

However, because of \epscha's greater distance (100--120~pc) compared to other nearby young groups, southern declination and position in the foreground of the Chamaeleon molecular cloud complex \citep[170--210~pc, age 2--4~Myr;][]{Luhman08} it has suffered from somewhat of an identity crisis in the literature. Over 50 members of a putative moving  group in the region have been proposed in the past fifteen years and there are several overlapping definitions of this group in the literature.  Many candidates lack radial velocities necessary for confirming membership in a moving group and several stars have no spectroscopy at all. If \epscha\ is a true coeval association then all of its members must have consistent age indicators \citep[e.g. X-ray and H$\alpha$ emission, lithium depletion, elevated colour-magnitude diagram position, low surface gravity;][]{Zuckerman04a} and space motions (via proper motions and radial velocities) consistent with being a comoving ensemble. 

In this contribution we attempt to clarify the situation by critically re-examining the membership of \epscha. The structure of this paper is as follows. In \S\ref{epslitcand} we outline the various members proposed in the literature. The   properties and kinematics of these stars are described in \S\ref{sec:obs} and \S\ref{sec:kinematics}, including new multi-epoch spectroscopy. Applying a kinematic and colour-magnitude analysis (\S\ref{sec:convergence}), we present the new membership of \epscha\ in \S\ref{sec:results}, with a discussion of its age, circumstellar discs and binaries. We conclude by discussing \epscha's relationship to the open cluster $\eta$ Chamaeleontis and their origin in the Scorpius-Centaurus OB association (\S\ref{sec:discussion}).

\section{$\epsilon$ Cha in the literature}\label{epslitcand}

\subsection{Early work}

The \epscha\ association has existed in the literature in various guises for fifteen years. \cite{Frink98} discovered several young \emph{ROSAT} sources between the Cha I and II dark clouds (Fig.\,\ref{fig:epscha}) with proper motions that placed them much closer to the Sun ($d\approx90$~pc, their ``subgroup 2''). \citet{Terranegra99} subsequently identified 13 stars between the clouds with similar proper motions, which they claimed formed a distinct kinematic association. They derived a distance of 90--110~pc and an isochronal age of 5--30~Myr.

While investigating the young open cluster \echa\ \citep{Mamajek99}, \citet{Mamajek00} also looked at stars in the vicinity of \epscha\ and HD 104237. They identified eight stars with congruent proper motions and photometry, and used several with radial velocities and parallaxes to derive a space motion for the group.

 \citet{Feigelson03} then obtained \emph{Chandra X-ray Observatory} snapshots of two fields around HD~104237, finding four low-mass companions to the Herbig Ae star at separations of 160--1700~au and three more spectroscopically-young mid-M stars at larger separations. \citet{Luhman04} soon added another three lithium-rich low-mass stars in the vicinity. These 12 stars constitute the `classical' membership of \epscha.

In their review of young stars near the Sun, \citet{Zuckerman04a} proposed a $\sim$10~Myr-old group surrounding (but not including) \epscha\ and HD~104237 which they called `Cha-Near'. It included six new candidates and 11 stars from the memberships of \citet{Terranegra99} and \citet{Mamajek00}. While \citeauthor{Zuckerman04a} considered Cha-Near and the aggregate around \epscha/HD 104237 to be separate groups, it is now apparent they are all members of the same spatially-extended association. 

Finally, while investigating the disk properties of Cha I members with the \emph{Spitzer Space Telescope},  \citet{Luhman08b} identified four stars with proper motions consistent with membership in the \epscha\ association defined by \citet{Feigelson03}, and several new \emph{ROSAT} sources not previously attributed to \epscha.

\begin{table*}
\centering
\begin{minipage}{0.82\textwidth}
\caption{Proposed members of the \epscha\ association from the literature}
\label{table:epscha}
\begin{tabular}{@{}rlcclccc@{}}
\hline
ID$^{\#}$ & Name & Right Ascension & Declination & Spec. & $V^{\dag}$ & Membership$^{\star}$ & Torres et al.\\
& & [J2000] & [J2000] & type$^{\dag}$ & [mag] & references & member?\\
\hline 
& HD 82879 & 09 28 21.1 	& $-$78 15 35~ & F6 & 8.99 & 8 & Y\\  
& CP$-$68 1388 & 10 57 49.3~ & $-$69 14 00~ & K1 &10.39 & 8 & Y\\ 
& VW Cha & 11 08 01.5~ & $-$77 42 29~ & K8 & 12.64 & 1 & \\
& TYC 9414-191-1 & 11 16 29.0~ & $-$78 25 21~ & K5 &  10.95 & 3 & \\
13 & 2MASS J11183572$-$7935548  & 11 18 35.7~ & $-$79 35 55~ & M4.5  & 14.91 & 7 & \\
14 & RX J1123.2$-$7924 & 11 22 55.6~ & $-$79 24 44~ & M1.5 & 13.71 & 7  & \\
& HIP 55746 & 11 25 18.1~ & $-$84 57 16~ & F5 & 7.6 & 6  & \\
15 & 2MASS J11334926$-$7618399 & 11 33 49.3~ & $-$76 18 40~ & M4.5 & \dots & 7  & \\
& \rxjeleventhirtyseven & 11 37 31.3~ & $-$76 47 59~ & M2.2  & \dots & 6  & \\
16 & 2MASS J11404967$-$7459394 & 11 40 49.7~ & $-$74 59 39~ & M5.5 & 17.28 & 7  & \\
& TYC 9238-612-1 & 11 41 27.7~ & $-$73 47 03~ & G5 & 10.7 & 6 & \\
17 & 2MASS J11432669$-$7804454 & 11 43 26.7~ & $-$78 04 45~ & M4.7 & 17.33 & 7  & \\
& \rxjelevenfortyseven  & 11 47 48.1~ & $-$78 41 52~ & M3.5 &  \dots & 6  & \\
18 & RX J1149.8$-$7850 & 11 49 31.9~ 	& $-$78 51 01~ & M0 &	12.9 & 7 & Y\\ 
19 & \rxjelevenfiftyfour\ & 11 50 28.3~ & $-$77 04 38~ & K4 & 12.0 & 1,2,6,7  & \\
& \rxjelevenfiftynine$^{\ddag}$ & 11 50 45.2$^{\ddag}$ & $-$74 11 13$^{\ddag}$ & M3.7 & 14.4 & 2  & \\
& 2MASS J11550485$-$7919108 &  11 55 04.9~ & $-$79 19 11~ & M3 & \dots & 10 & \\
& T Cha &	11 57 13.5~ &	$-$79 21 32~ & K0	&12.0 & 1,2,6 & Y\\ 
20 & RX J1158.5$-$7754B & 11 58 26.9~ & $-$77 54 45~ & M3 & 14.29 & 7 & Y\\ 
21 & \rxjelevenfiftyeight\ & 11 58 28.1~ & $-$77 54 30~ & K4 & 10.9 & 1,2,3,6,7  & \\
& HD 104036 & 	11 58 35.4~ &	$-$77 49 31~ & A7 & 6.73 & 3,6 & Y \\
1 & CXOU J115908.2$-$781232 & 11 59 08.0~ & $-$78 12 32~ & M4.75 &  & 4  & \\
2 & \epscha\ AB & 11 59 37.6~ & $-$78 13 19~ &	B9& 5.34 & 3,4 &Y\\ 
& RX J1159.7$-$7601 & 11 59 42.3~ &	$-$76 01 26~ & K4 & 11.31 & 1,2,3,6 & Y \\ 
3 & HD 104237C  & 12 00 03.6~ & $-$78 11 31~ & M/L & $\sim$25 & 4  & \\
4 & HD 104237B  & 12 00 04.0~ & $-$78 11 37~ & K/M & 15.1 & 4  & \\
5 & HD 104237A  & 12 00 05.1~ & $-$78 11 35~ & A7.75 & 6.73 & 2,3,4 & Y \\ 
6 & HD 104237D  & 	12 00 08.3~ & 	$-$78 11 40~ &M3.5 &	14.28 & 4 & Y\\ 
7 & HD 104237E & 	12 00 09.3~ & 	$-$78 11 42~ & K5.5 & 	12.08 & 4 & Y\\ 
10 & 2MASS J12005517$-$7820296 & 12 00 55.2~ & $-$78 20 30~ & M5.75 & \dots  & 5  & \\
& HD 104467 & 12 01 39.1~ & $-$78 59 17~ & G3 & 8.56 & 1,2,6 & Y\\ 
11 & 2MASS J12014343$-$7835472 & 12 01 43.4~ & $-$78 35 47~ & M2.25 &   \dots & 5  & \\
8 & USNO-B 120144.7$-$781926 & 12 01 44.4~ & $-$78 19 27~ & M5 &  \dots & 4  & \\
9 & CXOU J120152.8$-$781840 & 12 01 52.5~ & $-$78 18 41~ & M4.75 &  \dots & 4  & \\
& RX J1202.1$-$7853 & 12 02 03.8~ & $-$78 53 01~ & M0 & 	12.48 & 8 & Y \\ 
& \rxjtwelvetwo & 12 02 54.6~ & $-$77 18 38~ & M3.5 & 14.4 & 2,6  & \\
& RX J1204.6$-$7731 & 12 04 36.2~ & $-$77 31 35~ & M3 & 13.81 & 2,6 & Y \\ 
& TYC 9420-676-1 & 12 04 57.4~ & $-$79 32 04~ & F0 & 10.28 & 3 & \\
& HD 105234 & 12 07 05.5~ & $-$78 44 28~ & A9 & 7.4 & 3 & \\
12 & 2MASS J12074597$-$7816064 & 12 07 46.0~ & $-$78 16 06~ & M3.75 &  \dots & 5  & \\
& \rxjtwelveseven & 12 07 48.3~ & $-$79 52 42~ & M3.5 & 14.5 & 6  & \\
& HIP 59243 & 12 09 07.8~ & $-$78 46 53~ & A6 & 6.9 & 6 & \\
& HD 105923 & 12 11 38.1~ & $-$71 10 36~ & G8 &	9.16 & 8 & Y\\ 
& RX J1216.8$-$7753 & 12 16 45.9~ & $-$77 53 33~ & M4 & 13.88 & 11 & \\
& RX J1219.7$-$7403 & 12 19 43.5~ & $-$74 03 57~ & M0 & 13.08 & 2,6 & Y \\ 
& RX J1220.4$-$7407 & 12 20 21.9~ & $-$74 07 39~ & M0 & 12.85 & 2,6 & Y \\ 
& 2MASS J12210499$-$7116493 &12 21 05.0~ &$-$71 16 49~ & K7 & 12.16 & 9 & \\ 
& RX J1239.4$-$7502 & 12 39 21.2~ & $-$75 02 39~ & K3 & 10.30 & 2,6 & Y \\ 
& \rxjtwelvefortythree$^{\ddag}$ & 12 42 53.0$^{\ddag}$ & $-$74 58 49$^{\ddag}$ & M3.2 & 15.1 & 2  & \\
& CD$-$69 1055 &  12 58 25.6~ & $-$70 28 49~ & 	K0 & 9.95 &8 &Y\\ 
& CM Cha & 13 02 13.6~ & $-$76 37 58~ & K7 & 13.40 & 11 & \\
& MP Mus & 13 22 07.6~  & $-$69 38 12~ & K1 & 	10.35 & 8 & Y\\ 
\hline
\end{tabular}\\
($\#$): \epscha\ identification number \citep[][SIMBAD: {[}FLG2003{]} EPS CHA \#]{Feigelson03,Luhman04}\\
($\star$): Membership references: (1) \citet{Frink98}, (2) \citet{Terranegra99}, (3) \citet{Mamajek00}, (4) \citet{Feigelson03}, (5) \citet{Luhman04}, (6) \citet{Zuckerman04a}, (7) \citet{Luhman08b}, (8) \citet{Torres08}, (9) \citet{Kiss11}, (10) \citet{Kastner12}, (11) \citet{Lopez-Marti13}\\
($\ddag$): Updated coordinates to those presented by \citet{Alcala95}.\\
($\dag$): Spectral types from the literature (see Table\,\ref{table:bigtable}) and our WiFeS  observations. Indicative $V$ magnitudes from SIMBAD.\\
\end{minipage}
\end{table*}

\begin{figure}
   \centering
   \includegraphics[width=\linewidth]{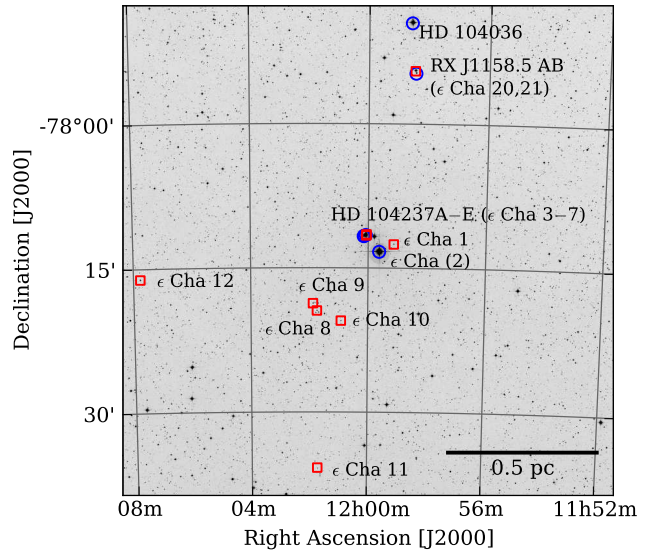} 
   \caption{POSS2-IR 1~deg$^{2}$ image centred on \epscha\ AB (11$^{\rm h}$59$^{\rm m}$37$^{\rm s}$.6, $-$78$^{\circ}$13$'$19$''$, J2000) with proposed association members from the literature (red squares) and kinematic members from \citet{Torres08} (blue circles). The scale uses the 111~pc distance to \epscha.} 
   \label{fig:epschazoom}
\end{figure}

\subsection{\citet{Torres08} compilation}

\citet{Torres08} reviewed \epscha\ as part of their ongoing program to identify new members of young, loose associations \citep[see also][]{Torres06}. Using radial velocities from the literature and their own observations, they proposed 24 high-probability members with congruent kinematics, photometry and lithium absorption. Their solution included 14 stars from previously described studies, six new members and four members of the open cluster \echa\ (RECX 1, 8, 12 and $\eta$ Cha itself). \citeauthor{Torres08} considered \echa\ to  be a part of \epscha\ and while the two groups have similar ages, distances and kinematics, in this work we consider only the 20 stars they classified as \epscha\ `field members'. We will discuss the relationship between $\eta$ and \epscha\ in greater detail in \S\ref{sec:epsetacha}. 

Of the candidates not proposed as members by \citet{Torres08}, they rejected only three\footnote{\citeauthor{Torres08} did not test the membership of VW Cha, the only \citet{Frink98} candidate not also included by \citet{Terranegra99}.}; \rxjelevenfiftyfour\ \citep[\epscha\ 19;][]{Terranegra99} had kinematics far from their convergent solution, HIP 55746 \citep{Zuckerman04a} was reclassified as a member of the AB Doradus association and \rxjelevenfiftyeight\ \citep[\epscha\ 21;][]{Terranegra99} was a poor kinematic match at its 90~pc \emph{Hipparcos} distance. The remaining stars lacked radial velocities necessary for the convergence method. Measuring  velocities for these stars is one of the key contributions of this work.

\begin{figure*} 
   \centering
   \includegraphics[width=\textwidth]{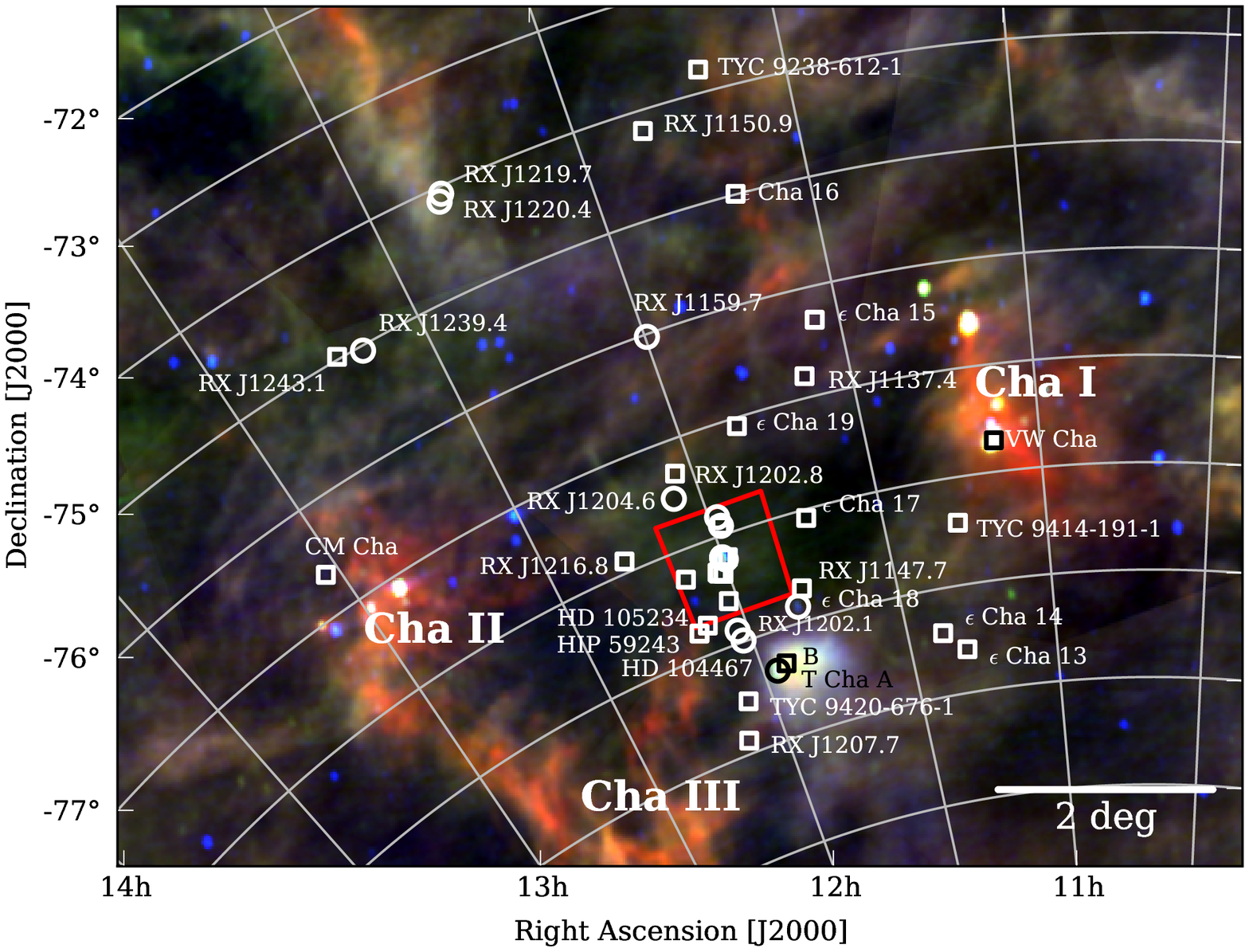} 
   \caption{\emph{Infrared Astronomical Satellite} (\emph{IRAS}) 100 \micron\ (red) / 60 \micron\ (green) / 25 \micron\ (blue) colour composite of the Chamaeleon region, with proposed \epscha\ members from the literature (squares) and \citet{Torres08} (circles). \echa\ members in the \citeauthor{Torres08} solution, HD~82879,  the likely AB Dor member HIP 55746 and several members north of $\delta=-72^{\circ}$ are not shown (see Fig. \ref{fig:etaeps}). The red square shows the 1~deg$^{2}$ extent of Fig.~\ref{fig:epschazoom}.}
   \label{fig:epscha}
\end{figure*}

\subsection{New candidates}

Several new members of \epscha\ have been proposed since the compilation of \citet{Torres08}. \citet{Kiss11} suggested  2MASS~J12210499$-$7116493 as a new member at an estimated kinematic distance of 98~pc. The star has strong Li\,\textsc{i} $\lambda$6708 absorption and a velocity only 1.5~\kms\ from the  \citeauthor{Torres08}  space motion\footnote{\citeauthor{Kiss11}  swapped the radial velocity of 2MASS~J1221$-$71 with the $\beta$ Pic member 2MASS J01071194$-$1935359 in their table 1. The listed space motions  for both stars are consistent with their correct velocities.}.  \citet{Kastner12} then  identified 2MASS J11550485$-$7919108 as a wide (0.2~pc), comoving companion to T~Cha \citep{Frink98}. Both stars are young and host  circumstellar material. Most recently, \citet{Lopez-Marti13} reanalysed the proper motions of young stars in Chamaeleon and proposed the new \epscha\ member RX J1216.8$-$7753 as well as reclassifying the Cha II star CM Cha as a potential member of the association. 

After tracing back in time the positions of nearby ($d<30$~pc) \emph{Hipparcos} stars with present-day space motions similar to local moving groups,  \citet{Nakajima12} proposed an additional seven members of the `Cha-Near' association of \citet{Zuckerman04a}. By considering only coarse youth indicators (variability, X-ray emission) their new `members' include the older pre-main sequence stars GJ 82 \citep[estimated age 35--300~Myr; ][]{Shkolnik09}, DK Leo \citep[$>$400~Myr; ][]{Shkolnik09}, GJ 755 \citep[$200\pm100$~Myr; ][]{Barrado-y-Navascues98}, HR 3499 \citep[$>$100~Myr; ][]{Wichmann03}, the $\beta$ Pic member AF Lep \citep{Torres08} and the binary EQ Peg, whose M3.5 primary has only a marginal lithium detection \citep{Zboril97}, implying an age greater than 10~Myr. Given their current locations, older ages and that none of the stars were closer than $\sim$20~pc from the centre of Cha-Near 10~Myr ago (the age assumed for the group), they are almost certainly not true members of Cha-Near or \epscha\ and we discuss them no further.

Excluding the \citeauthor{Nakajima12} candidates there are 52 putative members of the `\epscha\ association' in the literature. They are listed in Table~\ref{table:epscha} with spectral types and membership references. Their astrometry has been resolved against Two Micron All Sky Survey (2MASS) images and our own spectroscopic observations. For candidates proposed by \citet{Feigelson03}, \citet{Luhman04} and \citet{Luhman08b} we provide  the \epscha\ identification number (1--21) used in those studies. Figs.\,\ref{fig:epschazoom} and \ref{fig:epscha} show the location of \epscha\ candidates on the sky.  While the region around \epscha\ and HD 104237 (Fig.\,\ref{fig:epschazoom}) is well-studied, the majority of the dispersed population are isolated \emph{ROSAT} sources or incidental observations of stars around the Cha I cloud. Given the shallow depth of the flux-limited \emph{ROSAT} All Sky Survey, future studies may reveal  additional X-ray-faint members across the region. 

\section{Observations of candidate members}\label{sec:obs}

\subsection{WiFeS multi-epoch spectroscopy}

Many of the proposed \epscha\ members lack radial velocities necessary for confirming membership in a moving group. To remedy this and assess the youth of the candidates we observed 19 K and M-type stars from Table\,\ref{table:epscha} with the Wide Field Spectrograph \citep[WiFeS;][]{Dopita07} on the ANU 2.3-m telescope at Siding Spring.  To constrain any velocity variations and investigate circumstellar accretion \citep{Murphy11}, we observed each star 3--7 times over 60--480~d between 2010 February and 2011 June. We used the $R7000$ grating, which gave $\lambda/\Delta\lambda\approx7000$ and coverage from 5300--7100~\AA. Exposure times were 900--5400~s per epoch.   To estimate spectral types, surface gravities and reddenings we also obtained single-epoch $R3000$ spectra (5300--9600~\AA) for \epscha\ 13--17 and several \emph{ROSAT} candidates during 2011 July 30--31. All the observations were taken and reduced as described in \cite{Murphy10,Murphy12}. Briefly, we ran WiFeS in single-star mode with 2$\times$ spatial binning (1 arcsec spaxels) and used custom \textsc{iraf}, \textsc{figaro} and \textsc{python} routines to extract, wavelength-calibrate and combine the five image slices that contained the majority of the stellar flux. The $R3000$ spectra were corrected for telluric absorption and flux-calibrated using contemporaneous observations of the white dwarf EG 131. We took arc frames after each $R7000$ exposure and radial velocities were calculated by cross-correlation of the spectra against 5--7 K and M-type standards observed each night.

\begin{figure} 
   \centering
   \includegraphics[width=\linewidth]{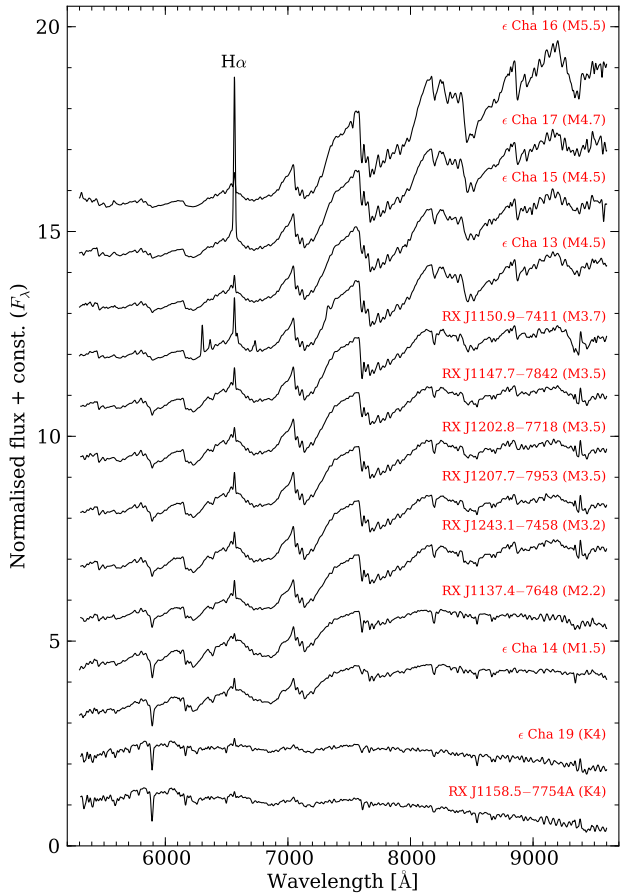} 
   \caption{Flux-calibrated WiFeS/$R3000$ spectra of the low-mass candidates listed in Table~\ref{table:epschaobs}. The spectra have been smoothed with a 10-px Gaussian kernel and normalised over the region 7400--7550~\AA\ for display.}
   \label{fig:r3000}
\end{figure}

The WiFeS/$R3000$ spectra are plotted in Fig.\,\ref{fig:r3000}. \rxjeleventhirtyseven\ and \rxjelevenfortyseven\ \citep{Zuckerman04a} have no previously published spectroscopy. The former is a 3~arcsec approximately equal-brightness visual binary that was unresolved in the typical 2--2.5~arcsec seeing of the $R3000$ observations. During a night of exceptional $\lesssim$1~arcsec seeing on 2011 May 16 we resolved the pair and extracted minimally-blended $R7000$ spectra. A full listing of WiFeS spectral types, Li\,\textsc{i} $\lambda$6708 and H$\alpha$ equivalent widths (EWs) and mean radial velocities are given in Table\,\ref{table:epschaobs}. Fifteen candidates have no previous velocity measurement.

\begin{table*}
\centering
\begin{minipage}{0.9\textwidth}
\caption{WiFeS observations of \epscha\ candidates from the literature}
\label{table:epschaobs}
\begin{tabular}{@{}rllcccccc|c@{}}
\hline
ID & Name & Spec.$^{\dag}$ & $E(B-V)^{\dag}$ & Li\,\textsc{i} EW & RV & $\sigma_{\rm RV}^{\ddag}$ & H$\alpha$ EW & $N_{\rm obs}$ & $\Delta t$ \\
 & & type & [mag] & [$\pm$50~m\AA] & [\kms] & [\kms] & [$\pm$1~\AA] && [days]  \\
\hline
13 &2MASS J11183572$-$7935548 & M4.5 & 0 & 600 & 19.3$^{\star}$ & 1.6$^{\star}$ & $[-30,-18]$ & 7 & 477 \\
14 & RX J1123.2$-$7924 & M1.5  & 0 & 150 & 2.7$^{\star}$ & 2.9$^{\star}$ & $-$2 & 6 & 477 \\
15 & 2MASS J11334926$-$7618399 & M4.5  & 0.15 & 650 & 16.7$^{\star}$ & 1.5$^{\star}$ & $-$6 & 5 & 413 \\
& \rxjeleventhirtyseven & M2.2  & 0 & 0 & $\sim$14& \dots& $-$1.5& 1 & \dots \\
16 & 2MASS J11404967$-$7459394 & M5.5 & 0 & 700 & 10.3~ & 1.0~ & $[-35,-11]$ & 4 & 411 \\
17 & 2MASS J11432669$-$7804454 & M4.7  & $<$0.1 & 700& 15.6~ & 1.0~ & $[-120,-60]$ & 6 & 480 \\
& \rxjelevenfortyseven & M3.5  & 0 & 650 & 16.1~ & 0.9~ & $[-7,-4]$& 5 & 409 \\
19 & \rxjelevenfiftyfour\ & K4 & 0 & 500 & 6.1$^{\star}$ & 1.6$^{\star}$ & $-$1 & 5 & 479 \\
& \rxjelevenfiftynine & M3.7  & 0 & 500 & 15.0 & 1.2 & $-$8 & 4 & 59 \\
21 & \rxjelevenfiftyeight\   & K4 & 0 & 500 & 19.9~ & 0.8~ & $-$0.5 & 4 & 480 \\
1 & CXOU J115908.2$-$781232 & \dots  &\dots & 650 & 15.1~ & 0.2~ & $-$5& 3 & 356 \\
10 & 2MASS J12005517$-$7820296 & \dots& \dots& 600& 10.7$^{\star}$& 1.3$^{\star}$& $[-20,-10]$& 6 & 411 \\
11 &2MASS J12014343$-$7835472 & \dots & \dots& 700 & 20.0~ & 0.6~ & $[-140,-70]$& 3 & 370\\
8 & USNO-B 120144.7$-$781926 &  \dots & \dots& 500 & 14.9~ & 1.1~ & $[-45,-20]$ & 4 & 60 \\
9 & CXOU J120152.8$-$781840 & \dots & \dots & 650 & 16.5~ & 1.1~ & $-$8 & 4 & 411 \\
& \rxjtwelvetwo & M3.5 & 0 & 300 & 17.1~ & 1.2~ & $[-12,-5]$ & 4 & 411 \\
12 & 2MASS J12074597$-$7816064 & \dots & \dots & 500 & 15.4$^{\star}$ & 2.3$^{\star}$ & $-$3.5 & 5 & 410 \\
& \rxjtwelveseven & M3.5 & 0 & 550 & 15.0~ & 0.7~ & $-4$ & 4 & 414\\
& \rxjtwelvefortythree & M3.2 & $<$0.1 & 600 & 13.5~ & 0.7~ & $[-7,-4]$& 4 & 60\\
\hline
\end{tabular}\\
($\dag$): Spectral types and reddening values determined from WiFeS $R3000$ spectra\\
($\star$): WiFeS $R7000$ time series shows a velocity trend indicative of binarity (see \S\ref{sec:binaries} and Table\,\ref{table:epschavels})\\
($\ddag$): Standard error of the mean, $\sigma_{\rm RV}=\sigma/\sqrt{N_{\rm obs}}$ \\
\end{minipage}
\end{table*}

\subsection{Spectral types and reddening}

We determined the spectral types in Table~\ref{table:epschaobs} using a selection of molecular indices from \citet{Riddick07} and visual comparison to $\eta$ and \epscha\ spectra from \citet{Lyo04a,Lyo08} and the \cite{Pickles98} library.  The average WiFeS/$R3000$ values agree with those previously determined at the 0.1--0.3~subtype level.  

\begin{figure}
   \centering
   \includegraphics[width=\linewidth]{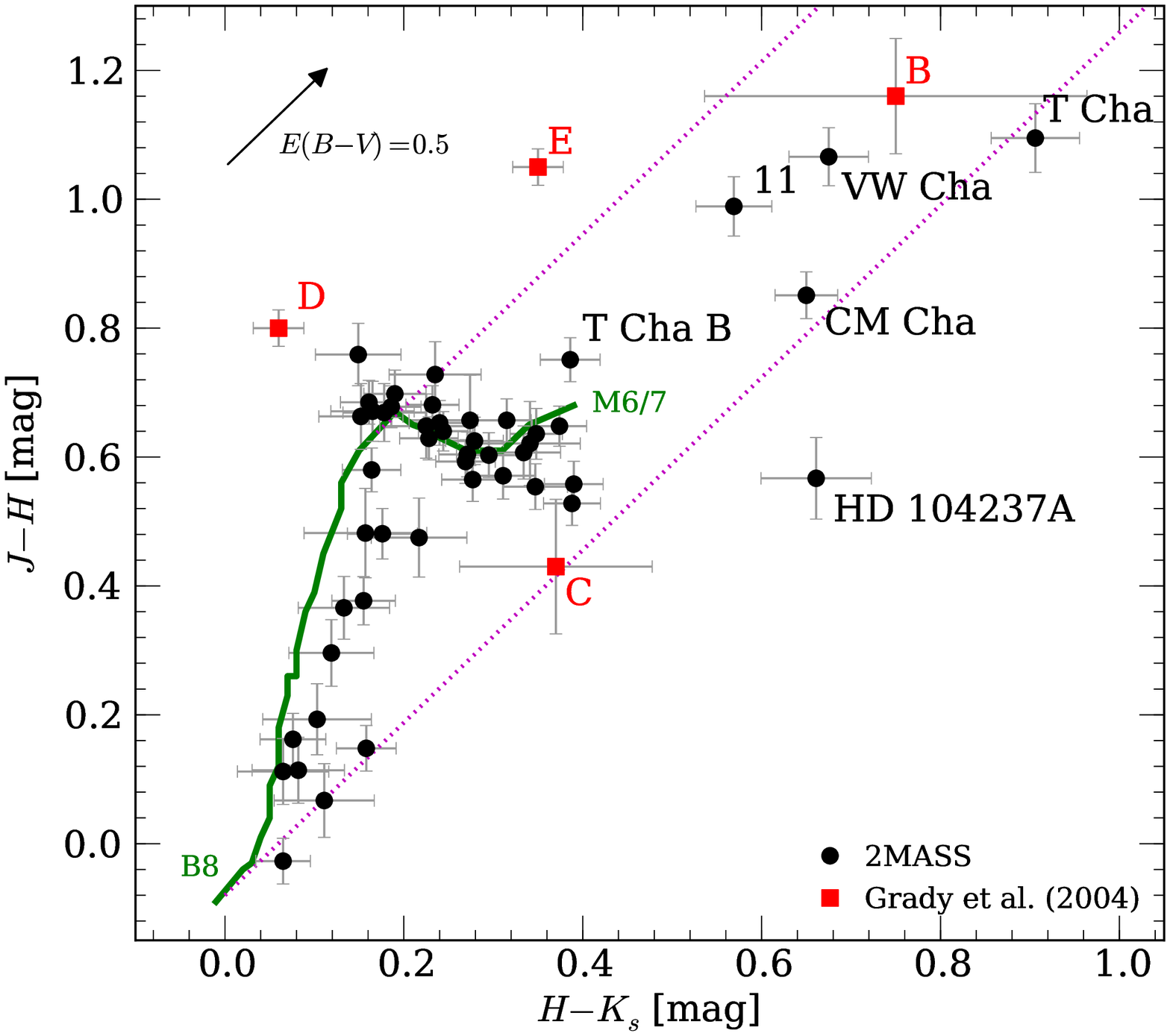}
   \caption{2MASS two-colour diagram for \epscha\ candidates (black circles), with VLT/NACO photometry for HD 104237B--E from \protect\citet{Grady04} (red squares) and the main sequence colours of \protect\citet[][solid line]{Kraus07c}. The \protect\cite{Schlegel98} reddening vector (arrow, dotted lines) was transformed to the 2MASS system using the relations of \protect\cite{Carpenter01}. The position of HD 104237D blueward of the locus of reddened stars is due to photometric errors \protect\citep{Grady04}.}
   \label{fig:2mass}
\end{figure}

There is evidence for a small amount of reddening between the Sun and the Chamaeleon cloud complex \citep{Knude98}. Compared to (unreddened) \echa\ spectra we estimate the WiFeS spectrum of 2MASS J11334926$-$7618399 (\epscha\ 15) is reddened by $E(B-V)\approx0.15$~mag, while 2MASS J11432669$-$7804454 (\epscha\ 17) and \rxjtwelvefortythree\ are reddened by no more than 0.1~mag.  With the exception of VW Cha (Cha I member?) and T~Cha AB \citep[possibly associated with the dark cloud Dcld 300.2$-$16.9;][]{Nehme08} the other candidates lie in regions of low-intensity dust emission (see Fig.\,\ref{fig:epscha}). This is also apparent in the 2MASS two-colour diagram (Fig.~\ref{fig:2mass}), where all but a handful of stars follow the zero-reddening locus. To supplement the WiFeS observations we estimated reddenings for all candidates using Fig.\,\ref{fig:2mass} and literature spectral types and photometry. The adopted values are given in Table\,\ref{table:bigtable}. An optically-thick circumstellar disc may also contribute to excess near-infrared emission. Several of the stars in the top-right corner of Fig.\,\ref{fig:2mass} host such discs (see \S\ref{sec:discs}.)

\subsection{Lithium depletion}

\begin{figure}
   \centering
   \includegraphics[width=\linewidth]{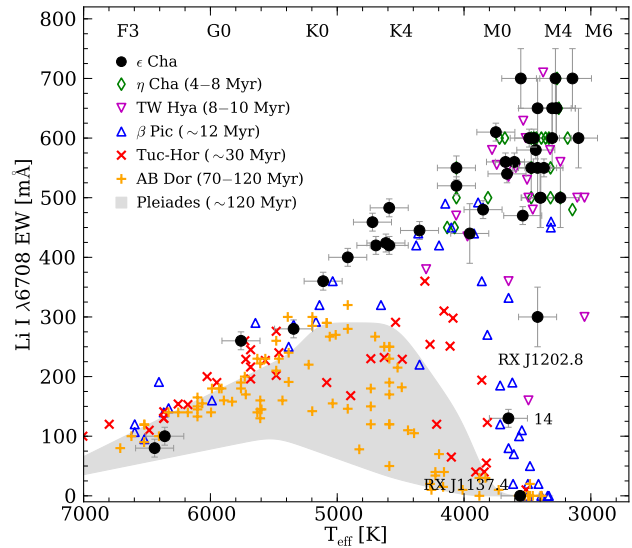}
   \caption{Li\,\textsc{i} $\lambda$6708 equivalent widths for \epscha\ candidates (black points, with errors), compared to members of young, nearby associations from \protect\citet{da-Silva09}. Members of the $\eta$ Cha open cluster and the envelope of equivalent widths observed in the Pleiades are also plotted. Following \citeauthor{da-Silva09}, effective temperatures were calculated from the spectral-type $T_{\rm eff}$ relation of \citet{Kenyon95}.}
   \label{fig:lithium}
\end{figure}

The amount of photospheric lithium depletion observed in low-mass stars can serve as a mass-dependent clock over pre-main sequence time-scales \citep[e.g.][]{Mentuch08,da-Silva09}. We plot in Fig.\,\ref{fig:lithium} the distribution of Li\,\textsc{i} $\lambda$6708 EWs assembled from the literature and WiFeS observations. The latter were obtained by fitting Gaussian line profiles, with the 50~m\AA\ error estimated from multiple observations and by varying the integration limits. No attempt was made to correct for contamination by the weak Fe\,\textsc{i} line at 6707.4 \AA\ \citep{Soderblom93} as its effect is negligible at these large EWs. We also see no evidence of strong continuum veiling, unsurprising given the low accretion rates (see \S\ref{epschactts}). The individual measurements and their sources are listed in Table\,\ref{table:bigtable}. Seven stars have no lithium measurements. Two are A-type stars which are not expected to show  Li\,\textsc{i} $\lambda$6708 absorption and the rest have no available spectra.

We also plot in Fig.\,\ref{fig:lithium} Li\,\textsc{i} $\lambda$6708 EW values  of young association members from \cite{da-Silva09}. The sensitivity of lithium depletion to both stellar age and mass is evident in the older groups. Several late-type members of TW Hya show significant depletion and the slightly older $\beta$ Pic ($\sim$12 Myr) presents a steep decline down to its depletion boundary  at a spectral type of $\sim$M4 \citep{Song02a}. Assuming that  the two \epscha\ candidates at $\sim$3500~K with low EWs are true members provides an upper age limit of 8--10 Myr, the age of TW Hya. The non-detection of lithium in RX J1137.4$-$7648 implies it is at least as old as members of the Tucana-Horologium  association ($\sim$30~Myr). Both RX J1137.4$-$7648 and \epscha\ 14 are also proper motion outliers (see next section). If these depleted stars are not members of \epscha\ then the remaining late-type candidates have lithium measurements consistent with an age no older than the 4--8~Myr \echa\ cluster.

\section{Kinematics of candidate members}\label{sec:kinematics}

\subsection{Proper motions}

\begin{figure*}
   \centering
   \includegraphics[width=0.495\textwidth]{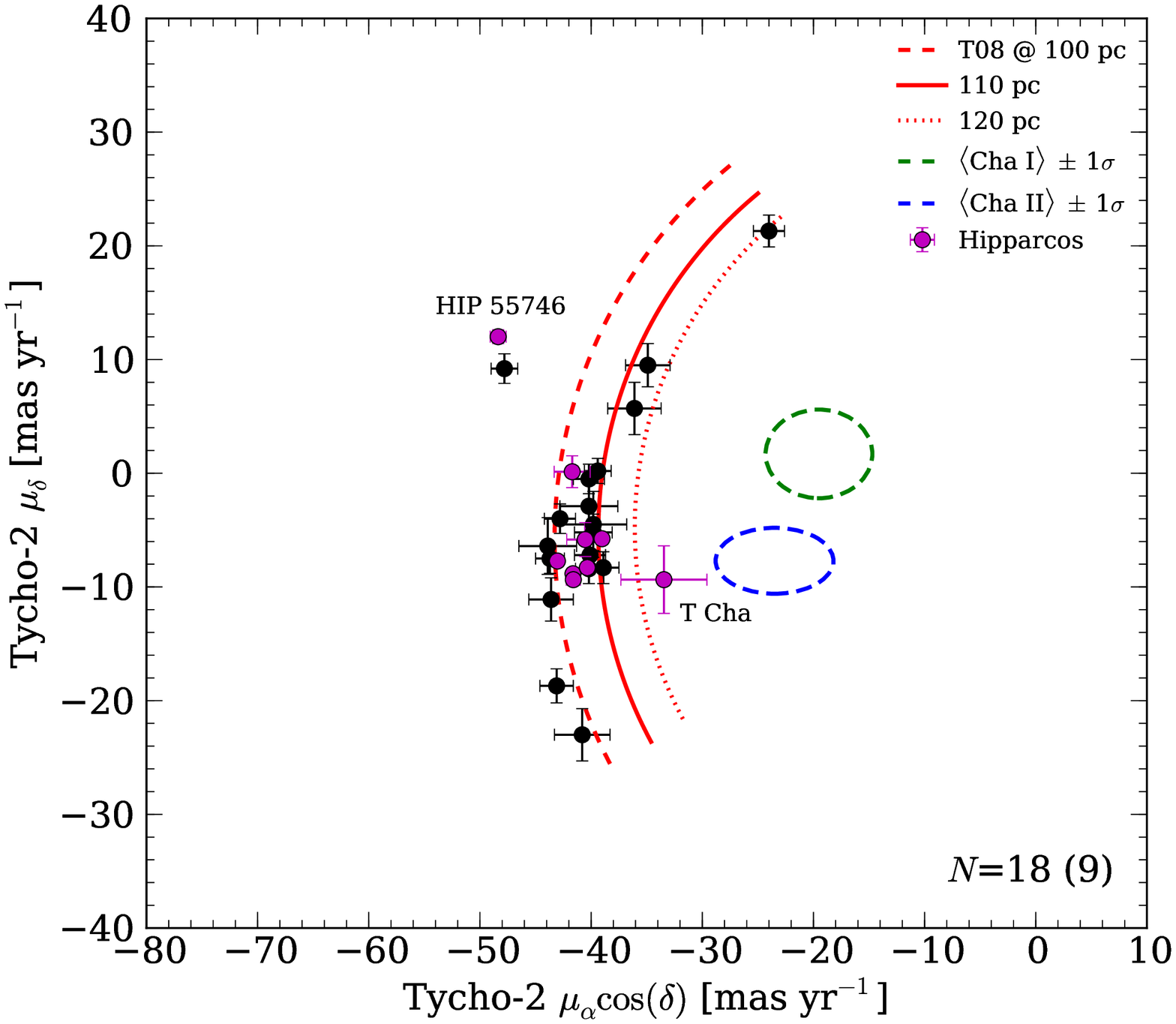}
   \includegraphics[width=0.495\textwidth]{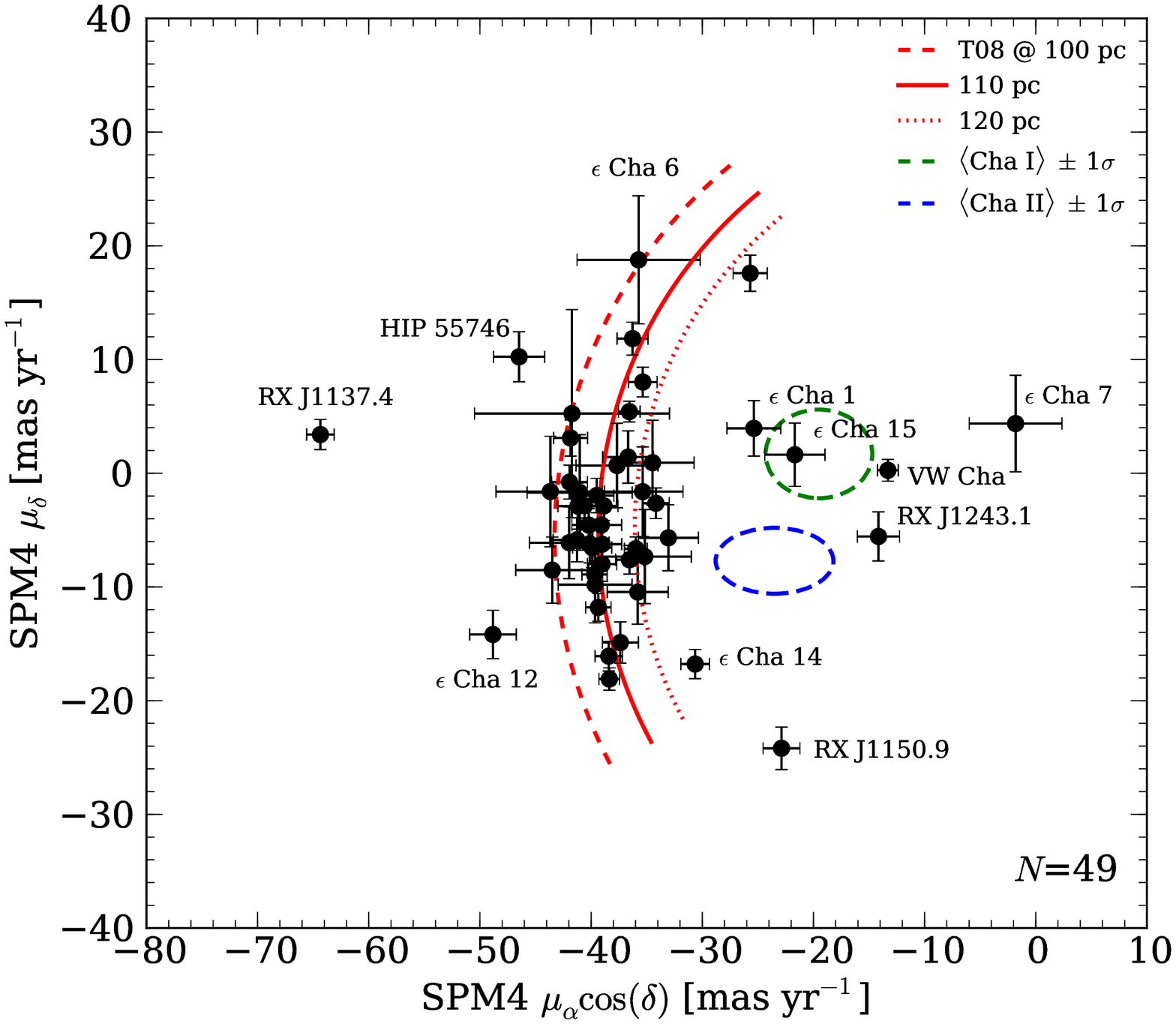}\\
   \includegraphics[width=0.495\textwidth]{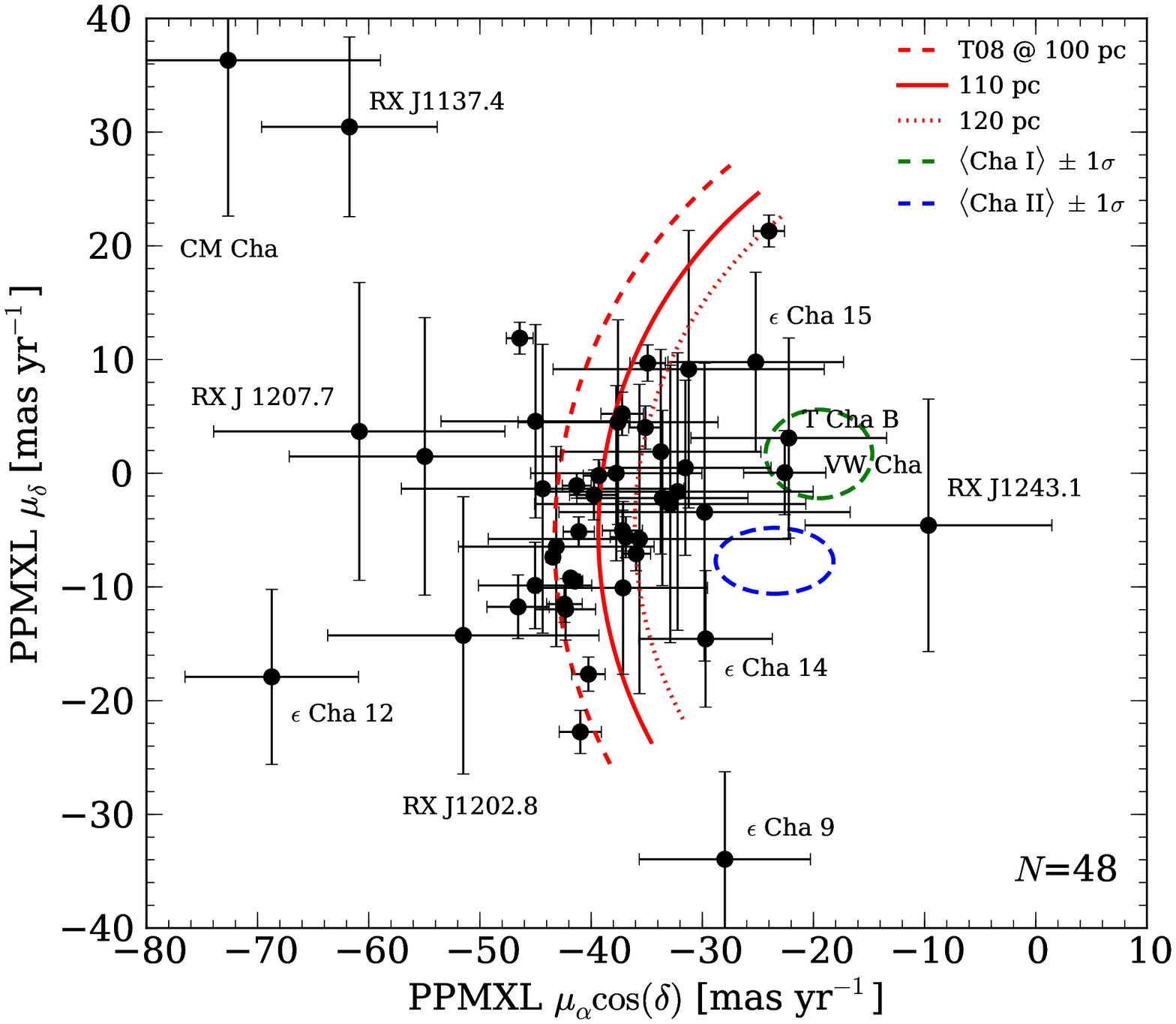}
   \includegraphics[width=0.495\textwidth]{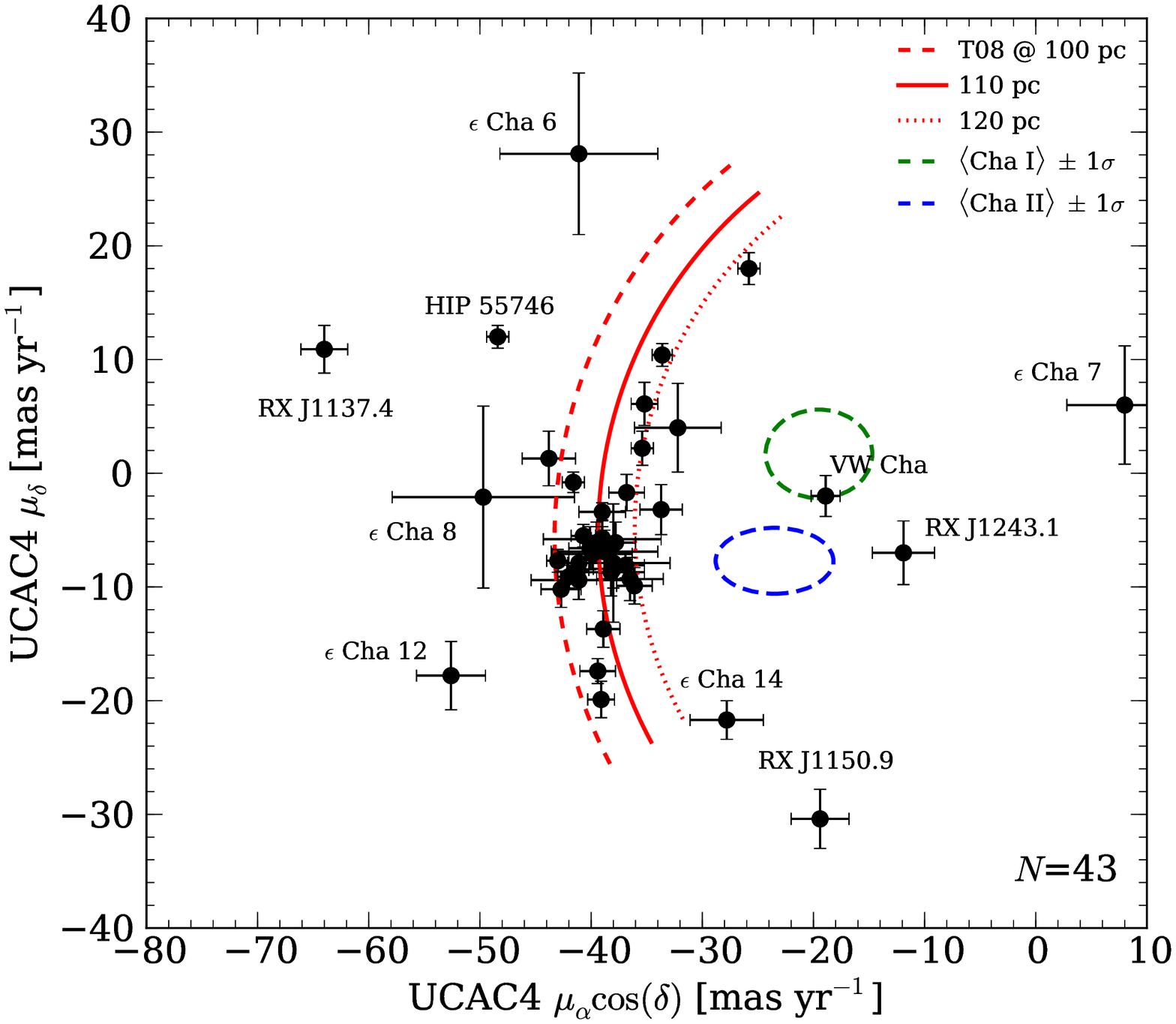}
   \caption{Proper motions of candidates in the \emph{Hipparcos}/Tycho-2 (top left), SPM4 (top right), PPMXL (bottom left) and UCAC4 (bottom right) catalogues.  The number of matches is given in the corner of each panel. The red lines show the \citet{Torres08} space motion projected onto the sky at $\delta=-79^{\circ}$ and $130^{\circ}<\alpha<210^{\circ}$ and distances of 100--120~pc.  Green and blue ellipses are the mean proper motions of Cha I and II sources from \citet{Lopez-Marti13}.}
   \label{fig:epschapm}
\end{figure*}

Nine stars have proper motions in the latest reduction of the \emph{Hipparcos} catalogue \citep{van-Leeuwen07} and a further ten were recovered in Tycho-2 \citep{Hog00}. We adopted these proper motions in the kinematic analysis of \S\ref{sec:convergence} with the exception of T Cha, which has a large error in \emph{Hipparcos} and is not in Tycho-2. For this star we used the higher-precision value from SPM4, the fourth iteration of Yale/San Juan Southern Proper Motion Catalog \citep{Girard11}. SPM4 contains absolute proper motions for over 103 million objects, including nearly all (49/52) of the proposed \epscha\ members. Only HD~104237B/C and 2MASS J12014343$-$7835472 \citep[underluminous due to an edge-on disk;][]{Luhman04} were not found in SPM4. To check for discordant proper motions we also queried the recent UCAC4 \citep{Zacharias13} and PPMXL \citep{Roeser10} catalogues but UCAC4 gave fewer matches (43/52) and in general PPMXL has less precise astrometry \citep[also see discussion in][]{Girard11}.  The results of these cross-matches are summarised in Fig.\,\ref{fig:epschapm}.  Proper motions from SPM4 and UCAC4 are somewhat correlated as they share some first epoch astrograph data \citep{Zacharias13}. As expected, the majority of stars cluster around the projection of the \epscha\ space motion on the sky at 100--120 pc. However, there are several stars with proper motions far from those expected of members.

\subsubsection{Outlying proper motions}

HIP~55746 was reclassified as a member of the AB Doradus association by \citet{Torres08}; its \emph{Hipparcos} proper motion and parallax are consistent with this new classification. Spectroscopically older RX~J1137.4$-$7648 is likely a nearby, unrelated system. CXOU J120152.8$-$781840 (\epscha\ 9) and 2MASS J12074597$-$7816064 (\epscha\ 12) were noted by \cite{Fang13} as having outlying PPMXL proper motions. With new UCAC4 and SPM4 astrometry, only \epscha\ 12 remains an outlier. We observed large ($\sim$13~\kms) radial velocity shifts (see \S\ref{sec:rv}) in the star which may indicate the presence of an unseen companion affecting the proper motion. RX~J1123.2$-$7924 (\epscha\ 14) is also suspected of having a close companion, though its lithium depletion implies an age older than \epscha. \rxjelevenfiftynine\ has a confirmed close companion \citep{Kohler01b} which may have altered its proper motion. \citet{Terranegra99} found \rxjtwelvefortythree\ was a kinematic but not spatial outlier to their proposed moving group. Its proper motion is near the mean of Cha II cloud members. VW Cha, CXOU J115908.2$-$781232 (\epscha\ 1) and 2MASS J11334926$-$7618399 (\epscha\ 15) may similarly be associated with the Cha I cloud. HD 104237D and E (\epscha\ 6 and 7) are probably affected by their location close to HD 104237A. 

\subsubsection{Discordant proper motions}\label{sec:discordpm}

\begin{table*}
\centering
\begin{minipage}{0.87\textwidth}
\caption{Candidates whose SPM4 and UCAC4 proper motions disagree by more than 2$\sigma$}
\label{table:badspm}
\begin{tabular}{lccccccc}
\hline
Name & \multicolumn{2}{c}{SPM4 [\masyr]} & \multicolumn{2}{c}{UCAC4 [\masyr]} & \multicolumn{2}{c}{PPMXL [\masyr]}\\
& $\mu_{\alpha}\cos\delta$ & $\mu_{\delta}$& $\mu_{\alpha}\cos\delta$ & $\mu_{\delta}$ &$\mu_{\alpha}\cos\delta$ & $\mu_{\delta}$ \\
\hline
VW Cha & $-13.3\pm0.9$ & $+0.3\pm1.0$ & $-18.9\pm1.3$ &$-2.0\pm1.8$ & $-22.6\pm3.7$ & $+0.1\pm3.7$\\ 
RX J1123.2$-$7924 (\epscha\ 14) & $-30.6\pm1.3$ & $-16.8 \pm1.3$ & $-27.8\pm3.3$ & $-21.7\pm1.7$ & $-29.7\pm6.0$ & $-14.6\pm6.0$ \\
\rxjeleventhirtyseven & $-64.4\pm1.2$ &$+3.4\pm1.3$ & $-64.0\pm2.1$ & $+10.9\pm2.1$ &$-61.7\pm7.9$ & $+30.5\pm7.9$\\ 
CM Cha & $-33.1\pm2.7$ & $-5.7\pm2.9$ & $-32.2\pm3.9$ & $+4.0\pm3.9$ & $-72.7\pm13.7$ & $+36.3\pm13.7$\\
\hline
\end{tabular}
\end{minipage}
\end{table*}

\begin{table*}
\centering
\begin{minipage}{0.85\textwidth}
\caption{Candidates whose \citet{Ducourant05} or \citet{Terranegra99} proper motions disagree with SPM4 by more than 2$\sigma$}
\label{table:d05t99}
\begin{tabular}{lcccccc}
\hline
Name & \multicolumn{2}{c}{SPM4 [\masyr]} & \multicolumn{2}{c}{\citet{Ducourant05} [\masyr]} & \multicolumn{2}{c}{\citet{Terranegra99} [\masyr]}\\
& $\mu_{\alpha}\cos\delta$ & $\mu_{\delta}$  & $\mu_{\alpha}\cos\delta$  & $\mu_{\delta}$  & $\mu_{\alpha}\cos\delta$ & $\mu_{\delta}$  \\
\hline
VW Cha & $-13.3\pm0.9$ & $+0.3\pm1.0$ & $-24\pm3$ & $+2\pm3$ & \dots & \dots\\
RX J1150.9$-$7411 & $-22.9\pm1.7$ & $-24.2\pm1.9$ & $+15\pm16$ & $+37\pm16$ &  $-39.0\pm4.2$ & $+4.4\pm2.9$ \\
RXJ1219.7$-$7403 & $-39.0\pm1.3$  & $-8.0\pm1.5$ & $-37\pm11$ & $-15\pm11$ & $-40.4\pm6.9$ & $-2.8\pm0.2$ \\ 
CM Cha & $-33.1\pm2.7$ & $-5.7\pm2.9$ & $-66\pm12$ & $+23\pm12$& \dots & \dots\\
\hline
\end{tabular}\\
\end{minipage}
\end{table*}

After comparing SPM4 and UCAC4 there were four stars with proper motions that differed by more than 2$\sigma$ in either component. Their astrometry is collated in Table\,\ref{table:badspm}, with values from PPMXL for comparison. \citet{Terranegra99} and \citet{Ducourant05} calculated proper motions for 15 and 20 of the candidates, respectively. Because of their different source observations these studies provide an independent check on the SPM4 measurements. The four stars whose proper motions disagreed with SPM4 by more than 2$\sigma$ are listed in Table\,\ref{table:d05t99}. 

The \citeauthor{Ducourant05} proper motion for VW Cha agrees with both UCAC4 and PPMXL. We adopted the UCAC4 value, while for RX J1123.2$-$7924 (\epscha\ 14), RX J1219.7$-$7403  and \rxjeleventhirtyseven\ we retained the SPM4 proper motions. CM~Cha is a special case. Although the \citeauthor{Ducourant05} proper motion matches PPMXL within the (large) errors we have adopted the higher-precision SPM4 values. If the larger proper motion is correct then CM Cha may be even closer to the Sun than \epscha\ \citep[see discussion in][]{Lopez-Marti13}. The motion of \rxjelevenfiftynine\ measured by both studies differs significantly from SPM4 and UCAC4. While \citeauthor{Ducourant05} used an incorrect position from \citet{Alcala95} and gave the proper motion of an unrelated star,  the \citeauthor{Terranegra99} proper motion for \rxjelevenfiftynine\ is very close to other candidates. We chose this value over SPM4 but note that it may be influenced by a close companion \citep{Kohler01b}.  The proper motions of all stars and their sources are listed in Table~\ref{table:bigtable}.

\subsection{Radial velocities}\label{sec:rv}

Candidate radial velocities from our WiFeS observations and the literature are listed in Table~\ref{table:bigtable}.  For 2MASS J11550485$-$7919108 (T~Cha B) we adopted the value for T~Cha itself, $14.0\pm1.3$~\kms\ \citep{Guenther07}. Since T~Cha exhibits large ($\Delta RV\sim 10$~\kms) aperiodic velocity variations on daily time-scales \citep{Schisano09} we caution that this velocity may not be representative. However, it agrees with velocities reported by \citet{Torres06} ($16.3\pm5.8$~\kms) and \citet{Franchini92} ($14.6\pm2.1$~\kms). For HD~104237E (\epscha\ 7) we adopted the velocity of the D component (\epscha\ 6), 13.4~\kms. \citet{Grady04} stated that both stars were a good velocity match to HD~104237A (14~\kms), although they did not explicitly give a velocity for the E component.  

\subsubsection{Spectroscopic binaries}\label{sec:binaries}

\begin{table}
\caption{Candidates suspected of spectroscopic binarity}
\label{table:epschavels}
\begin{tabular}{lccc}
\hline
\textbf{Star} & RV & $\sigma$(RV) & \citet{Kohler01b}\\
Epoch& [\kms] & [\kms] & companion? \\
\hline
\multicolumn{3}{l}{\bf 2MASS J12005517$-$7820296 (\epscha\ 10)} & \dots\\
2010 May 04 & $+$15.0 & 1.1\\
2011 Jan 09 & $+$7.2 & 0.8\\
2011 Feb 24 & $+$9.0 & 2.0\\
2011 May 09 & $+$13.4 & 1.1\\
 2011 May 17 & $+$12.0 & 1.7\\
2011 Jun 19 & $+$7.5 & 1.0\\
\hline
\multicolumn{3}{l}{\bf 2MASS J12074597$-$7816064 (\epscha\ 12)} & \dots\\
2010 May 05 & $+$23.4 & 1.5\\ 
2011 Jan 08 & $+$10.7 & 0.8 \\ 
2011 Jan 13 & $+$15.2 & 1.9 \\ 
2011 May 17 & $+$16.7 & 1.1 \\ 
2011 Jun 19 & $+$11.0 & 0.5 \\
\hline
\multicolumn{3}{l}{\bf 2MASS J11183572$-$7935548 (\epscha\ 13)} & \dots\\
2010 Feb 25 & $+$19.8 & 2.6\\
2010 May 02 & $+$15.8 & 0.5 \\
2010 Dec 20 & $+$14.3 & 1.9\\
2011 Feb 11 & $+$17.9 & 1.2\\
2011 Feb 12 & $+$17.8 & 2.5\\
2011 May 16 & $+$26.7 & 1.2 \\
2011 Jun 17 & $+$23.2 & 2.7\\
\hline
\multicolumn{3}{l}{\bf RX J1123.2$-$7924 (\epscha\ 14)} & N (\#42)\\
2010 Feb 25 & $-$2.6 & 3.1\\
2010 May 02 & $-$5.0 & 0.1\\
2010 Dec 20 & $-$0.4 & 1.7\\
2011 Feb 10 & $+$2.4 & 0.4\\
2011 May 16 & $+$7.3 & 1.8\\
2011 Jun 17 & $+$14.1 & 2.1\\
\citet{Covino97} & $+$10.0 & 2.0\\
\hline
\multicolumn{3}{l}{\bf 2MASS J11334926$-$7618399 (\epscha\ 15)} & \dots\\
2010 May 02 & $+$13.9 & 0.6 \\
2010 Dec 20 & $+$14.0 & 1.5 \\ 
2011 Feb 11 & $+$15.1 & 0.7 \\ 
2011 May 16 & $+$20.4 & 1.5 \\ 
2011 Jun 19 & $+$20.1 & 0.6 \\
\hline
\multicolumn{3}{l}{\bf \rxjelevenfiftyfour\ (\epscha\ 19)} & N (\#47) \\
2010 Feb 25 & $+$9.0 & 3.5 \\
2010 May 06 & $+$9.7& 1.7\\
2011 Feb 10 & $+$6.2 & 0.4\\
2011 May 17 & $+$1.0 & 1.7\\
2011 Jun 19 & $+$4.5 & 1.4\\
\citet{Covino97} & SB?\\
\citet{Guenther07} & $-$3.3 & 1.0\\
\hline
\multicolumn{3}{l}{\bf \rxjelevenfiftyeight\ (\epscha\ 21)} & Y (\#50, 0.07\arcsec) \\
2010 Feb--2011 Jun & $+$19.9 & 0.8\\
\citet{Covino97} & $+$13.1 & 2.0 \\
\citet{James06} & $+$10.2 & \dots\\
\hline
\bf \rxjtwelvefortythree & & & Y (\#73, 0.3\arcsec)\\
2011 Apr--Jun & $+$13.5 & 0.7\\
\citet{Covino97} &$+$7.0 & 2.0\\ 
\hline
\bf RX J1202.1$-$7853 & & & N (\#55)\\
\citet{Guenther07} & $+$17.1 & 0.2 \\
\citet{Covino97} & $+$5 & 2\\
\hline
\bf RX J1220.4$-$7407 & & & Y (\#65, 0.3\arcsec)\\
\citet{Guenther07} & $+$12.3 & 0.4\\
\citet{Covino97} & $+$18 & 2\\
\hline
\end{tabular}
\end{table}

The multi-epoch WiFeS observations revealed six candidates (marked in Table\,\ref{table:epschaobs}) with velocity variations indicative of binarity. Two more candidates, \rxjelevenfiftyeight\ (\epscha\ 21) and \rxjtwelvefortythree, have mean velocities that differ significantly from those in the literature.  The velocities of these eight stars are presented in Table\,\ref{table:epschavels}. The individual errors in the table are the standard deviations of cross-correlations against 4--8 K and M-type standards observed each night. We also found that the \citet{Torres08} members RX~J1202.1$-$7853 and RX~J1220.4$-$7407 have velocities in \citet{Guenther07} and \citet{Covino97} that differ outside their errors. These stars are also listed in Table\,\ref{table:epschavels}. 

\citet{Kohler01b} conducted a speckle and direct-imaging survey for close (0.13--6 arcsec) companions around 82 \emph{ROSAT} sources in Chamaeleon, including six of the candidates in Table~\ref{table:epschavels}. We confirm the binarity of \rxjelevenfiftyeight\ (\epscha\ 21), \rxjtwelvefortythree\ and RX~J1220.4$-$7407. \citet{Kohler01b} also found close companions around HIP~55746 (probable AB Dor member) and \rxjelevenfiftynine.  Four WiFeS measurements of this star over 59 d showed no significant velocity variation, unsurprising given the cadence of the observations. Despite finding no close companions around RX J1123.2$-$7924 (\epscha\ 14) the star is clearly a spectroscopic binary.  We have also confirmed suspicions about the binarity of \rxjelevenfiftyfour\ (\epscha\ 19) raised by \cite{Covino97}, with both our WiFeS time-series and an earlier measurement by \cite{Guenther07} showing a clear velocity variation. With only two measurements we cannot  confirm the spectroscopic binarity of RX J1202.1$-$7853, though the large velocity difference ($12.1\pm2$~\kms) is compelling. 

\subsubsection{Other known or suspected binaries}

RX~J1204.6$-$7731 is a double-lined spectroscopic binary \citep{Torres08} and HD 104467 is a suspected single-line system \citep{Cutispoto02}. \rxjeleventhirtyseven\ and \epscha\ are  approximately equal-brightness visual binaries and VW Cha is a hierarchical triple \citep{Correia06}. \citet{Bohm04} detected a 1.7~\msun\ companion to HD 104237A in an eccentric 19.8~d orbit (separation $\sim$0.1 au) which they called HD 104237b. It is likely the source of the late-type spectral features in HD 104237A reported by \citet{Feigelson03} and is distinct from HD 104237B, which is resolved at $\sim$150~au (1.3 arcsec).

\section{Kinematic membership analysis}\label{sec:convergence}

\subsection{Initial observational isochrone}

\begin{figure}
   \centering
   \includegraphics[width=\linewidth]{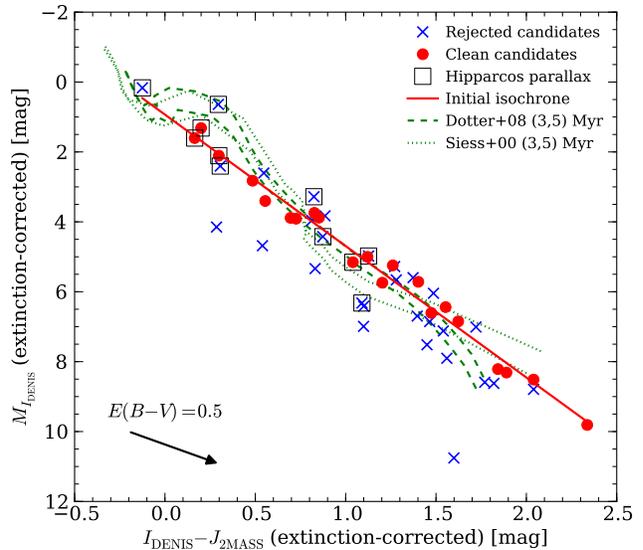} 
   \caption{The initial isochrone (solid line) used in the kinematic analysis. A linear fit was performed on the extinction-corrected photometry (red circles) after excluding binaries,  proper motion and Li\,\textsc{i} outliers and NIR excess sources (blue crosses). \citet{Siess00} (dotted lines) and \citet{Dotter08} (dashed lines) theoretical isochrones are shown for comparison.}
   \label{fig:isochrone}
\end{figure}

To compare the candidates' photometry to kinematic distances derived in \S\ref{sec:membanalysis} an absolute isochrone is required. We plot in Fig.\,\ref{fig:isochrone} extinction-corrected DENIS \citep{Epchtein99} and 2MASS \citep{Skrutskie06} photometry with 3 and 5~Myr isochrones from \citet{Dotter08} and \citet{Siess00}. Stars  without \emph{Hipparcos} parallaxes were assigned a distance of 110~pc and we adopted the reddening relations $A_{I}=0.601A_{V}$ and $A_{I-J}=0.325A_{V}$ from \citet{Schlegel98}. There is significant variation between the observed photometry and the isochrones, and also between different model sets. Rather than use a theoretical isochrone, which would require assuming an age and could introduce model-dependent photometric distances, we adopted a similar strategy to \citet{Torres06} and used the candidates themselves to define an ad hoc, empirical isochrone. After excluding known or suspected binaries, proper motion outliers and those stars with 2MASS colour excesses or low Li\,\textsc{i} $\lambda$6708 equivalent widths, the 21 remaining candidates were well-fitted by the linear regression:
\begin{equation}\label{eqn:iso}
M_{I} = 3.760\times(I-J) + 0.93
\end{equation}
This relation was used in the kinematic analysis in the next section. As noted by \citet{Torres06}, defining an isochrone assuming membership in a group then using it to test membership appears to be circular reasoning. However, the high-quality candidates used to define Eqn.\,\ref{eqn:iso} all have photometry, proper motions, radial velocities and lithium measurements consistent with membership in \epscha\ and are highly likely to represent the true association isochrone. Moreover, several stars have \emph{Hipparcos} parallaxes which anchor the distance scale of the fainter candidates. 

Although binaries were not included in the calculation of Eqn.\,\ref{eqn:iso}, before running the kinematic analysis we corrected the photometry of \rxjelevenfiftynine, \rxjelevenfiftyeight\ (\epscha\ 21), RX J1220.4$-$7407, \rxjtwelvefortythree, HIP 55746, \epscha\  and VW Cha  to that of the primary using published flux ratios. For \rxjeleventhirtyseven\ we assumed an equal-mass system. More details and $IJHK_{s}$ photometry for all candidates is given in Table\,\ref{table:phot}.

\subsection{Member selection} \label{sec:membanalysis}

With the best-available proper motions, velocities and photometry we tested  membership of the candidates using a similar technique to the convergence method of \citet{Torres06,Torres08}. First, for each star without a good \emph{Hipparcos} parallax\footnote{For \emph{Hipparcos} candidates with a parallax error greater than 10 per cent we adopted the kinematic distance in lieu of the trigonometric.} we found the distance which minimised the difference in space motions:
\begin{equation}
K = \sqrt{(U-U_{0})^2 + (V-V_{0})^2 + (W-W_{0})^2}
\end{equation}
where $K=K(\alpha,\delta, \mu_{\alpha}, \mu_{\delta}, RV; d)$ and $(U_{0},V_{0},W_{0})$ is the mean space motion \citep[initially from][]{Torres08}\footnote{Throughout this work $(U,V,W)$ are a right-handed triad with $U$ (and $X$) increasing towards the Galactic centre.}. With this \emph{kinematic} distance (or a parallax) and the extinction-corrected photometry we then calculated the difference in absolute magnitude ($\Delta M_{I}$) between the star and the isochrone (Eqn.~\ref{eqn:iso}). Candidates were retained as possible members if $K<3.5$~\kms\ (a difference of 2~\kms\ in each dimension and similar to the observed velocity dispersion in young groups) and $|\Delta M_{I}|<1$~mag (the approximate rms variation around the initial isochrone). To exclude background members of Cha I and II we additionally required $d<150$~pc. From these kinematic members a new isochrone and  $(U_{0},V_{0},W_{0})$ were calculated. The process was iterated until the membership  and space motion no longer changed significantly.

Due to the high-quality initial space motion from \citet{Torres08} the final list of 25 kinematic members converged on the second iteration while final distances and velocities took four iterations (for $\Delta d=1$~pc). Both the mean space motion (Table~\ref{table:uvw}) and distance ($110\pm7$~pc) agree with those previously determined by \citet{Torres08}. The revised values should be more accurate as they were derived from a larger number of members and by not subsuming \echa\ into the \epscha\ solution.  

\begin{figure}
   \centering
   \includegraphics[width=\linewidth]{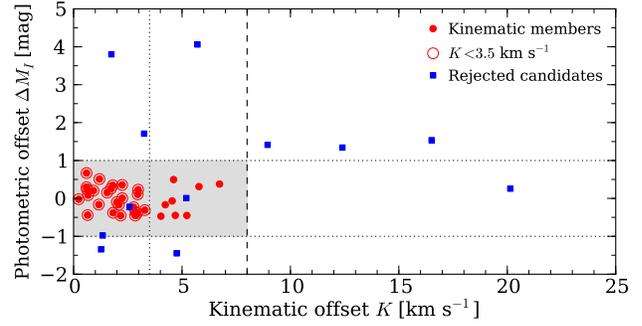} 
   \caption{Results of the convergent analysis, showing the difference in space motions ($K$) and the offset from the final isochrone ($\Delta M_{I}$). Dotted lines mark the $\pm$1~mag and 3.5~\kms\ limits for calculating the association space motion (open circles), while the dashed line is the limit for selecting final kinematic members (filled circles).}
   \label{fig:conv}
\end{figure}

\begin{figure}
   \centering
   \includegraphics[width=\linewidth]{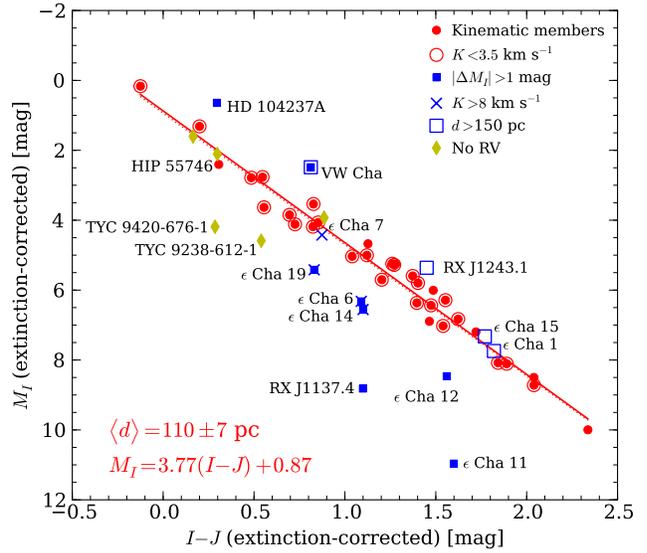} 
   \caption{Colour-magnitude diagram of \epscha\ after the membership analysis. Candidates rejected in the analysis are labelled (blue squares, crosses). \epscha\ 11, HD 104237A and its companions \epscha\ 6 and 7 are confirmed members with bad photometry or astrometry (see Appendix~\ref{sec:candidates}). The five stars without radial velocities (yellow diamonds) are plotted at their best-fitting kinematic distance or  \emph{Hipparcos} parallax, if available.}
   \label{fig:conv_cmd}
\end{figure}

\begin{figure}
   \centering
   \includegraphics[width=\linewidth]{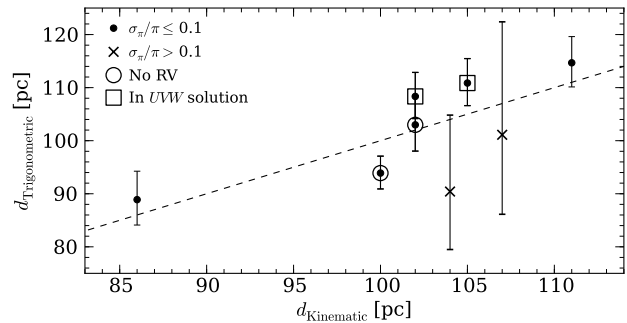} 
   \caption{Comparison of trigonometric and kinematic distances. Kinematic distances for those  stars without radial velocities were estimated from proper motions alone. The dashed line is the 1:1 relation. T Cha (relative parallax error $\sim$50 per cent) is not plotted.}
   \label{fig:hipparcos}
\end{figure}

\begin{table*}
\begin{minipage}{0.8\textwidth}
\caption{Heliocentric velocities and positions of $\eta$ and $\epsilon$ Chamaeleontis.}
\label{table:uvw}
\begin{tabular}{@{}lccccccccccccccc@{}}
\hline
& $U$ & $\sigma_{U}$ & $V$ & $\sigma_{U}$ &$W$ & $\sigma_{W}$& $X$ & $\sigma_{X}$ & $Y$ & $\sigma_{Y}$ &$Z$ & $\sigma_{Z}$\\
\dots & \multicolumn{6}{c}{\hrulefill\ [\kms] \hrulefill} &  \multicolumn{6}{c}{\hrulefill\ [pc] \hrulefill}\\
\hline
\epscha\ (this work) & $-$10.9 & 0.8 & $-$20.4 & 1.3 & $-$9.9 & 1.4 & 54 & 3 & $-$92 & 6 & $-$26& 7\\
\citet{Torres08} & $-$11.0 & 1.2 & $-$19.9 & 1.2 & $-$10.4 & 1.6 & 50 &\dots & $-$92 & \dots & $-$28 & \dots\\
\hline
\echa\ (updated)$^{\ddag}$ & $-$10.2 & 0.2 & $-$20.7 & 0.1 & $-$11.2 & 0.1 & 33.4 & 0.4 & $-$81.0 & 1.0 & $-$34.9 & 0.4\\
\hline
\end{tabular}\\
($\ddag$): Derived from the weighted mean \emph{Hipparcos}/Tycho-2 positions and proper motions of $\eta$ Cha, RS Cha, RECX~1 and HD 75505 (08$^{\rm h}$40$^{\rm m}$48.24$^{\rm s}$, $-$79$^{\circ}$ 00$'$ 24.8$''$; $\mu_{\alpha}\cos\delta=-29.35\pm0.13$~\masyr, $\mu_{\delta}=27.41\pm0.13$~\masyr), the weighted mean \emph{Hipparcos} parallaxes of $\eta$ Cha and RS Cha \citep[$94.3\pm1.1$~pc;][]{van-Leeuwen07} and the weighted mean radial velocities of RECX 1,3,4--6,9,10--13 ($18.3\pm0.1$~\kms; A. Brandeker, unpublished private communication).\\
\end{minipage}
\end{table*}

\begin{figure*}
   \centering
   \includegraphics[width=\textwidth]{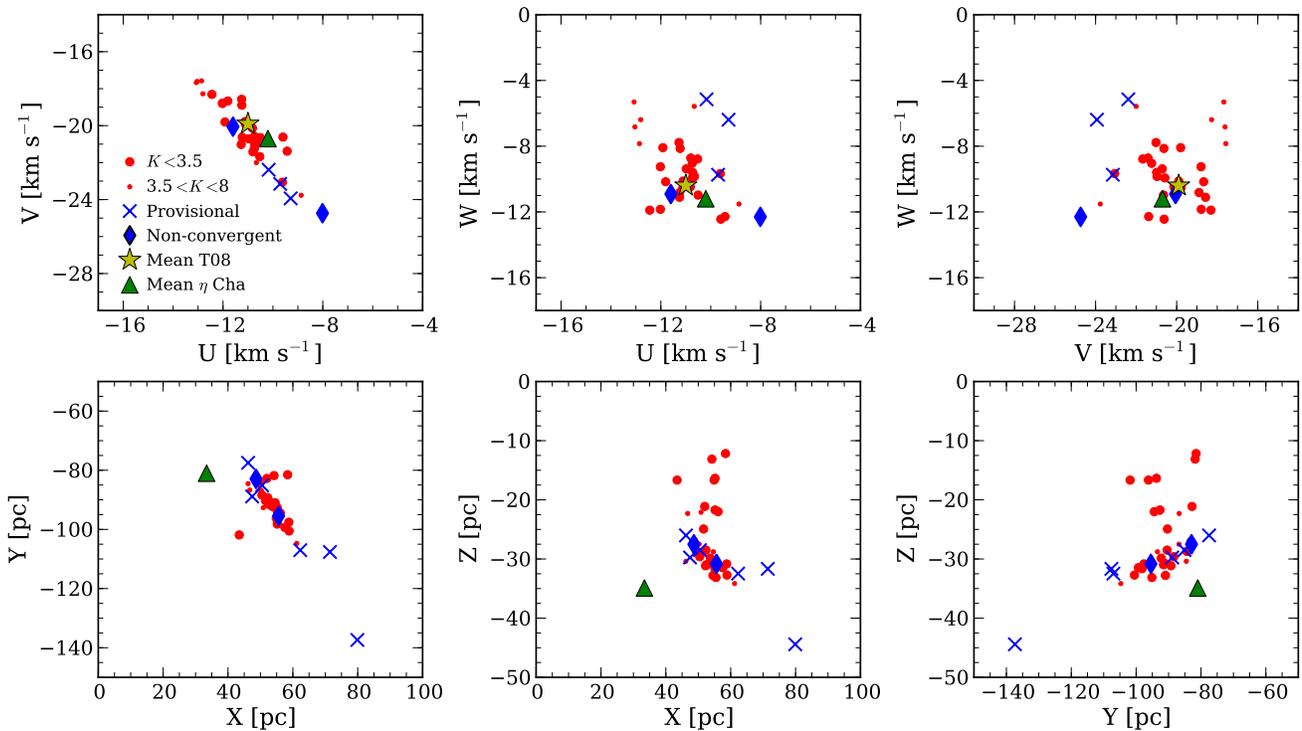} 
   \caption{Heliocentric velocities and positions of kinematic members (red points), confirmed members not selected by the convergent analysis (HD 104237 and \epscha\ 11; blue diamonds) and provisional candidates requiring confirmation (crosses). \epscha\ 1 is shown at the 165~pc distance implied by its SPM4 proper motion. The previous \citet{Torres08} and updated \echa\ space motions are given by the yellow star and green triangle, respectively.}
   \label{fig:uvwxyz}
\end{figure*}

Fig.\,\ref{fig:conv} shows the final distribution of $K$ and $\Delta M_{I}$ values. There are several candidates with $|\Delta M_{I}|<1$~mag but velocity differences in excess of the 3.5~\kms\ limit. To account for non-systemic radial velocities and the lower-resolution WiFeS observations we selected as \emph{a posteriori} kinematic members those candidates with $K<8$~\kms. The mean space motion and isochrone were \emph{not} updated after this step. Six of the eight stars this criteria added are known or suspected spectroscopic binaries.  The four candidates with $K>8$~\kms\ are either members with obviously erroneous proper motions (\epscha\ 6 and 7) or under-luminous spectroscopic binaries (\epscha\ 14 and 19). Placing them near the isochrone would require unrealistic distances of 170--220~pc. \epscha\ 14 also has a low (130~m\AA) lithium measurement.  The three rejected candidates inside the shaded region  have estimated distances greater than 150~pc. Fig.\,\ref{fig:conv_cmd} presents the new \epscha\  colour-magnitude diagram (CMD) and the twelve candidates rejected by our technique. As expected of true members, the tight clustering of \emph{kinematic} distances in the CMD almost perfectly replicates the mean photometric isochrone of Eqn.~\ref{eqn:iso}.

Kinematic distances for those candidates observed by \emph{Hipparcos} are compared to their trigonometric parallaxes in Fig.~\ref{fig:hipparcos}. All agree within the 2$\sigma$ errors. Only two parallaxes (\epscha, HD 104036) were used in the final space motion solution. The other candidates had large parallax errors (in which case the kinematic distance was adopted) or failed one of the selection criteria. 

\section{Results}\label{sec:results}

\subsection{Final membership of \epscha}

After considering the results of the kinematic, colour-magnitude diagram and lithium analysis our final membership for \epscha\ contains 35 stars; 29 from the kinematic solution plus  2MASS J12014343$-$7835472 (\epscha\ 11) and HD~104237A--E.  The heliocentric positions and velocities of these stars are plotted in Fig.\,\ref{fig:uvwxyz} and listed in Table~\ref{table:members}. Six candidates (two selected as kinematic members) are provisional members requiring further observation. They are also plotted in Fig.\,\ref{fig:uvwxyz} and listed in Table~\ref{table:provmembers}. The 11 stars unlikely to be members of \epscha\ (two initially selected as kinematic members) are summarised in Table~\ref{table:nonmembers} with their suggested membership assignments. All of the provisional and non-members are discussed in greater detail in Appendix~\ref{sec:candidates}, with several noteworthy confirmed members from Table~\ref{table:members}. 

\begin{table*}
\centering
\begin{minipage}{\textwidth}
\caption{Confirmed members of the \epscha\ association}
\label{table:members}
\begin{tabular}{@{}rlllcccccccccc@{}}
\hline
ID$^{\#}$ & Name & Spec. & Dist.$^{\star}$ & $K^{\dagger}$ & $U$ & $V$ & $W$ & $X$ & $Y$ & $Z$ & $T_{\rm eff}$ & $L_{\rm bol} $ & Age$^{\ddag}$\\
& & Type & [pc] & \multicolumn{4}{c}{\hrulefill\ [\kms] \hrulefill} & \multicolumn{3}{c}{\hrulefill\ [pc] \hrulefill} & [K] & [$L_{\odot}$] & [Myr]\\
\hline
(22) & CP$-$68 1388 & K1 & 112 & 2.2  & $-$11.3 & $-$21.0 & $-$7.8 & 43.5 & $-$101.9 & $-$16.7 & 5080 & 1.4 & 7.9/13.2\\
13 & 2MASS J11183572$-$7935548 & M4.5 & 101 & 4.2  & $-$8.9 & $-$23.8 & $-$11.5 & 46.2 & $-$84.5 & $-$30.4 & 3198 & 0.084 & 2.5/4.4\\
16 & 2MASS J11404967$-$7459394 & M5.5 & 101 & 5.8 & $-$13.1 & $-$17.7 & $-$5.3 & 46.7 & $-$86.7 & $-$22.3 & 3058 & 0.011 & 13.8/\dots\\
17 & 2MASS J11432669$-$7804454 & M4.7 & 117 & 1.7 & $-$10.5 & $-$21.7 & $-$8.8 & 55.2 & $-$98.2 & $-$31.6 & 3169 & 0.041 & 5.7/\dots \\
(23) & RX J1147.7$-$7842 & M3.5 & 106 & 2.9  & $-$9.6 & $-$20.6 & $-$12.4 & 50.5 & $-$88.4 & $-$29.6 & 3343 & 0.24 & 0.9/2.2\\
18 & RX J1149.8$-$7850 & M0 & 110 & 1.8  & $-$11.2 & $-$18.9 & $-$10.8 & 52.5 & $-$91.5 & $-$31.0 & 3850 & 0.44 & 2.1/2.9\\
(24) & RX J1150.9$-$7411 & M3.7 & 108 & 4.6  & $-$10.7 & $-$22.0 & $-$5.6 & 50.9 & $-$92.7 & $-$22.1 & 3314 & 0.092 &  3.9/5.2\\
(25) & 2MASS J11550485$-$7919108 &M3 & 115 & 1.8  & $-$11.2 & $-$20.6 & $-$8.1 & 55.4 & $-$95.1 & $-$33.1 & 3415 & 0.092 & 6.6/6.6\\
(26) & T Cha & K0 & 108 & 0.9 & $-$11.1 & $-$19.8 & $-$10.5 & 52.2 & $-$89.2 & $-$31.1 & 5250 & 1.2 & 12.7/20.7\\
20 & RX J1158.5$-$7754B & M3 & 119 & 2.2 & $-$11.3 & $-$18.6 & $-$11.1 & 57.5 & $-$99.3 & $-$31.5 & 3415 & 0.15 & 2.9/3.9 \\
21 & RX J1158.5$-$7754A & K4 & 104 & 4.7 & $-$13.0 & $-$17.6 & $-$6.8 & 50.3 & $-$86.8 & $-$27.5 & 4590 & 0.75 & 7.2/\dots\\
(27) & HD 104036 & A7 & 108H & 3.3 & $-$12.4 & $-$18.3 & $-$11.9 & 52.4 & $-$90.5 & $-$28.5 & 7850 & 20.0 & 2.8/7.2\\
2 & $\epsilon$ Cha AB & B9 & 111H & 2.8 &  $-$12.0 & $-$18.8 & $-$11.8 & 53.7 & $-$92.3 & $-$29.9 & 10500 & 99.9 & 2.6/2.8\\
(28) & RX J1159.7$-$7601 & K4 & 107 & 0.6 & $-$11.1 & $-$19.9 & $-$10.1 & 51.6 & $-$90.3 & $-$24.9 & 4590 & 0.50 & 14.3/14.4\\
 3  & HD 104237C & M/L & 114H & \dots & \dots & \dots & \dots &55.6 & $-$95.4 & $-$30.8 &  \dots & \dots & \dots\\
  4  & HD 104237B & K/M & 114H & \dots & \dots & \dots & \dots & 55.6 & $-$95.4 & $-$30.8 & \dots & \dots & \dots\\
5 & HD 104237A & A7.75 & 114H & 1.3 & $-$11.6 & $-$20.1 & $-$10.9 & 55.6 & $-$95.4 & $-$30.8 & 7648 & 38.6 & 3.3/3.9\\
6 & HD 104237D & M3.5 &114H & \dots & \dots & \dots & \dots & 55.6 & $-$95.4 & $-$30.8 & 3343 & 0.11 & 3.3/4.6\\
7 & HD 104237E & K5.5 & 114H & \dots & \dots & \dots & \dots & 55.6 & $-$95.4 & $-$30.8 & 4278 & 0.74 & 3.2/5.2\\
10 & 2MASS J12005517$-$7820296 & M5.75 & 126 & 4.5 & $-$12.8 & $-$18.3 & $-$6.4 & 61.2 & $-$104.7 & $-$34.2 & 3024 & 0.032 & 2.6/\dots\\
(29) & HD 104467 & G3 &102 & 2.1 & $-$12.0 & $-$18.8 & $-$9.3 & 49.6 & $-$84.4 & $-$28.7 & 5830 & 4.0 & 8.1/12.3\\
11 & 2MASS J12014343$-$7835472 & M2.25 & 100 & 5.7 & $-$8.0 & $-$24.7 & $-$12.3 & 48.6 & $-$82.9 & $-$27.5 & 3524 & 0.0029 & \dots\\
8 & USNO-B 120144.7$-$781926 & M5 & 100 & 1.5 & $-$10.8 & $-$21.4 & $-$8.7 & 48.6 & $-$83.1 & $-$27.1 & 3125 & 0.027 & 7.5/8.9\\
9 & CXOU J120152.8$-$781840 & M4.75 & 121 & 3.0 & $-$9.4 & $-$21.4 & $-$12.3 & 58.9 & $-$100.5 & $-$32.7 & 3161 & 0.042 & 5.3/7.1\\
(30) & RX J1202.1$-$7853 & M0 & 110 & 2.9 & $-$9.6 & $-$23.1 & $-$9.7 & 53.6 & $-$91.0 & $-$30.8 & 3850 & 0.43 & 2.2/3.0\\
(31) & RX J1204.6$-$7731 & M3 & 112 & 4.0 &  $-$12.9 & $-$17.6 & $-$7.8 & 54.7 & $-$93.4 & $-$28.8 & 3415 & 0.22 & 1.4/2.5\\
(32) & RX J1207.7$-$7953 & M3.5 & 111 & 1.2 &  $-$10.5 & $-$20.6 & $-$11.0 & 54.5 & $-$91.0 & $-$32.8 & 3343 & 0.11 & 3.2/4.5\\
(33) & HD 105923 & G8 & 112 & 0.7 &  $-$10.7 & $-$21.0 & $-$9.6 & 54.9 & $-$96.2 & $-$16.7 & 5520 & 3.3 & 5.6/10.5\\
(34) & RX J1216.8$-$7753 & M4 & 118 & 0.6 & $-$10.8 & $-$20.1 & $-$10.5 & 58.8 & $-$97.6 & $-$30.8 & 3270 & 0.17 & 1.2/2.9\\
(35) & RX J1219.7$-$7403 & M0 & 112 & 0.6 &  $-$11.0 & $-$20.7 & $-$9.4 & 56.1 & $-$94.4 & $-$22.0 & 3850 & 0.28 & 5.5/5.9\\
(36) & RX J1220.4$-$7407 & M0 & 110 & 2.2 &  $-$11.9 & $-$19.8 & $-$8.1 & 55.2 & $-$92.7 & $-$21.7 & 3850 & 0.36 & 3.3/4.0\\
(37) & 2MASS J12210499$-$7116493 & K7 &110 & 2.0 & $-$11.8 & $-$18.7 & $-$10.2 & 55.2 & $-$93.7 & $-$16.4 & 4060 & 0.53 & 3.0/4.5\\
(38) & RX J1239.4$-$7502 & K3 & 100 & 1.2 & $-$10.7 & $-$21.2 & $-$9.0 & 52.0 & $-$82.8 & $-$21.1 & 4730 & 0.96 & 6.5/\dots\\
(39) & CD$-$69 1055 & K0 & 99 & 0.6 & $-$10.7 & $-$21.0 & $-$9.8 & 54.2 & $-$81.8 & $-$13.1 &  5250 & 1.5 & 9.2/15.6\\
(40) & MP Mus & K1 & 101 & 0.2 & $-$10.7 & $-$20.6 & $-$9.9 & 58.4 & $-$81.5 & $-$12.2 & 5080 & 1.3 & 8.5/14.1\\
\hline
\end{tabular}\\
(\#): \epscha\ identification number (SIMBAD: [FLG2003] EPS CHA \#). Bracketed values are new members confirmed in this study, ordered by increasing RA.\\
($\star$): Suffix `H' denotes a trigonometric distance from \emph{Hipparcos}. All other distances are kinematic.\\
($\dagger$): Difference between the star and mean \epscha\ space motion, see \S\ref{sec:membanalysis}.\\
($\ddag$): Ages estimated from the \citet{Dotter08} and \citet{Siess00} models, respectively.\\
\end{minipage}
\end{table*}

\begin{table*}
\begin{minipage}{\textwidth}
\caption{Provisional members of the \epscha\ association requiring confirmation}
\label{table:provmembers}
\begin{tabular}{@{}rlllcccccccccc@{}}
\hline
ID & Name & Spec. & Dist.$^{\star}$ & $K^{\dagger}$ & $U$ & $V$ & $W$ & $X$ & $Y$ & $Z$ & $T_{\rm eff}$ & $L_{\rm bol} $ & Age$^{\ddag}$\\
& & Type & [pc] & \multicolumn{4}{c}{\hrulefill\ [\kms] \hrulefill} & \multicolumn{3}{c}{\hrulefill\ [pc] \hrulefill} & [K] & [$L_{\odot}$] & [Myr]\\
\hline
& TYC 9414-191-1 & K5 & 105 & \dots & \dots & \dots & \dots & 47.4 & $-$88.9 & $-$29.7 & 4350 & 1.2 & 1.6/3.1 \\
1 & CXOU J115908.2$-$781232 & M4.75 & 165 & 5.2  & $-$10.2 & $-$22.4 & $-$5.1 & 79.9 & $-$137.4 & $-$44.4 & 3161 & 0.055 & 3.7/5.8\\
 & RX J1202.8$-$7718$^{\#}$ & M3.5 &128 & 3.0 & $-$9.7 & $-$23.1 & $-$9.7 & 62.3 & $-$107.0 & $-$32.5 & 3343 & 0.14 & 2.2/\dots\\
& HD 105234 & A9 & 103H & \dots & \dots & \dots & \dots & 50.5 & $-$85.1 &$-$28.5 & 7390 & 9.4 & 10.0/\dots \\
& HIP 59243 & A6 & 94H & \dots & \dots & \dots & \dots & 46.2 & $-$77.5 & $-$26.0 & 8350 & 15.8 & 7.9/9.0\\
 & CM Cha$^{\#}$ & K7 & 133 & 5.2 & $-$9.3 & $-$23.9 & $-$6.4 & 71.4 & $-$107.6 & $-$31.7 & 4060 & 0.49 & 3.6/5.1\\
\hline
\end{tabular}\\
(\#): Selected as kinematic member (see Appendix~\ref{sec:candidates}).\\
($\star$): Suffix `H' denotes a trigonometric distance from \emph{Hipparcos}. All other distances are kinematic.\\
($\dagger$): Difference between the star and mean \epscha\ space motion, see \S\ref{sec:membanalysis}.\\
($\ddag$): Ages estimated from the \citet{Dotter08} and \citet{Siess00} models, respectively.\\
\end{minipage}
\end{table*}

\begin{table}
\caption{Candidates from the literature rejected as \epscha\ members}
\label{table:nonmembers}
\begin{tabular}{llll}
\hline
ID & Name & Reason$^{\dagger}$ & Membership\\
\hline
& HD 82879$^{\#}$/82859 & Binary & ? Young\\
& \rxjeleventhirtyseven\ & $\Delta M_{I}$, dist, Li & Old field\\
&HIP 55746$^{\#}$ & \dots & AB Dor\\
15 & 2MASS J11334926$-$7618399 & dist & Cha I\\
&VW Cha & $\Delta M_{I}$, dist & Cha I\\
& \rxjtwelvefortythree & dist & Cha II\\
14 & RX J1123.2$-$7924  & $\Delta M_{I}$, Li & Octans?\\
12 & 2MASS J12074597$-$7816064 & $\Delta M_{I}$ & ? Young\\
19 & \rxjelevenfiftyfour\ & $\Delta M_{I}$ & ? Young\\
& TYC 9238-612-1 & $\Delta M_{I}$ & ?\\
& TYC 9420-676-1 & $\Delta M_{I}$ & ?\\
\hline
\end{tabular}
(\#): Selected as kinematic member (see Appendix~\ref{sec:candidates}).\\
($\dagger$): $\Delta M_{I}$: Photometry $>$1~mag from isochrone, dist = bad kinematic distance, Li = incongruous Li\,\textsc{i} $\lambda$6708 equivalent width.
\end{table}

\subsection{Age of \epscha}\label{sec:ages}

\begin{figure}
   \centering
   \includegraphics[width=\linewidth]{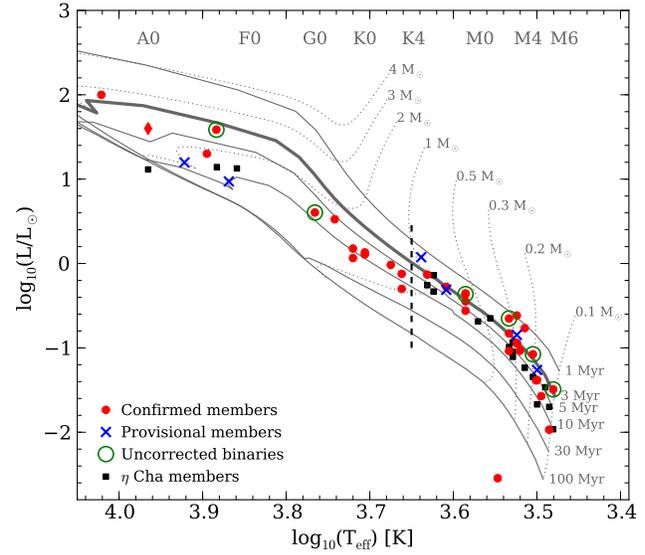} 
   \caption{HR diagram of confirmed \epscha\ members (red circles), provisional members (crosses) and \echa\ members (squares). Uncorrected \epscha\ binaries are circled. The filled diamond is \epscha\ B \citep{Feigelson03}. Pre-main sequence mass tracks and isochrones from \citet{Dotter08} are plotted for comparison. The dashed line at $\log_{10}T_{\rm eff}=3.65$ is the limit of the low-mass sample (see text).}
   \label{fig:hrd}
\end{figure}

Fig.\,\ref{fig:hrd} shows the HR diagram for confirmed and provisional members of \epscha. The effective temperature of each star was obtained from its spectral type using the temperature scales of \citet{Kenyon95} for spectral types earlier than M1 and \citet{Luhman03} for M1--M6.  For TYC 9414-191-1 we adopted a spectral type of K5 ($A_{V}=0.9$) which is consistent with its $B-V$ colour and position in Fig.\,\ref{fig:2mass}. Luminosities were calculated using de-reddened $J$-band magnitudes and the bolometric corrections (BC) of \citet{Kenyon95}. We adopted $M_{\rm bol, \odot}=4.64$, appropriate for the \citeauthor{Kenyon95} BC scale.  Pre-main sequence isochrones and evolutionary tracks from \citet{Dotter08} are plotted for comparison.\footnote{Although isochrones are only provided for $t>250$~Myr, the tracks contain the full pre-main sequence evolution. We created isochrones from 0.9--100~Myr ($\Delta \log t=0.1$~Myr) by interpolating the tracks over 0.1--5~\msun.} The four early type members fall around the 3--5~Myr isochrones, as do the majority of late-type stars. However, the solar-type members appear systematically older in this diagram, with ages of 5--10~Myr. Such mass-dependent age differences are unexpected and probably result from systematic errors in the models and/or temperature/luminosity scales. 

\begin{figure}
   \centering
   \includegraphics[width=\linewidth]{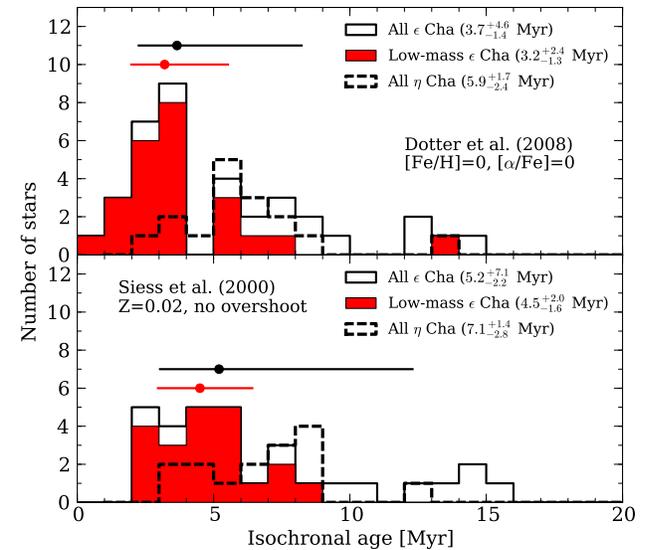} 
   \caption{Ages derived from the \protect\cite{Dotter08} (top panel) and \protect\citet{Siess00} (bottom panel) isochrones.  Stars with $\log_{10}T_{\rm eff}<3.65$ are shown in red, the solid line is the envelope of all members. Dashed lines are \echa\ members. Median ages are given in the legend and by the horizontal error bars (\epscha\ only), with their 68 per cent confidence intervals.}
   \label{fig:ages}
\end{figure}

Individual ages were obtained by interpolating the isochrones at the position of each star.  The resulting age distribution is given in the top panel of Fig.\,\ref{fig:ages}. The median age of all members is $3.7_{-1.4}^{+4.6}$~Myr.\footnote{As a measure of dispersion we adopted the 16$^{\rm th}$ and 84$^{\rm th}$ percentiles in the cumulative age distribution.} If we restrict the sample to $\log T_{\rm eff}<3.65$  (K4 spectral type and later) to exclude the systematically older solar-type stars, the dispersion is much reduced and the median age is $3.2_{-1.3}^{+2.4}$~Myr. Excluding uncorrected binaries lowers the median low-mass age and dispersion by only 0.1~Myr. This age agrees with previous estimates using fewer stars \citep{Fang13} and different model grids \citep{Feigelson03,Luhman04}. Monte Carlo simulations along the 3~Myr model isochrone with realistic errors ($\sigma_{T_{\rm eff}}=100$~K, $\sigma_{\log L/L_\odot} = 0.05$) show that the observed dispersion around the median age is consistent with measurement errors alone, i.e. the low-mass members of \epscha\ appear coeval.  

As a comparison we also calculated ages using the isochrones of \citet{Siess00}. The age distribution (Fig.\,\ref{fig:ages}, bottom panel) is very similar and both sets of models together indicate a median low-mass age of 3--5~Myr. This makes \epscha\ one of the youngest groups in the solar neighbourhood ($\lesssim$150~pc) and its members ideal targets for high-sensitivity studies of young stars, discs and nascent planetary systems. Temperatures, luminosities and ages for all confirmed or provisional members are listed in Tables~\ref{table:members} and \ref{table:provmembers}.

\subsubsection{Age of \echa}

\begin{figure}
   \centering
   \includegraphics[width=\linewidth]{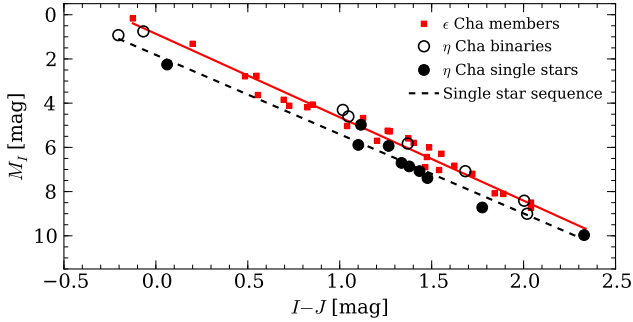} 
   \caption{Colour-magnitude diagram with confirmed members of \epscha\ (red squares) and the eight binary \protect\citep[][open circles]{Lyo04} and 10 single members of \echa\ (filled circles). As predominantly equal-mass systems the binaries follow a sequence $\sim$0.75~mag above the single stars. }
   \label{fig:etacha_comparison}
\end{figure}

As they are commonly cited as being coeval \citep[e.g.][]{Torres08} it is worthwhile to compare the ages of $\eta$ and $\epsilon$ Cha. We conducted a similar analysis using the 18 core members of \echa, adopting spectral types from \citet{Luhman04a}, a distance of 94~pc and correcting the photometry of the equal-mass binaries RECX~1, 9 and 12. The results suggest a small (1--3 Myr) age difference, with \echa\ having median ages of $5.9_{-2.4}^{+1.7}$ Myr \citep{Dotter08} and $7.1_{-2.8}^{+1.4}$~Myr \citep{Siess00}. These agree with previous estimates \citep{Lawson01,Luhman04a}. A small age difference is also apparent in the CMD (Fig\,\ref{fig:etacha_comparison}), where the sequence of single (or high mass-ratio binary) \echa\ members lies slightly below the \epscha\ sequence. Comparison against theoretical models again suggests an age difference of 1--3 Myr.

\begin{figure}
   \centering
   \includegraphics[width=\linewidth]{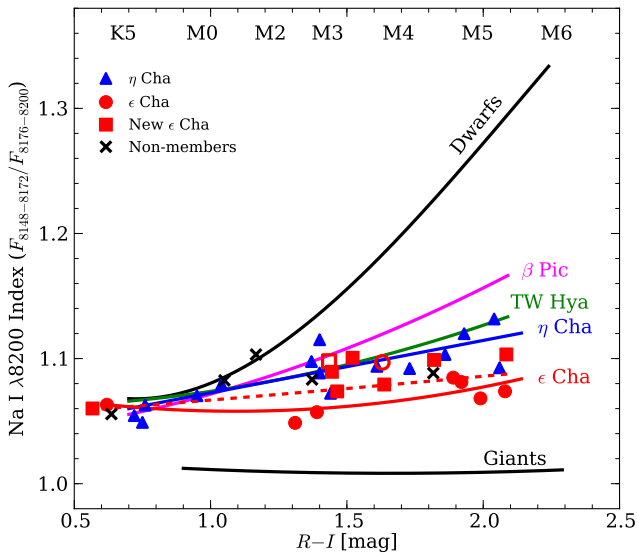} 
   \caption{Na\,\textsc{i} $\lambda$8200 gravity indices of young associations. Solid lines are mean trends from \citet{Lawson09b}, with $\eta$ and \epscha\ members from \citet{Lyo04a,Lyo08} (filled circles, triangles) and new \epscha\ members observed with WiFeS (filled squares). Open symbols are the non-member 2MASS J12074597$-$7816064 (\epscha\ 12) and provisional member \rxjtwelvetwo. The dashed line is the new fit to all \epscha\ members.}
   \label{fig:gravity}
\end{figure}

The strength of the Na\,{\sc i} $\lambda$8183/8195 doublet is highly dependent on surface gravity in mid-to-late M-type stars. It can therefore be used as an age proxy for pre-main sequence stars contracting towards their main sequence radii. \citet{Lawson09b} used Na\,{\sc i} strengths to rank the ages of several young associations. Their mean trends (Fig.\,\ref{fig:gravity}) agree with the isochronal and lithium depletion age rankings of the groups. Furthermore, they resolved $\eta$ and \epscha\ in gravity, with the former appearing several Myr older. We computed the same Na\,{\sc i} index after smoothing the WiFeS $R$3000 spectra to the $R\approx800$ resolution of the \citeauthor{Lawson09b}\ data and resampling to the same wavelength scale. Despite the large scatter, the revised \epscha\ trend is still clearly younger than \echa, confirming the previous HRD and CMD analyses. 

\subsection{Circumstellar discs and accretion in \epscha}\label{sec:discs}

\begin{figure*}
   \centering
   \includegraphics[width=0.245\textwidth]{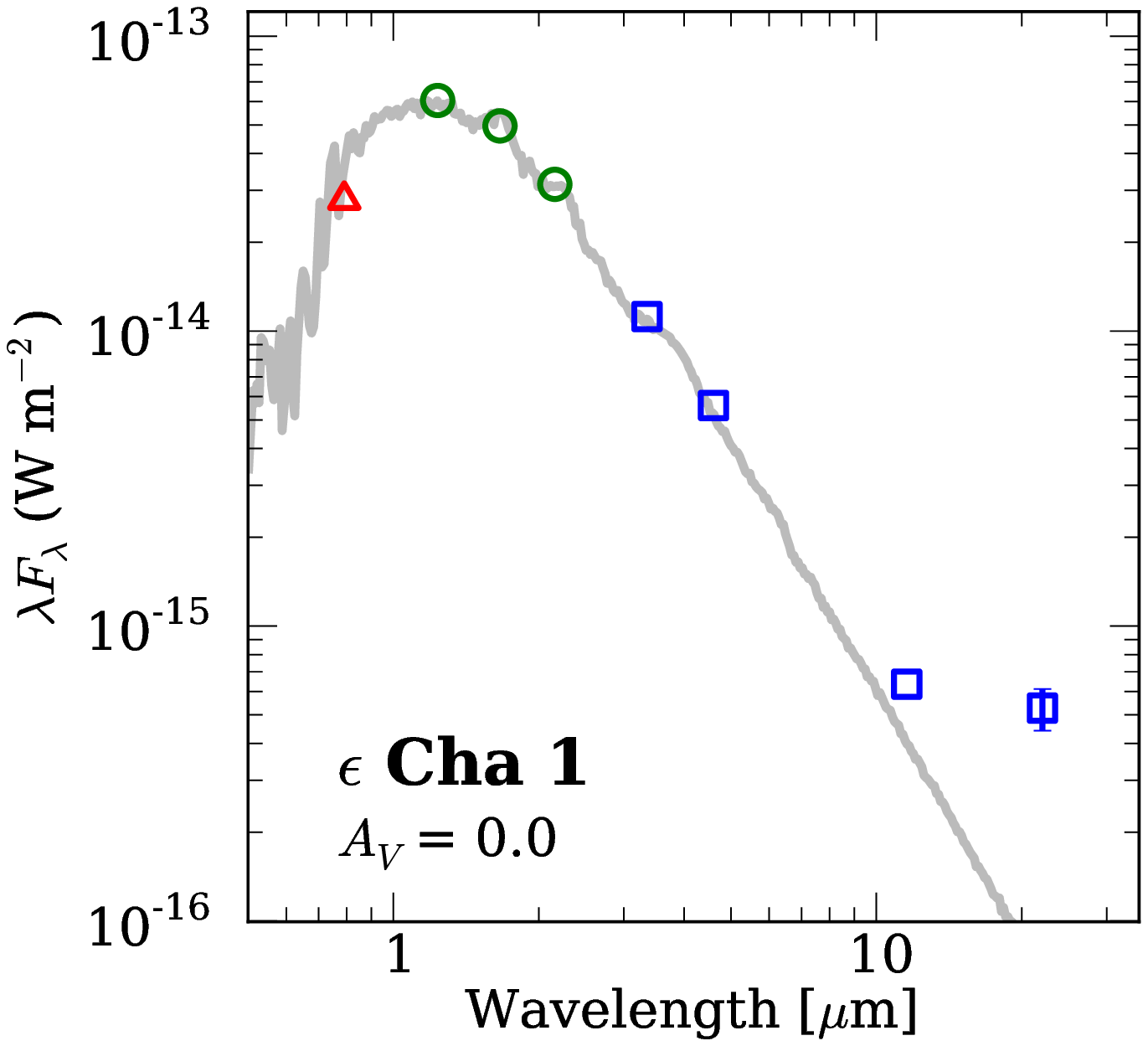}
   \includegraphics[width=0.245\textwidth]{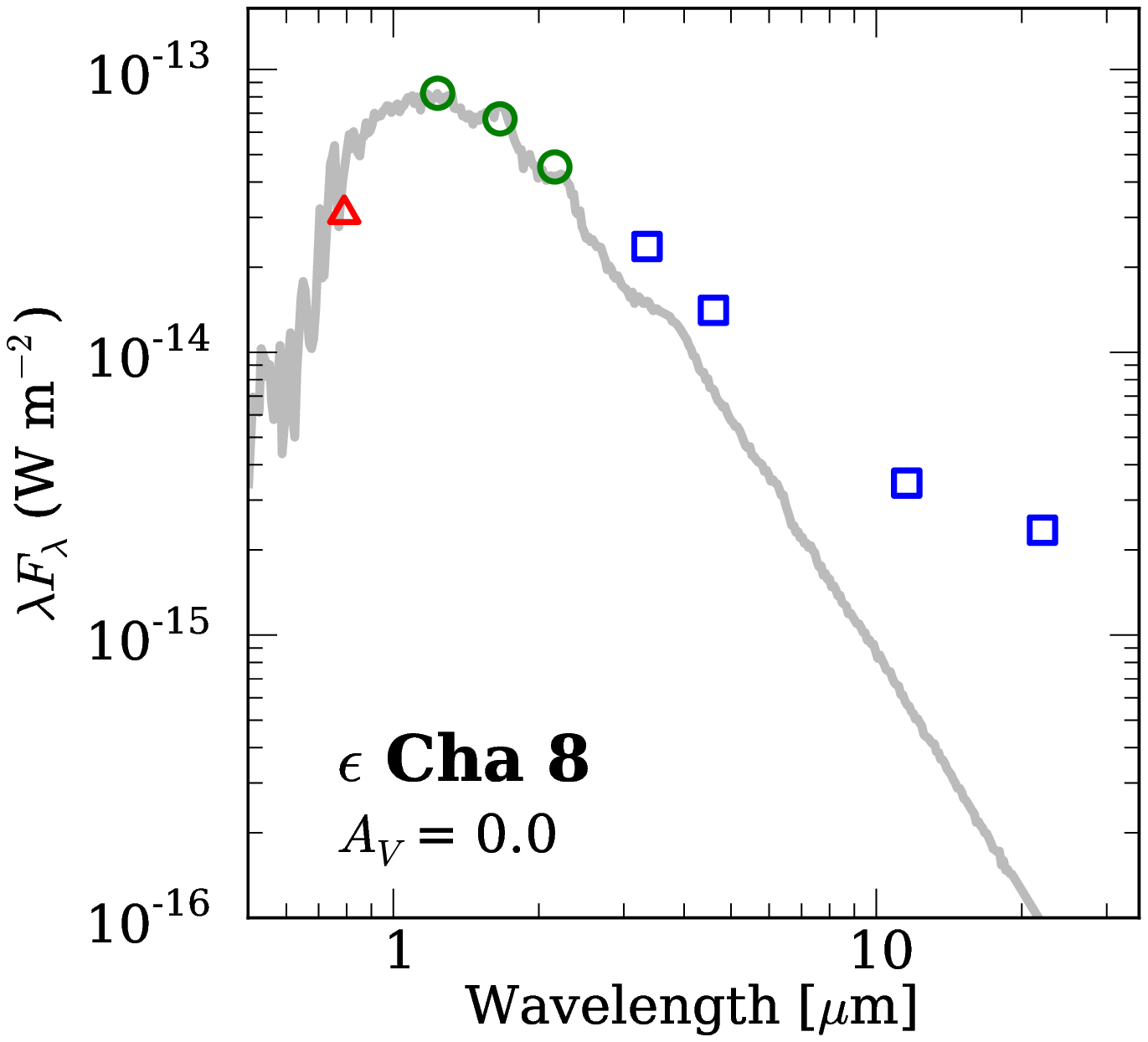}
   \includegraphics[width=0.245\textwidth]{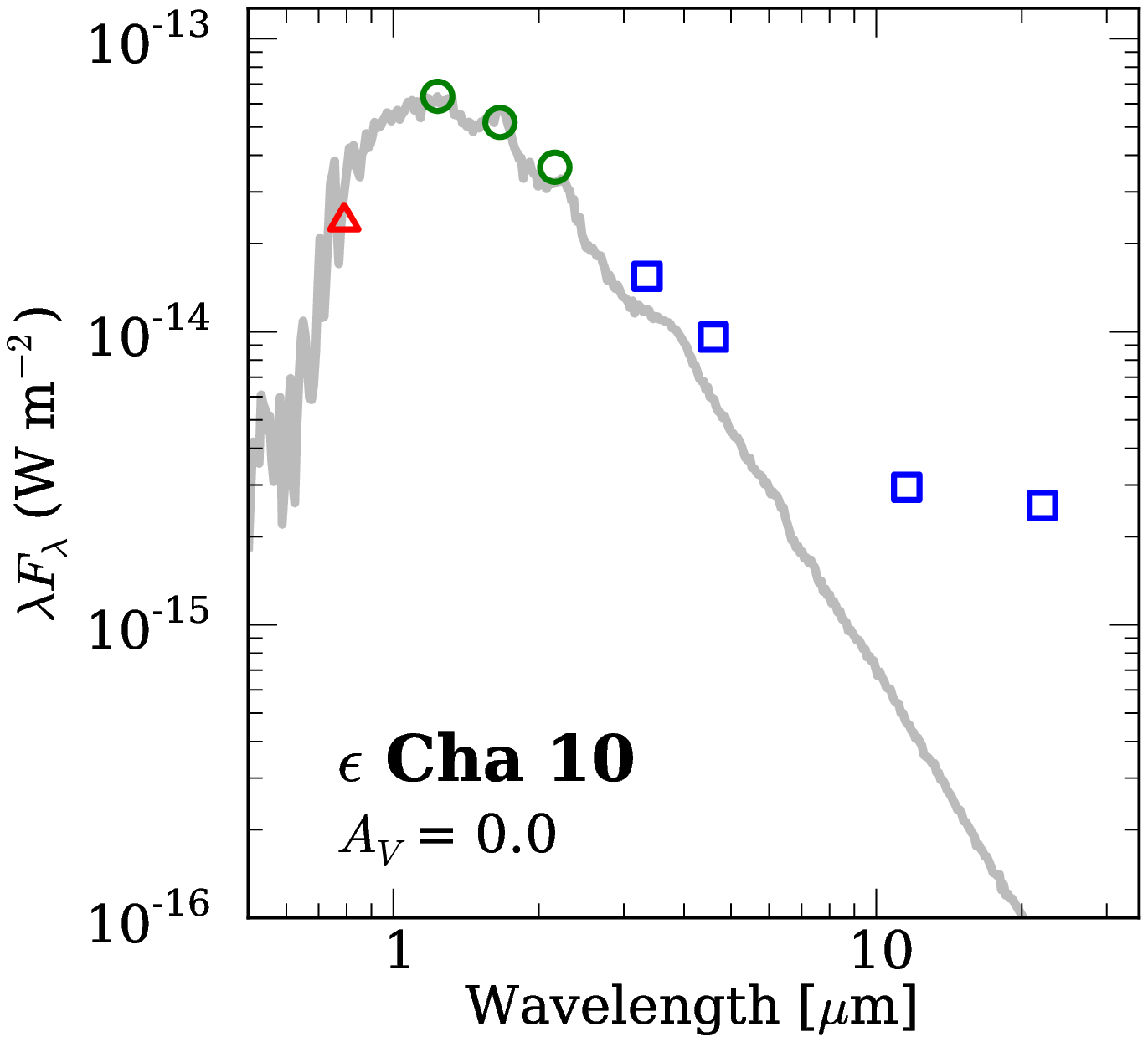}
   \includegraphics[width=0.245\textwidth]{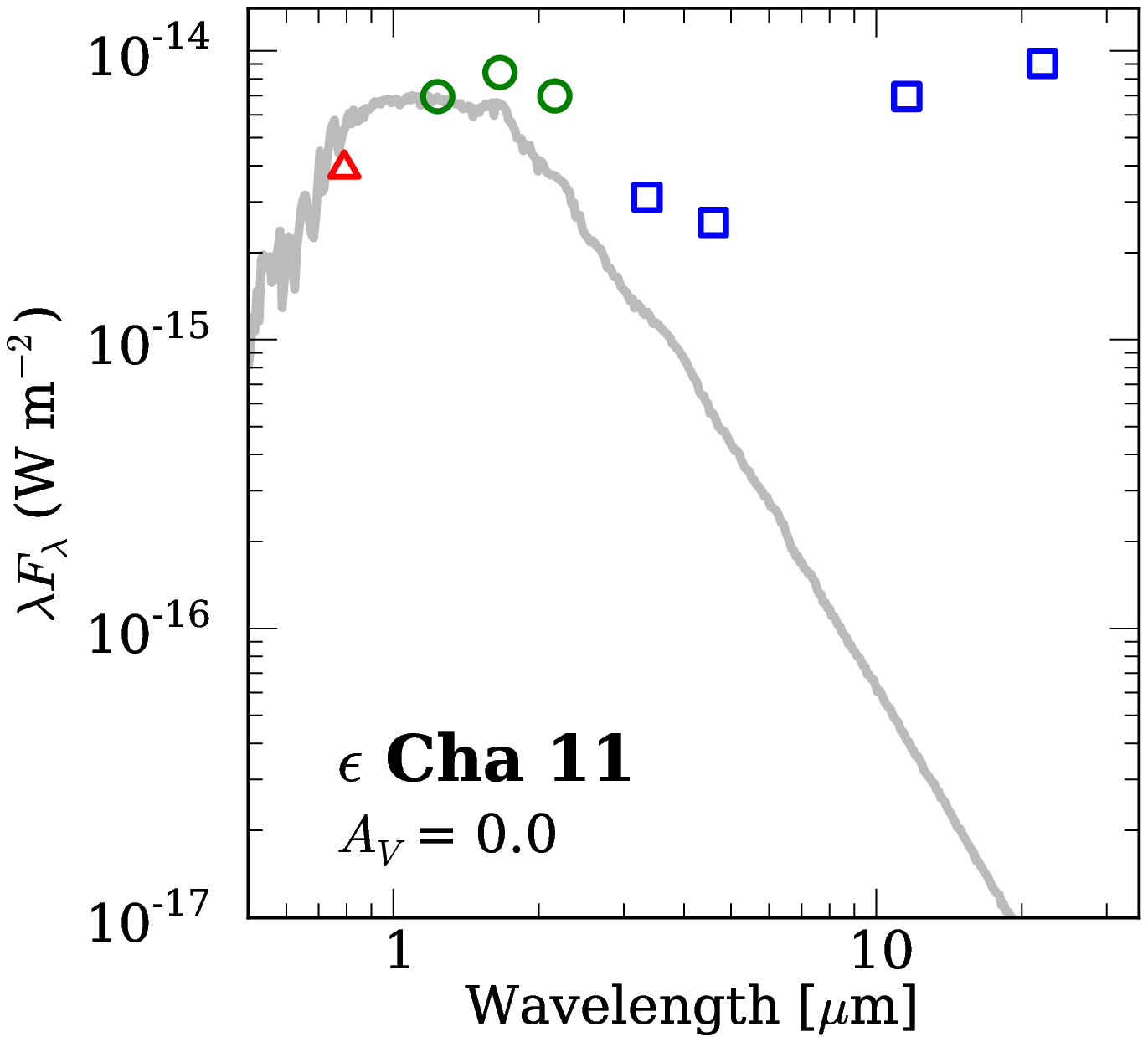}\\
      \includegraphics[width=0.245\textwidth]{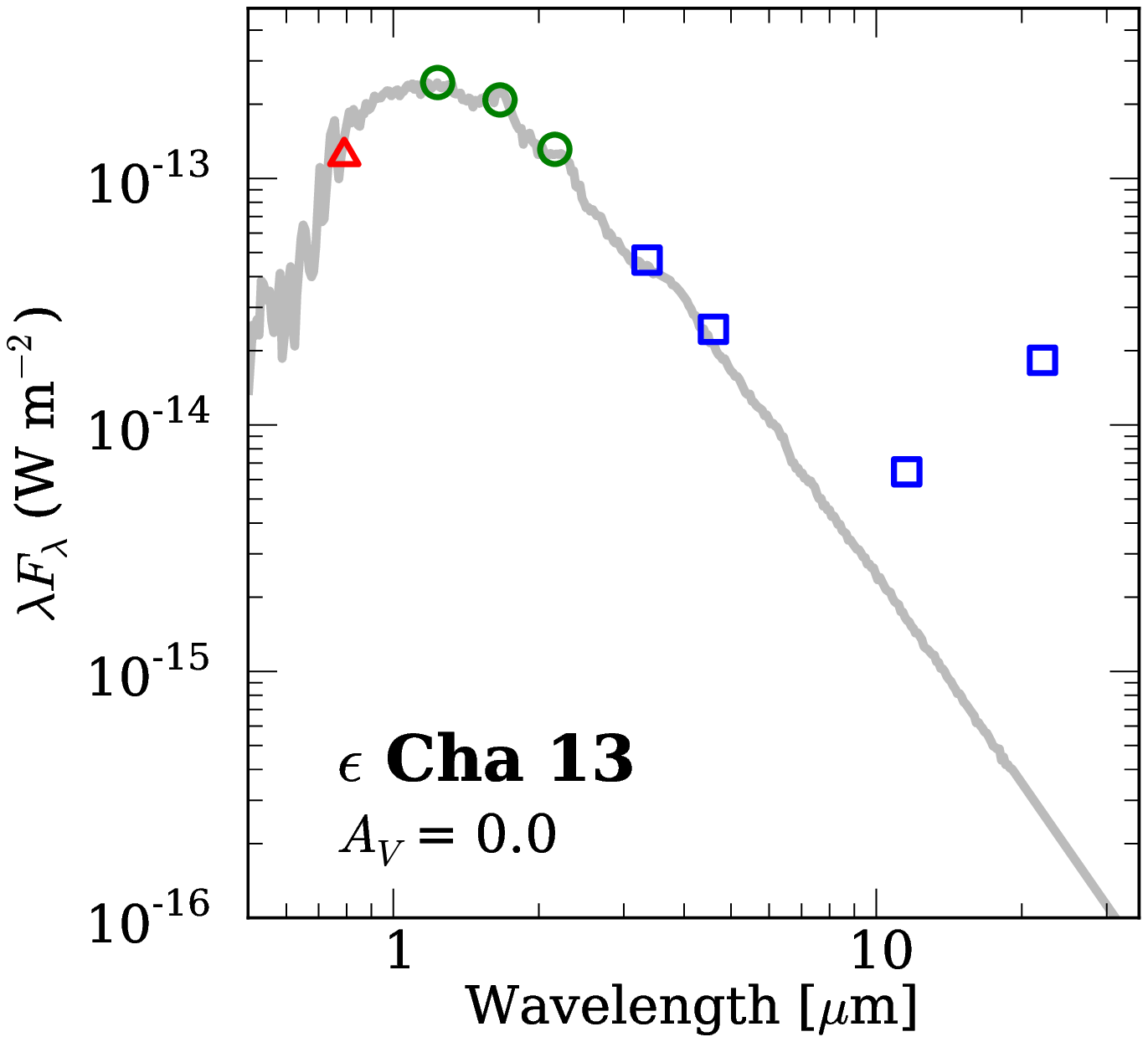}
   \includegraphics[width=0.245\textwidth]{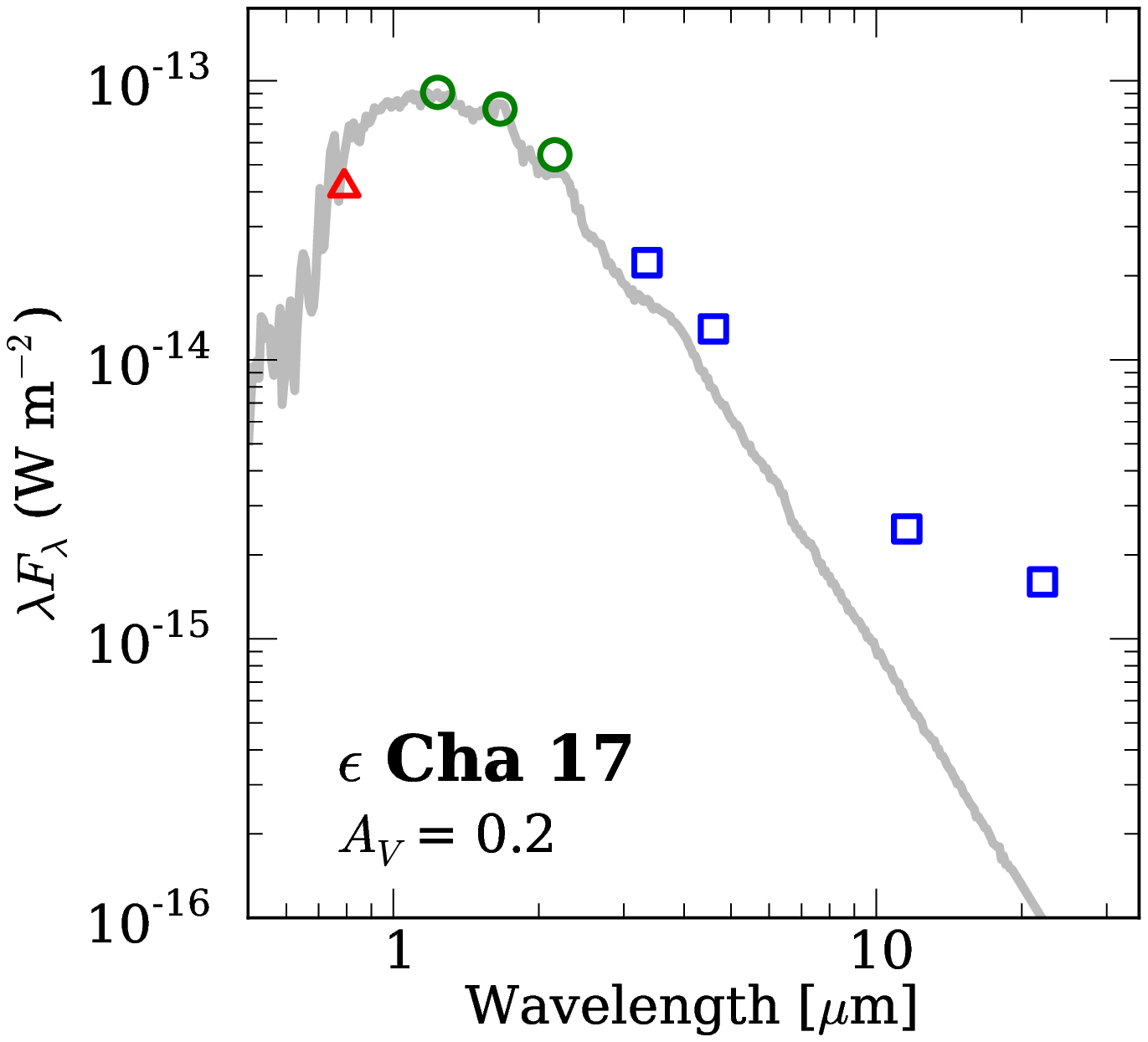}
      \includegraphics[width=0.245\textwidth]{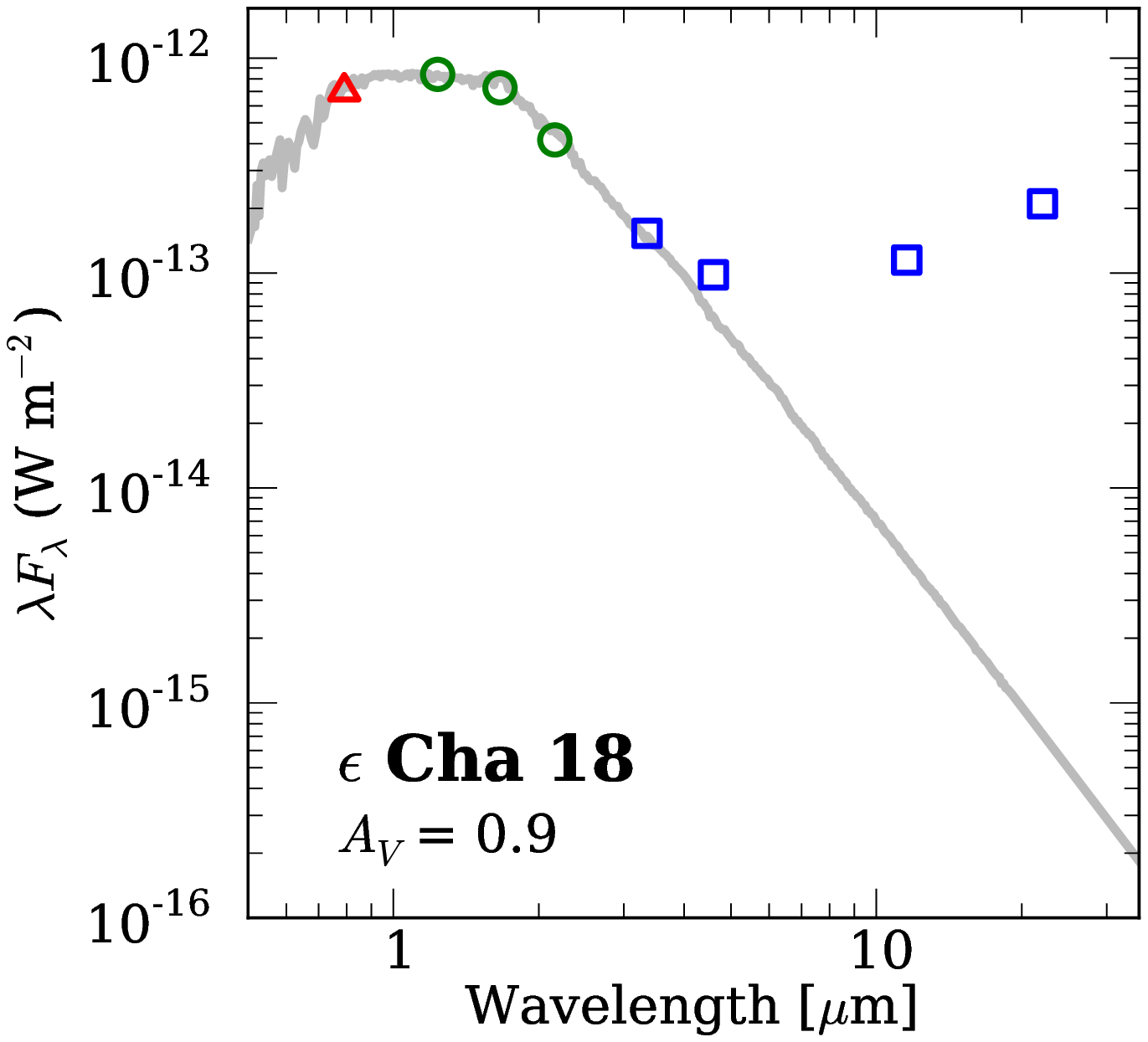}
   \includegraphics[width=0.245\textwidth]{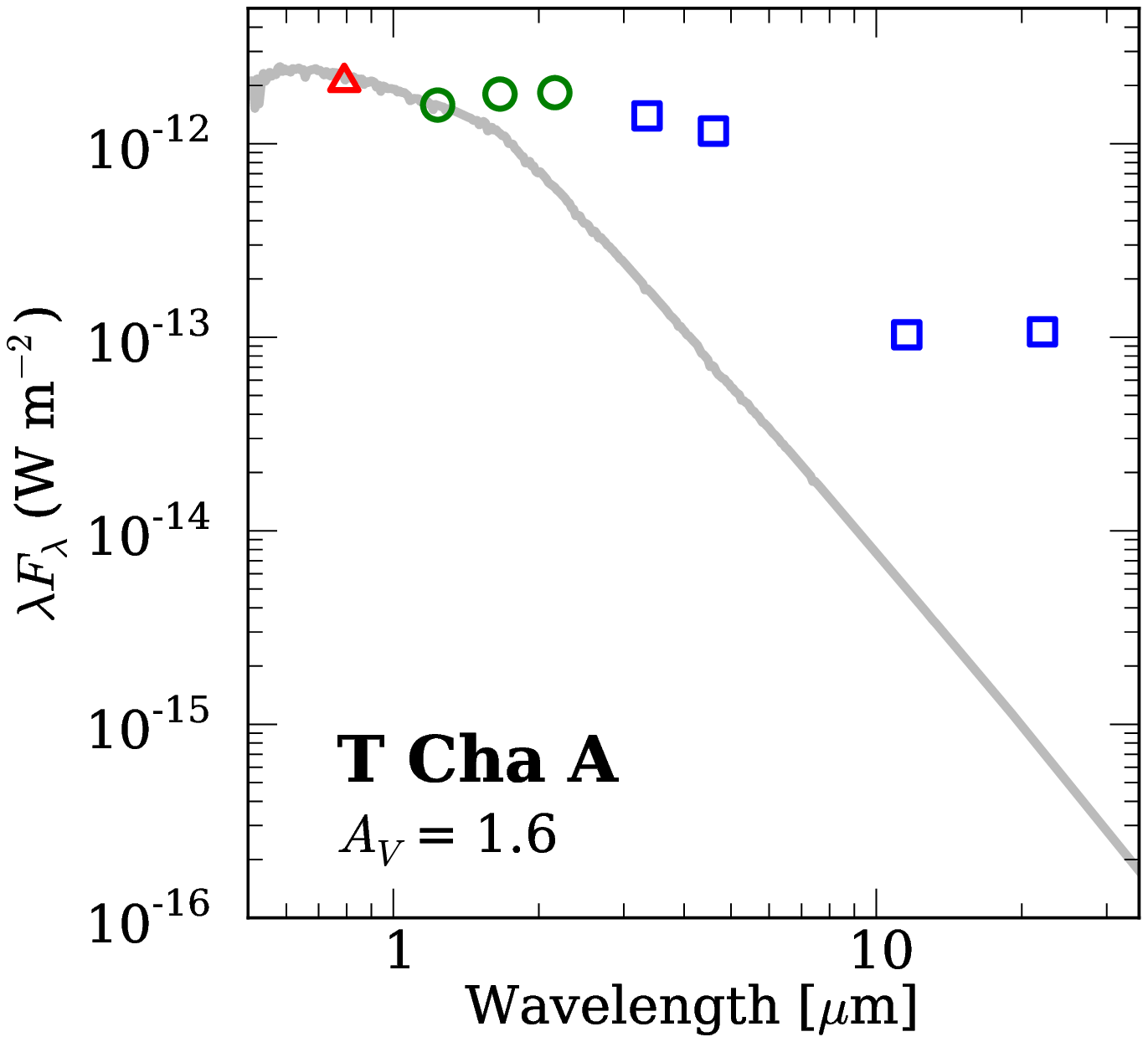}\\
   \includegraphics[width=0.245\textwidth]{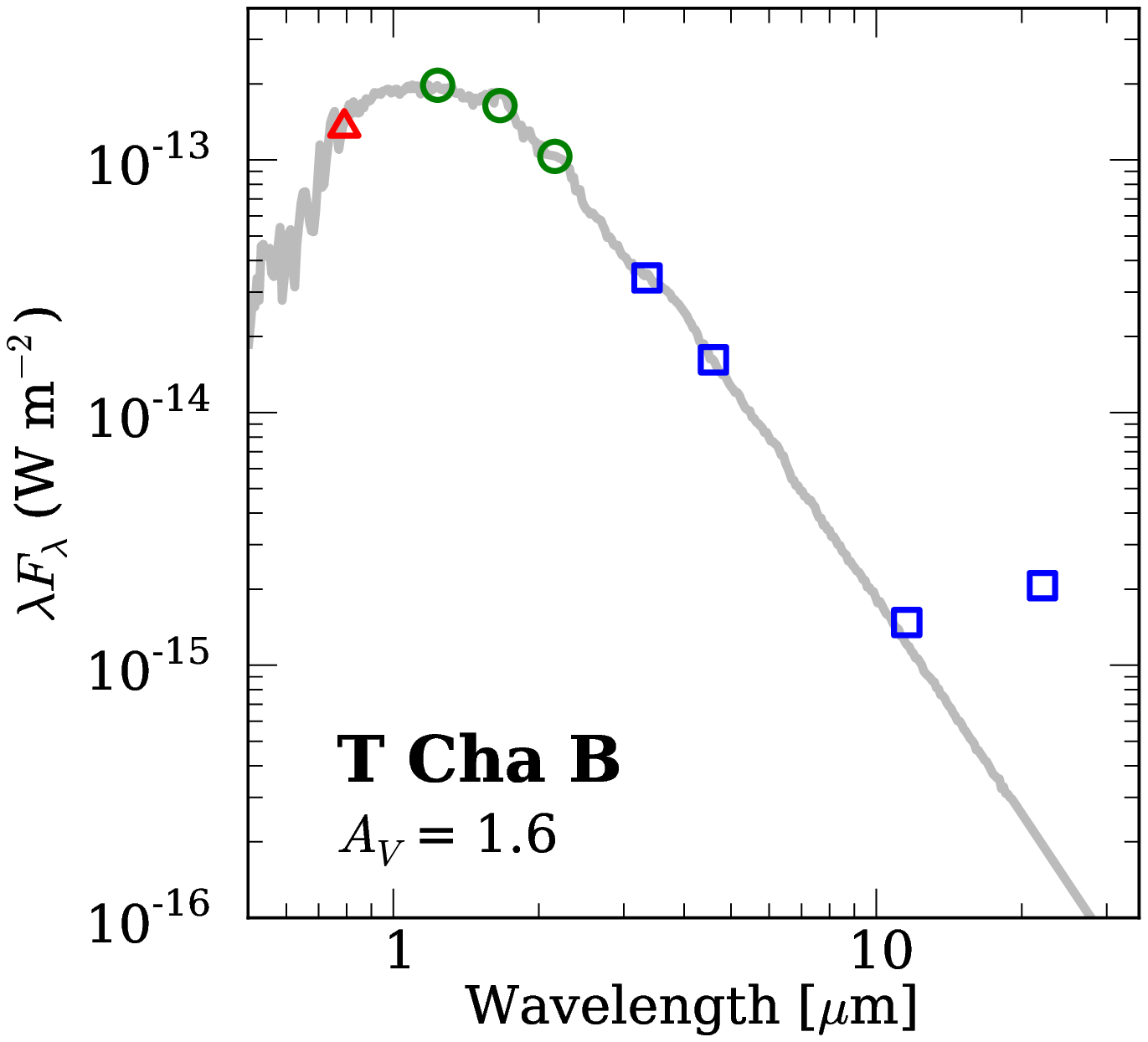}
   \includegraphics[width=0.245\textwidth]{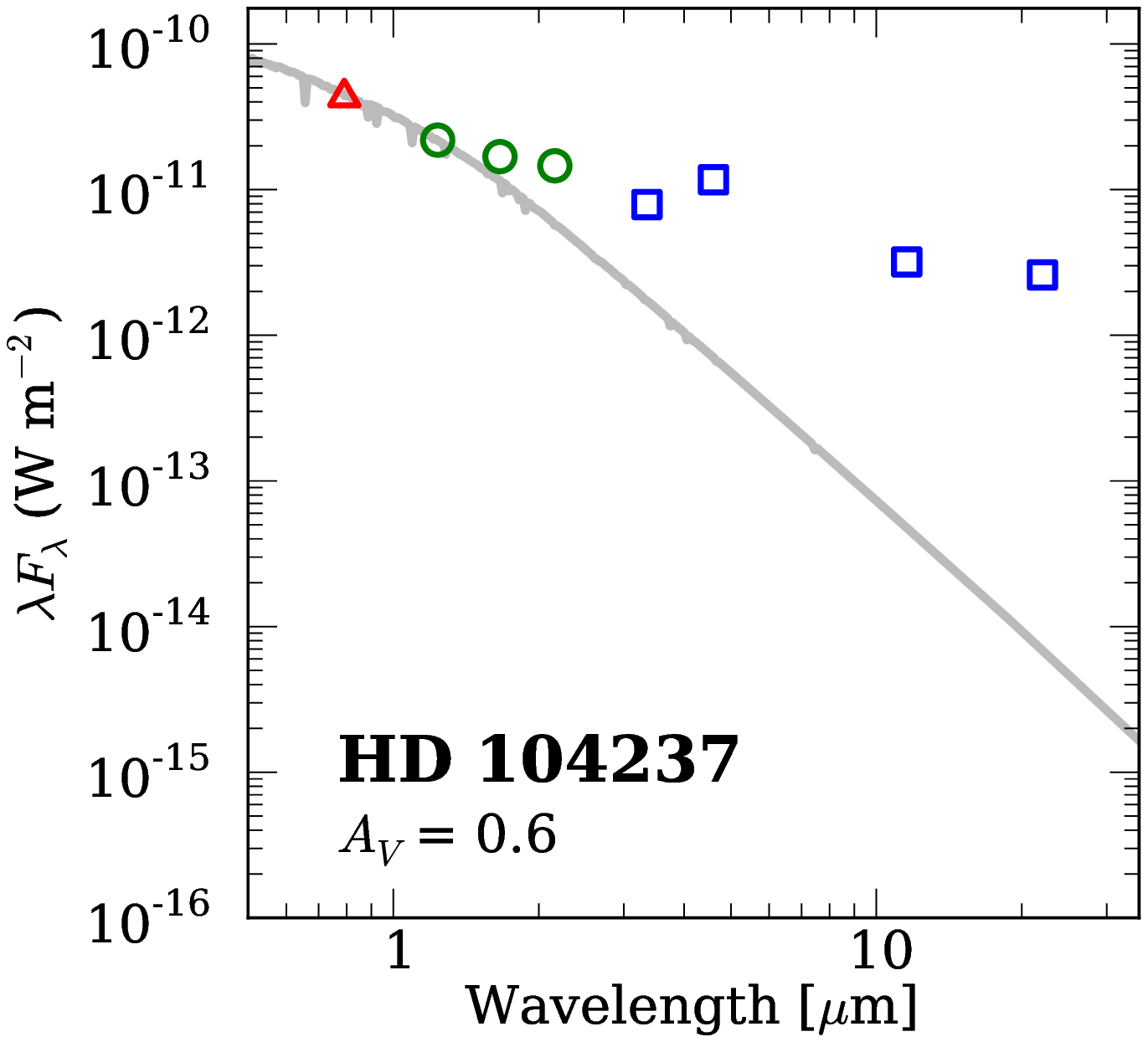}
      \includegraphics[width=0.245\textwidth]{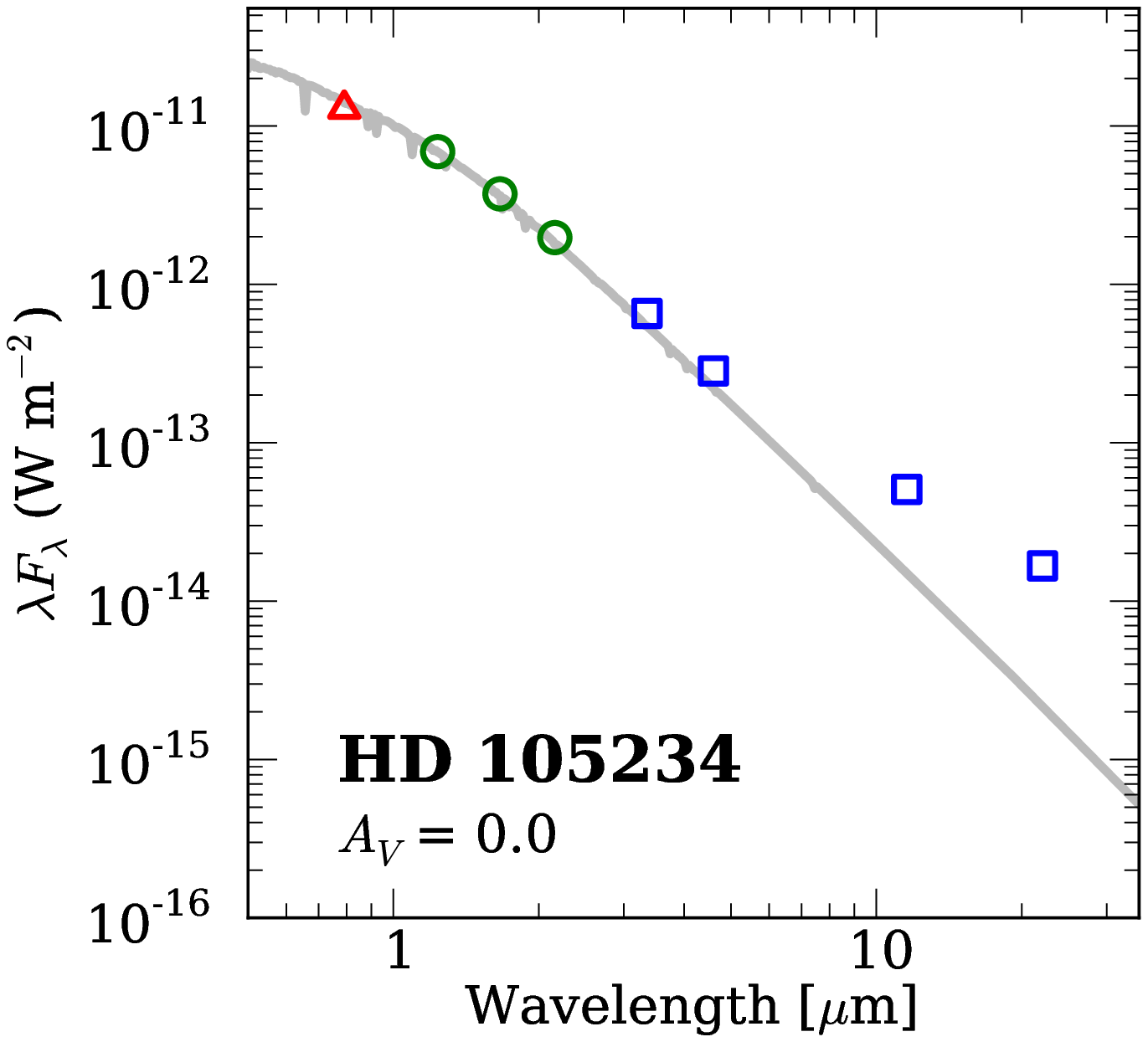}
   \includegraphics[width=0.245\textwidth]{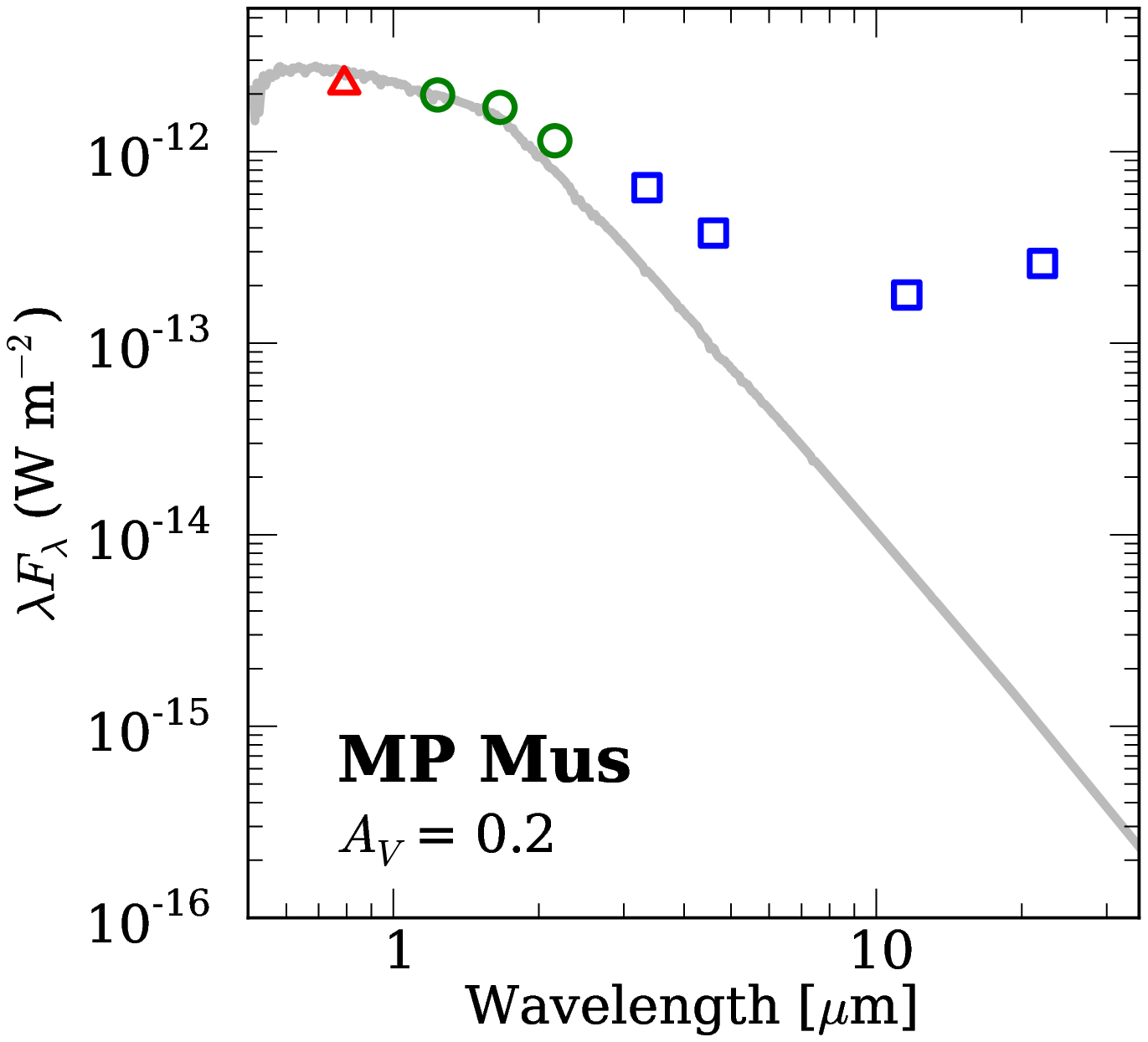}\\
   \includegraphics[width=0.245\textwidth]{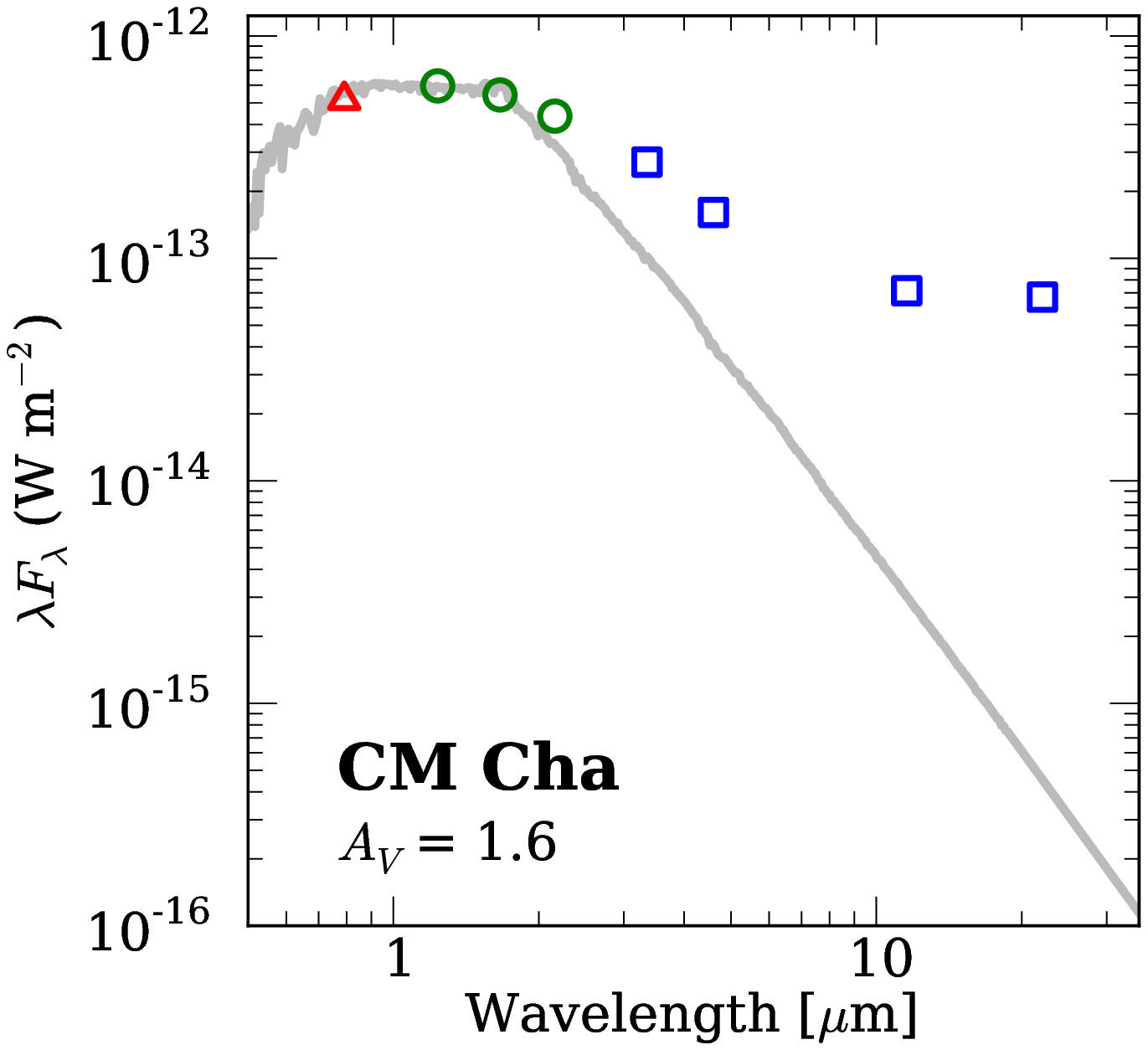}
   \includegraphics[width=0.245\textwidth]{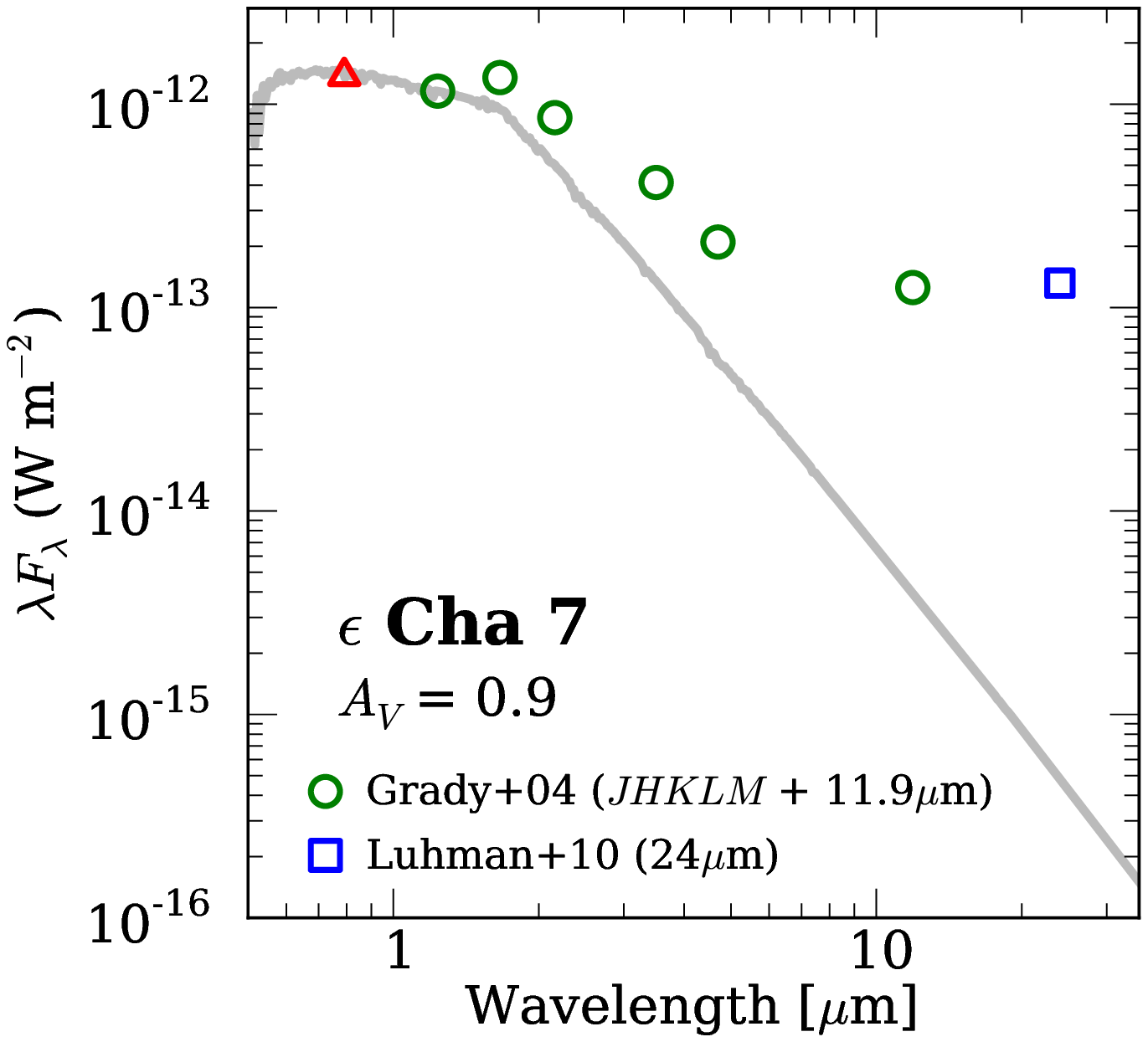}
      \includegraphics[width=0.245\linewidth]{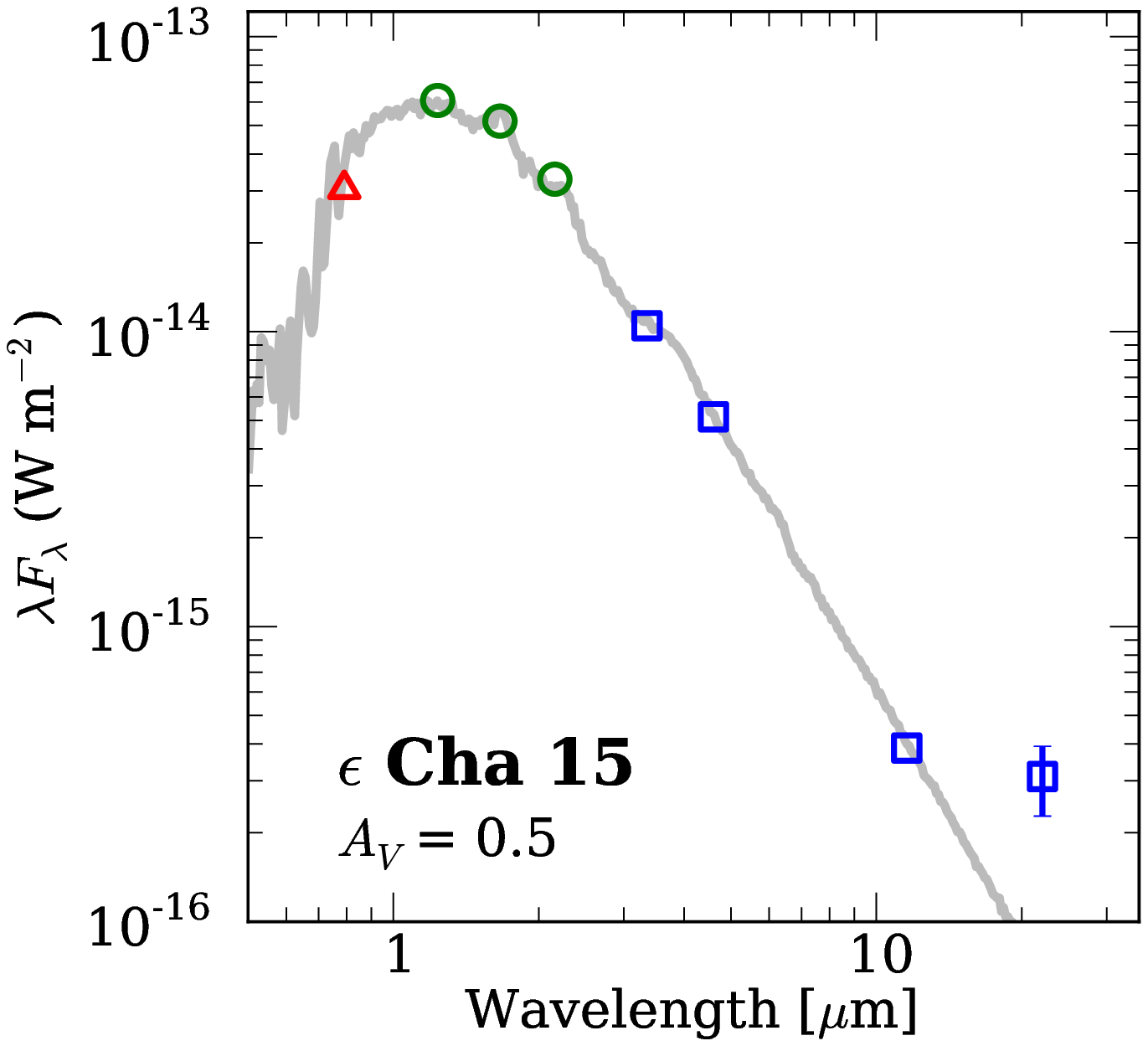}
   \includegraphics[width=0.245\linewidth]{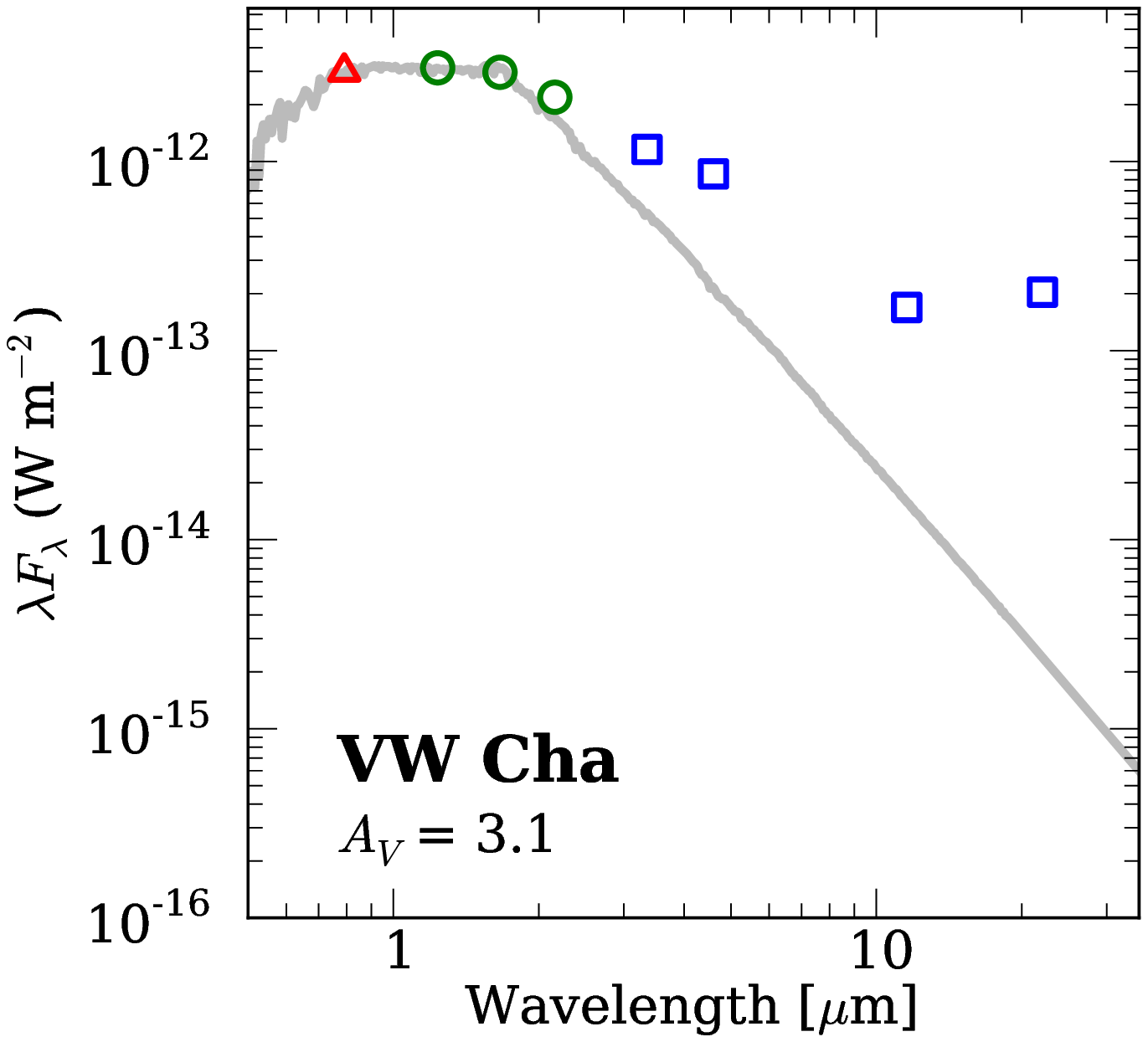}
     \caption{De-reddened DENIS ($i$, red triangle), 2MASS ($JHK_{s}$, green circles) and \emph{WISE} ($W1$--$W4$, blue squares) 0.7--22~\micron\ spectral energy distributions of \epscha\ members with an infrared excess. Isophotal wavelengths and zero magnitude fluxes were taken from \citet{Fouque00}, \citet{Cohen03} and \citet{Jarrett11}. Flux errors are within the plotted points unless shown. The photospheric flux (grey line) is approximated by a solar metallicity, $\log g = 4.5$ MARCS model \citep{Gustafsson08} with similar effective temperature scaled to the $J$-band magnitude of each star. Photometry for \epscha\ 7 (HD 104237E) comes from \citet{Grady04} and \citet{Luhman10}. \epscha\ 15 and VW Cha are kinematic members of Cha I.}
          \label{fig:epschased}
\end{figure*}

To check for circumstellar disc emission we queried the recent  \emph{Wide-field Infrared Survey Explorer} (\emph{WISE}) All Sky Data Release \citep{Wright10}. The components of HD~104237 were not resolved individually at the 6--12~arcsec resolution of \emph{WISE}. Spectral energy distributions (SEDs) of the 13 members with excess emission at wavelengths shorter than 22~\micron\ are plotted in Fig.\,\ref{fig:epschased}. Also plotted is the SED of HD 104237E (\epscha\ 7), which has a clear disc excess from 3--24~\micron\ photometry reported by \citet{Grady04} and \citet{Luhman10}. The former also reported strong excess $K_{s}$ and $L'$ emission from HD 104237B (\epscha\ 4). The variety of SED morphologies in Fig.\,\ref{fig:epschased} is typical of rapidly-evolving intermediate-age discs \citep{Williams11} and the manifold physical processes (photo-evaporation, accretion, grain growth, dynamical interactions with companions) shaping them. A similar range of discs are observed in \echa\ \citep{Sicilia-Aguilar09}. While a detailed analysis of the disc and dust properties of \epscha\ members is outside the scope of this work, we briefly describe some germane results from the literature below.

\citet{Fang13} presented \emph{Spitzer Space Telescope} 7--35~\micron\  spectroscopy of ten members (\epscha\ 1,2,5--12). Their spectrum of  CXOU J115908.2$-$781232 (\epscha\ 1) and preliminary \emph{WISE} photometry showed no excess emission, whereas in the final \emph{WISE} release there is a clear excess at 22~\micron. USNO-B 120144.7$-$781926 (\epscha\ 8) and 2MASS J12005517$-$7820296 (\epscha\ 10) show signs of reduced disc heights due to dust settling and the disc around HD 104237E (\epscha\ 7) may be undergoing partial dissipation in its inner regions, leaving the outer disc intact. \citeauthor{Fang13} attributed the under-luminosity of 2MASS J12014343$-$7835472 (\epscha\ 11) to a flared disk seen at moderately-high inclination ($\sim$85$^{\circ}$), in which the central star is seen in only scattered light.  2MASS J11432669$-$7804454 (\epscha\ 17) has a transitional disc with a strong 10~\micron\ silicate feature \citep{Manoj11}. The provisional member HD 105234 is surrounded by a warm, gas-poor debris disc \citep{Currie11b}. Unlike most debris discs it also has numerous solid-state features. MP Mus was classified by \citet{Mamajek02} as a 7--17~Myr-old member of LCC but its optically-thick accretion disc is more consistent with the younger age of \epscha. Such discs are rare around stars older than 5--10~Myr (see \S\ref{sec:discevo}) and MP Mus was the only accretor with a $K$-band excess detected by that study from 110 solar-type members of Sco-Cen. The membership of this star is discussed in greater detail in \S\ref{sec:lcc}.

\begin{figure}
   \centering
   \includegraphics[width=0.495\linewidth]{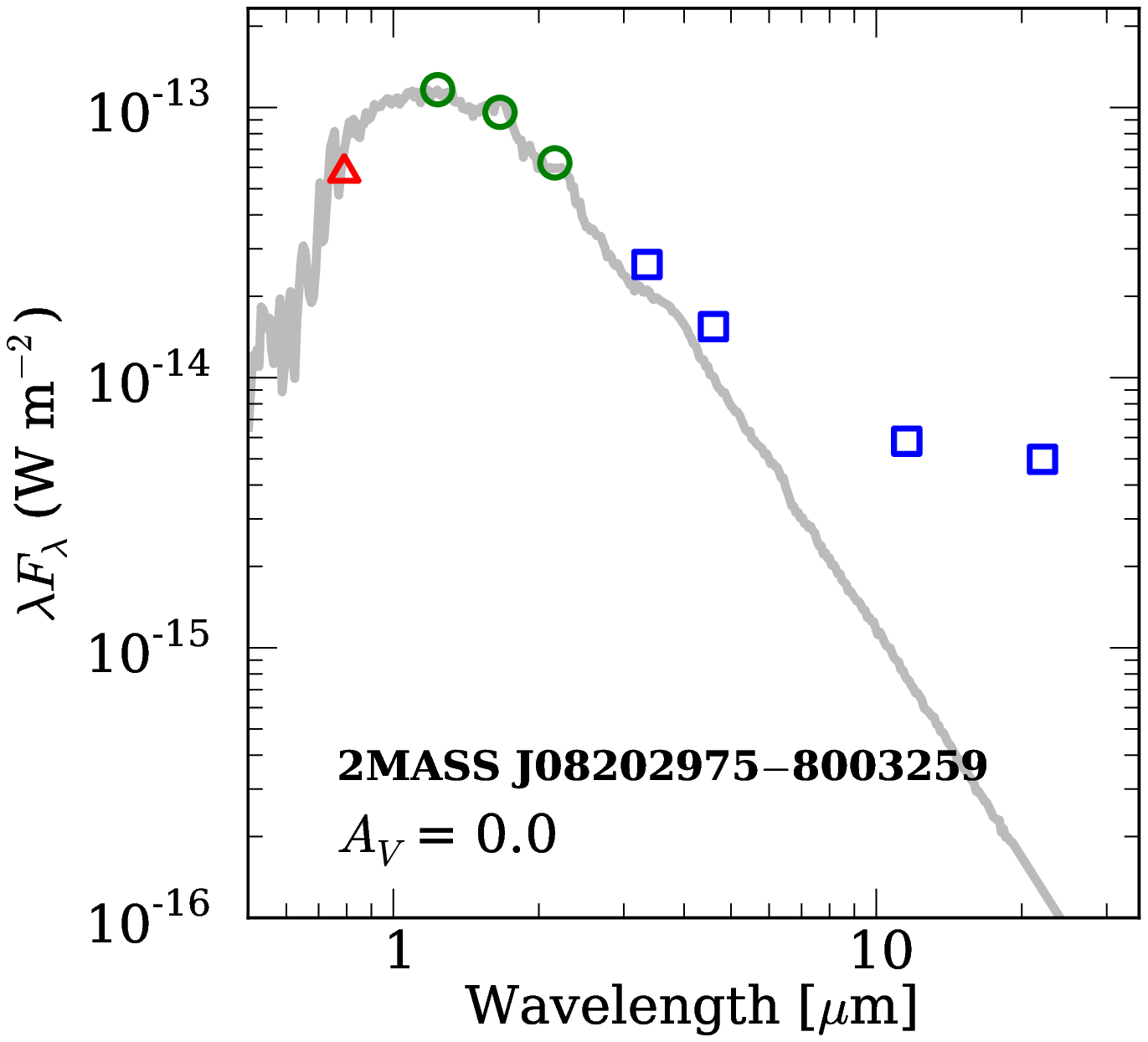}
   \includegraphics[width=0.495\linewidth]{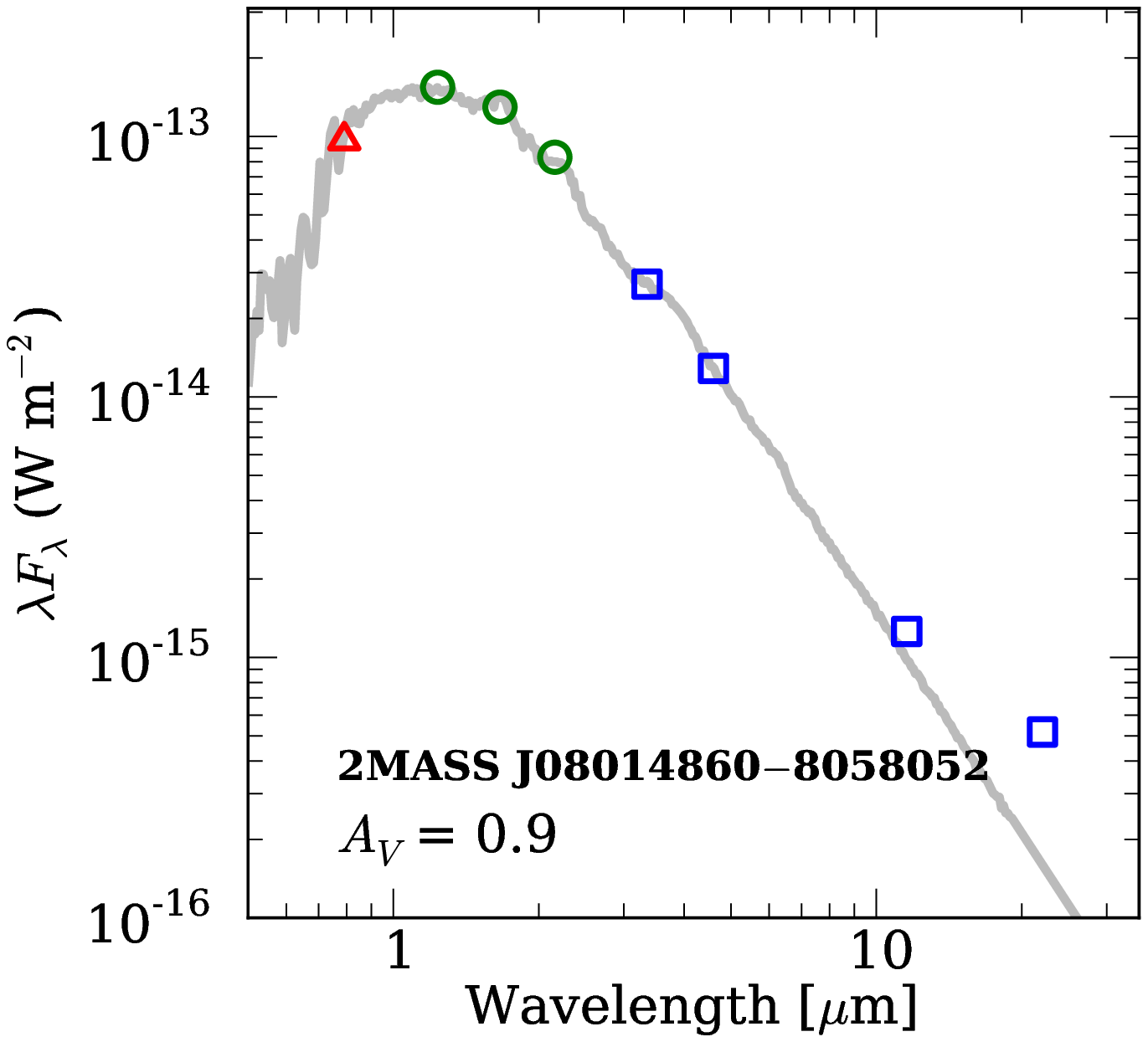}
     \caption{Same as Fig.\,\ref{fig:epschased}, for the \echa\ halo members  2MASS J08202975$-$8003259 and 2MASS J08014860$-$8058052. After reconsidering its photometry and optical spectrum 2M J0801$-$80 has a  spectral type of M4 and $A_{V}\approx0.9$ \citep[c.f.][]{Murphy10}.}
     \label{fig:etachased}
\end{figure}

Two candidates reclassified as Cha I members have infrared excesses. The SEDs of 2MASS J11334926$-$7618399 (\epscha\ 15) and VW Cha are plotted in Fig.\,\ref{fig:epschased}. We also constructed SEDs for the new halo members of \echa\ \citep{Murphy10}. The two members with detectable excesses are shown in Fig.\,\ref{fig:etachased}. 

\subsubsection{Circumstellar disc frequency}\label{sec:discevo}

\citet{Fang13} reported a disc frequency of $55_{-15}^{+13}$ per cent (6/11) for the classical members \citep{Feigelson03,Luhman04} in the core of \epscha. Considering all 41 confirmed and provisional members there are 12 stars with excesses at wavelengths shorter than $\sim$8~\micron\ (corresponding to the \emph{Spitzer} IRAC bands). This yields a disc fraction of $29_{-6}^{+8}$ per cent\footnote{Uncertainties for small samples are calculated for 68 per cent confidence intervals following the prescription of \citet{Cameron11}.}.

\begin{figure}
   \centering
   \includegraphics[width=\linewidth]{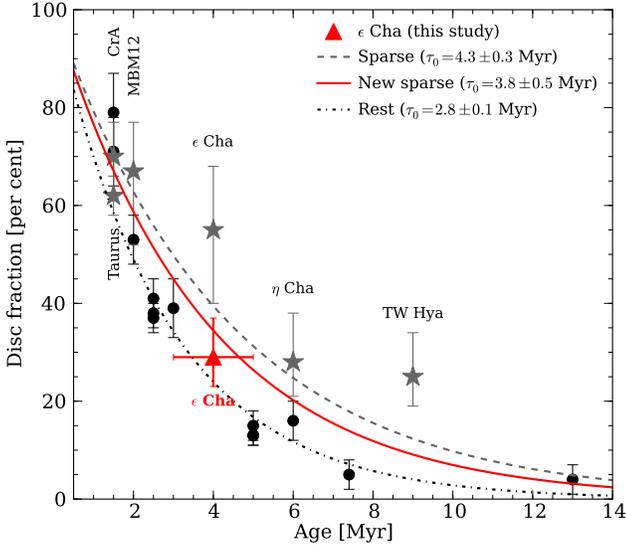} 
   \caption{Disc frequencies for several clusters and star-forming regions from \citet{Fang13}. Sparse associations (TW Hya, $\eta$ Cha, $\epsilon$ Cha, MBM12, Cr A, Taurus) are plotted as filled stars. Lines show the exponential decay ($f_{\rm disk}=e^{-t/\tau_{0}}$) models fitted to each environment. Our updated value for \epscha\ ($29^{+8}_{-6}$ per cent) is given by the red triangle.}
   \label{fig:discevo}
\end{figure}

Based on the elevated disc fraction of \epscha\ and other nearby groups, \citet{Fang13} proposed that circumstellar disc evolution proceeds more slowly in sparse associations. We plot in Fig.\,\ref{fig:discevo} the results of that study compared to our new \epscha\ disc fraction. Fitting exponential decay models ($f_{\rm disk}=e^{-t/\tau_{0}}$, with $f_{\rm disk,t=0}=1$) to these data, they estimated that disc lifetimes in sparse associations were longer ($\tau_{0}=4.3\pm0.3$~Myr) than in denser environments ($\tau_{0}=2.8\pm0.1$~Myr). The latter time-scale agrees with similar fits by \citet{Mamajek09} (2.5~Myr) and \citet{Fedele10} (3~Myr). With a larger and more complete membership, the updated \epscha\ disc fraction now agrees with the general (dense) relation within the uncertainties.

Re-fitting the sparse associations, we derive a shorter characteristic time-scale of $\tau_{0}=3.8\pm0.5$~Myr, but one which is still $\sim$2$\sigma$ longer than in  denser environments. The fit is very sensitive to the adopted disc fractions of \echa\ and TW Hya. The former ($28^{+10}_{-7}$ per cent) includes the seven dispersed halo members proposed by \citet{Murphy10} but should be considered an upper limit as more halo members likely remain undiscovered. Additional members of TW Hya continue to be proposed \citep[e.g.][and references therein]{Schneider12} and the membership of \epscha\ is almost certainly incomplete, especially at lower masses. Although Fig.\,\ref{fig:discevo} provides some evidence for longer disc lifetimes in sparse associations (particularly in \echa\ and TW Hya), given the incompleteness of these groups and the large uncertainties involved it may be premature to draw firm conclusions until larger, more complete samples are available. 

\begin{figure}
   \centering
   \includegraphics[width=\linewidth]{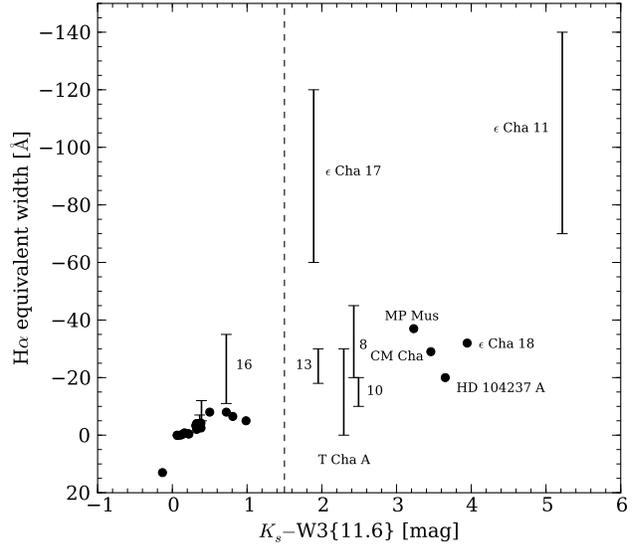} 
   \caption{Equivalent width (negative values denote emission) of the H$\alpha$ line versus the 2MASS/\emph{WISE} $K_{s}-W3$ colour. With the exception of T Cha, all of the multi-epoch measurements (solid bars) are from our WiFeS $R7000$ observations. We classify as accretors those stars with $K_{s}-W3>1.5$ (dashed line), as well as HD 104237E (\epscha\ 7).}
   \label{fig:halpha}
\end{figure}

\subsubsection{Accretor frequency}

The strength of H$\alpha$ emission is commonly used to identify stars actively accreting from circumstellar discs. The 32 members of \epscha\ with H$\alpha$ equivalent width (EW) measurements are plotted in Fig\,\ref{fig:halpha}, excluding HD 104237D and E which were not resolved by \emph{WISE}. Equivalent widths  were measured from the WiFeS $R$7000 spectra by direct integration of the line profile. We estimate an uncertainty of 1~\AA\ (rising to $\sim$3~\AA\ for the broadest lines), primarily due to uncertainties defining the pseudo-continuum around the broad H$\alpha$ lines at this modest spectral resolution. By the EW criteria of \citet{Fang09} ten stars (all with $K_{s}-\textrm{W3}>1.5$) are accreting\footnote{The equivalent width of 2MASS J11404967$-$7459394 (\epscha\ 16) flared to $-$35~\AA\ on 2011 June 19 from its quiescent level around $-11$~\AA. The star has no infrared excess and its H$\alpha$ emission is likely chromospheric.}. HD 104237E is also accreting based on its disc excess (Fig.\,\ref{fig:epschased}) and broad ($v_{10}\approx500$~\kms) H$\alpha$ line \citep{Feigelson03}.  This is an accretor fraction of $32_{-7}^{+9}$ per cent (11/34). \citet{Fedele10} proposed a similar exponential time-scale ($\tau_{0}=2.3$~Myr) for accretion in young groups. At ages of 3--5~Myr their  relation predicts accretor fractions of $\sim$10--30~per cent, marginally  consistent with our estimate.

\subsubsection{Notable T Tauri systems}\label{epschactts}

\begin{figure}
   \centering
   \includegraphics[width=\linewidth]{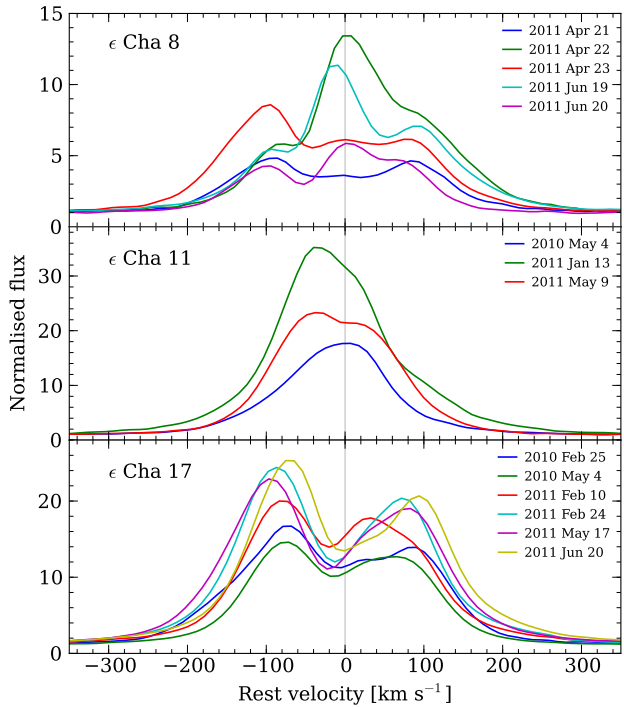} 
   \caption{WiFeS/$R7000$ velocity profiles for \epscha\ members with complex H$\alpha$ emission lines; USNO-B 120144.7$-$781926 (\epscha\ 8), 2MASS J12014343$-$7835472 (\epscha\ 11) and 2MASS J11432669$-$7804454 (\epscha\ 17). All of the profiles are shifted to zero radial velocity. }
   \label{fig:ctts}
\end{figure}

The multi-epoch WiFeS observations revealed three accretors with broad, variable H$\alpha$ emission. Velocity profiles of those stars are plotted in Fig.\,\ref{fig:ctts}. They show asymmetric, multi-component emission and velocity widths at 10 per cent of peak flux ($v_{10}$) in excess of the $v_{10}=200$--$270$~\kms\ accretion threshold \citep{Jayawardhana03}. Note the large \emph{daily} variation in the triple-peaked profiles of \epscha\ 8.  All three stars also exhibited He\,\textsc{i} $\lambda$5876/6678 emission and \epscha\ 11 displayed strong forbidden [O\,\textsc{i}] $\lambda$6300/6363 and [N\,\textsc{ii}] $\lambda$6584 emission. Using the $v_{10}$ velocity widths and the relation of \citet{Natta04}, the  accretion rates in these stars are $\sim$$10^{-10}$--$10^{-8.5}$~\msun~yr$^{-1}$, 1--2 orders of magnitude larger than those in \echa\ \citep{Lawson04,Murphy11} and the TW Hya association \citep{Muzerolle00}.  

\begin{figure} 
   \centering
   \includegraphics[width=\linewidth]{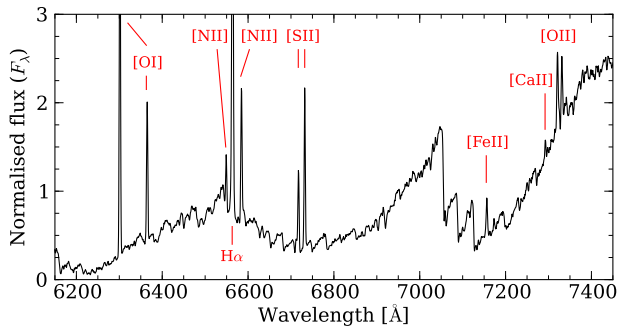} 
   \caption{WiFeS/$R3000$ spectrum of 2MASS J11183572$-$7935548 (\epscha\ 13). The star exhibits strong H$\alpha$ and the forbidden line emission typical of a much younger Classical T Tauri (CTT) star.}
   \label{fig:epscha13}
\end{figure}

The spectrum of 2MASS J11183572$-$7935548 (\epscha\ 13) is unique amongst \epscha\ members, with strong H$\alpha$, Na\,\textsc{i}~D, He\,\textsc{i} and forbidden [O\,\textsc{i}], [O\,\textsc{ii}] $\lambda$7320/7331, [N\,\textsc{ii}] $\lambda$6548/6583, [S\,\textsc{ii}] $\lambda$6716/6731, [Ca\,\textsc{ii}] $\lambda$7291/7324 and [Fe\,\textsc{ii}] $\lambda$7155 emission lines all present (Fig.\,\ref{fig:epscha13}). While persistent over our 2010--11  observations the emission is obviously variable as only the H$\alpha$ line was present in the discovery spectrum of \citet{Luhman07}. The star has a transitional disc with a large implied inner hole \citep{Manoj11}, consistent with its narrower H$\alpha$ line ($v_{10}\approx170$~\kms) and smaller accretion rate (10$^{-11}$~\msun~yr$^{-1}$). Forbidden emission typically arises in low-density, accretion-driven outflows or winds \citep{Appenzeller89}.   Such an outflow may have been responsible for clearing the inner region of \epscha\ 13's disc.

\subsection{Binaries in \epscha}

Six members of \epscha\ are confirmed spectroscopic or visual binaries (\rxjelevenfiftyeight, RX J1220.4$-$7407, \rxjelevenfiftynine, RX J1204.6$-$7731, HD 104237A and \epscha) and  four stars are suspected of having a spectroscopic companion (\epscha\ 10, 13, RX  J1202.1$-$7853 and HD 104467). These give binary fractions of $21_{-6}^{+10}$ per cent (confirmed) and $36_{-8}^{+10}$ per cent (including suspected) when compared to the 18  single members with two or more radial velocity measurements. The core of \echa\ has a similar binary fraction of 27--44 per cent \citep{Lyo04}, but lacks any systems with separations greater than 20 au \citep{Brandeker06}.

\subsubsection{Wide binaries}

There are also several wide systems in \epscha\ with projected separations of 10$^{3}$--10$^{4}$~au. We have already discussed HD 104237A--E (160--1700~au), \rxjelevenfiftyeight B (1700~au, hierarchical triple), T Cha AB \citep[0.2~pc; ][]{Kastner12} and the non-member HD 82879 (F4+F6, 2900~au). In addition to these systems the M0 members RX J1219.7$-$7403 and RX J1220.4$-$7407 are separated by only 4.5~arcmin (0.14~pc at 110~pc) in the north of the association. The latter also has a 0.3~arcsec companion \citep{Kohler01b}. Given their congruent radial velocities, proper motions and isolation they are highly likely to be physically associated.  Several other wide pairs (HD 105234/HIP 59243, RX J1149.8$-$7850/\rxjelevenfortyseven\ and HD 104467/RX J1202.1$-$7853) have incompatible kinematics or distances.

\begin{figure*}
   \centering
   \includegraphics[width=\textwidth]{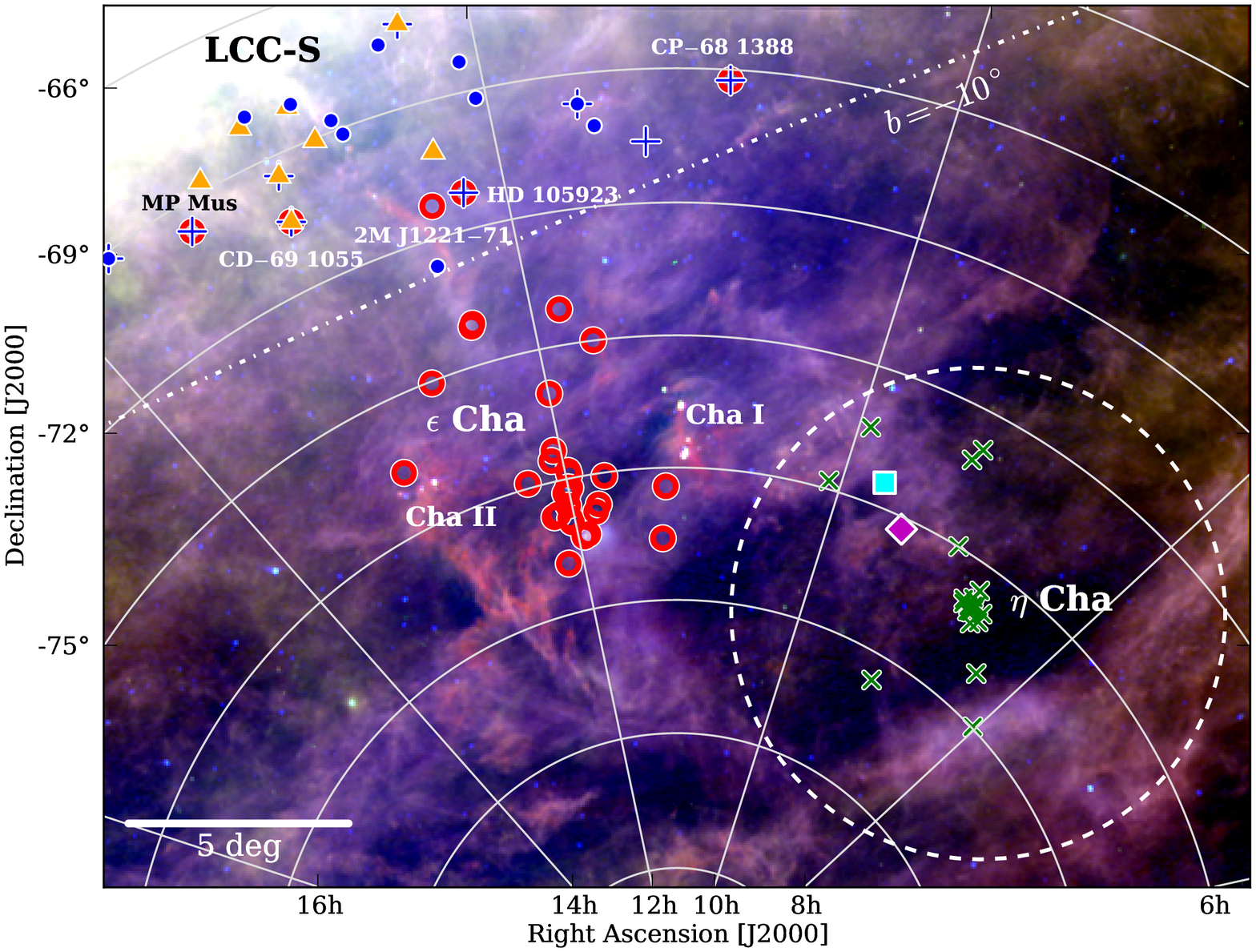} 
   \caption{\emph{IRAS} 100/60/25 \micron\ colour composite with \echa\ members (green crosses) and confirmed or provisional members of \epscha\ (open circles). Stars discussed in the text are labelled. The  diamond and square are the wide systems HD 82879/82859 and RX J0942.7-7726AB, respectively. LCC members from \citet{Preibisch08} (blue pluses), \citet{de-Zeeuw99} (filled circles) and \citet{Song12} (triangles) are shown with the $b=-10^{\circ}$ `boundary' of the subgroup (dot dashed line). The dashed line is the 5.5~deg radius around \echa\ surveyed for new members by \citet{Murphy10}.}
   \label{fig:etaeps}
\end{figure*}

\section{Discussion}\label{sec:discussion}

\subsection{Relationship to $\eta$ Chamaeleontis}\label{sec:epsetacha}

The core of \epscha\ is $\sim$10~deg (25~pc in space) from the young open cluster $\eta$ Cha \citep{Mamajek99}.  Fig.\,\ref{fig:etaeps} shows the distribution of their members on the sky. The groups share similar ages, distances and kinematics, which led \citet{Torres08} to subsume the four members of \echa\ with measured radial velocities at the time into their \epscha\ solution. We did not consider these stars  \epscha\ members in this study. From the analysis of \S\ref{sec:ages} it is now clear that the two groups are distinct, with subtly different properties.  

\begin{figure}
   \centering
   \includegraphics[width=\linewidth]{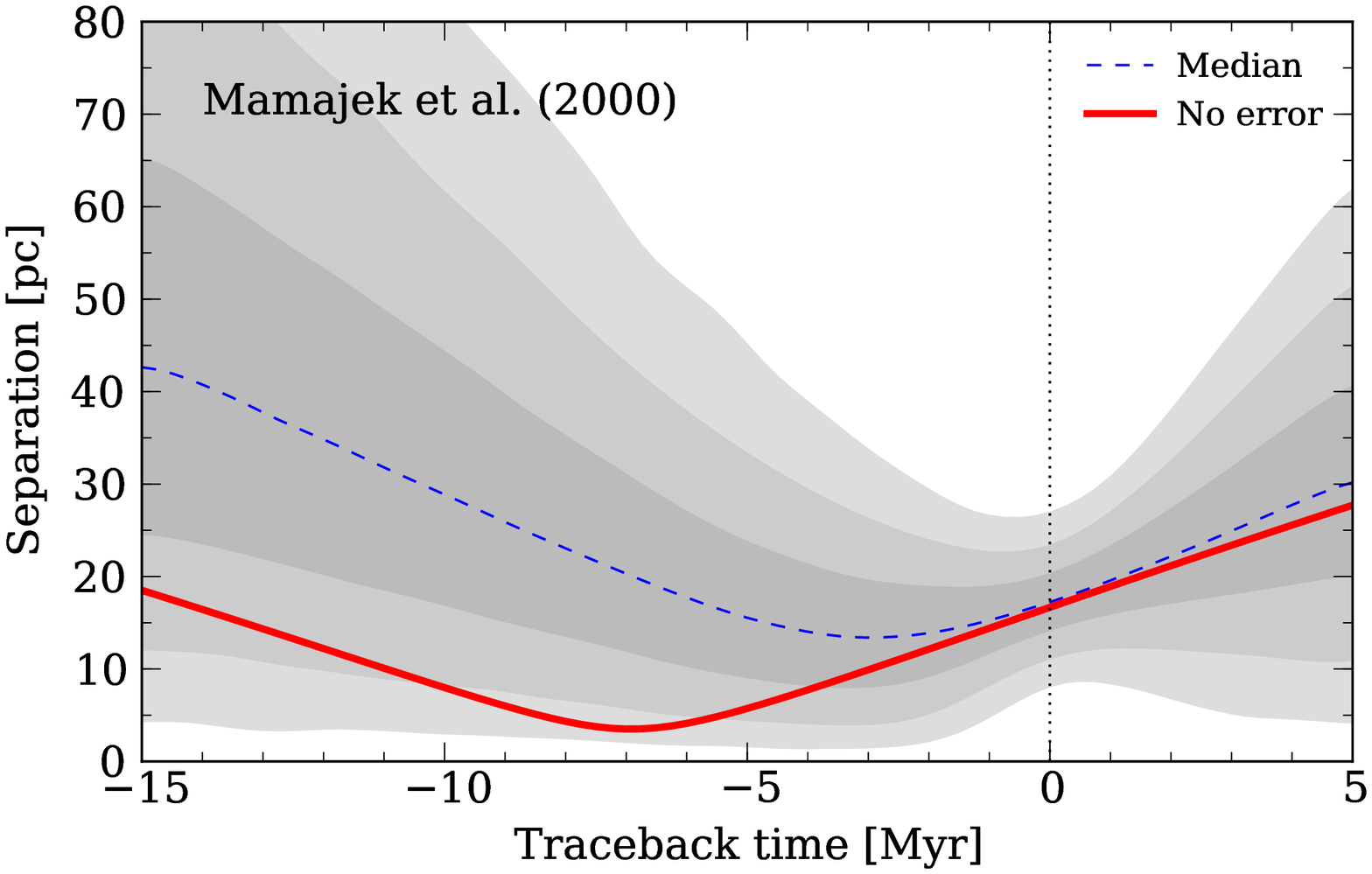}\\
   \includegraphics[width=\linewidth]{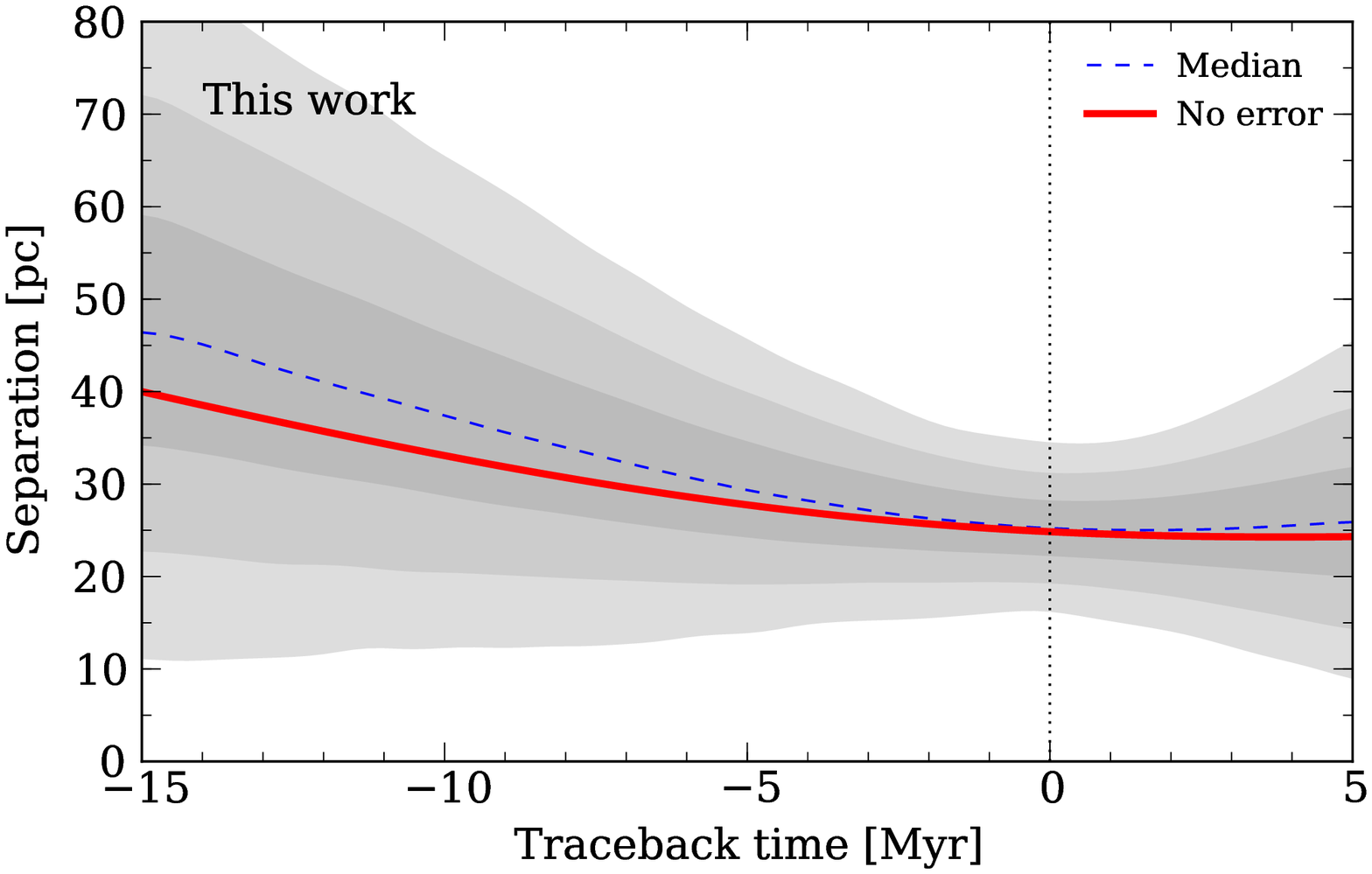}
   \caption{Separation between $\eta$ Cha and the core of \epscha\ as a function of time. Top: Initial positions and velocities from \citet{Mamajek00}. Bottom: updated values from Table~\ref{table:uvw}. The solid line assumes no errors on the position or velocity of either association  while the shaded regions are the 16--84, 2--98 and 0.1--99.9 per cent contours of the cumulative distribution ($\pm$1--3$\sigma$) from 5000 realisations. In both panels the initial centre of \epscha\ is assumed normally distributed in $XYZ$ with $\sigma=3$~pc.}
   \label{fig:traceback}
\end{figure}

Using the best-available space motions and positions from \citet{Mamajek00}, \citet{Jilinski05} showed that the centres of $\eta$ and \epscha\ reached a minimum separation of $\sim$3~pc some 6--7~Myr ago. They concluded that the groups were born together or very close to one another in the outskirts of Sco-Cen. However, that study did not account for the large errors in the groups' space motions. Following the prescription of \citet{Makarov04}, our epicyclic traceback analysis (Fig.\,\ref{fig:traceback}, top panel) replicates the \citeauthor{Jilinski05} result (red line), but shows that such a small separation is unlikely at any epoch given the quoted velocity uncertainties. Around 6--7~Myr ago fewer than 2 per cent of realisations  resulted in a separation closer than 3~pc. With improved kinematics and positions for both groups (Table~\ref{table:uvw}), it is clear that $\eta$ and \epscha\ were unlikely to have been much closer than their current separation and were $\sim$30~pc apart at the time of their birth  (Fig.\,\ref{fig:traceback}, bottom panel). This is further evidence the two groups are distinct entities.

Any model for the birth of $\eta$ and \epscha\ must also account for their different physical characteristics.  For example, to explain \echa's  apparently primordial deficit of wide ($>$20 au) binaries \citep{Brandeker06} and low-mass objects, \citet{Becker13} speculated that the cluster was born in a small, highly magnetised cloud which prevented fragmentation on large scales. \epscha\ was likely born under more quiescent conditions in a less-dense environment, as suggested by its larger spatial extent, the still-bound HD 104237A--E and a number of wider (10$^{3}$--10$^{4}$~au) systems. Moreover, the older wide binaries HD 82879/82859 and RX J0942.7-7726AB \citep{Murphy12} as well as several young stars whose origins are as-yet undetermined (e.g. Table~\ref{table:nonmembers}) are hints that the star-formation history of the region was more complex than the formation of two small groups in isolation a few Myr apart. 

\subsection{Star formation across the greater Sco-Cen region}\label{sec:lcc}

Immediately north of \epscha\ lies the Lower Centaurus Crux (LCC) subgroup of the Scorpius-Centaurus OB association (Fig.\,\ref{fig:etaeps}). LCC shows signs of spatio-kinematic substructure \citep{Preibisch08} whereby the north of the subgroup appears older, richer and more distant  ($\sim$20~Myr, 120~pc) than the south ($\sim$10~Myr, 110~pc). Recent work has also found a N-S gradient in the $W$ velocity component (E. Mamajek, private communication). These trends may be evidence for a wave of star formation starting in the north 15--20~Myr ago, spreading across the Galactic plane to form the southern part of the subgroup as well as the $\beta$~Pic and TW Hya associations $\sim$10~Myr ago \citep{Mamajek01} and ending with the birth of $\eta$ and \epscha\ as recently as 3--5~Myr ago. 

\begin{table*}
\begin{minipage}{0.9\textwidth}
\caption{Observed and predicted kinematics of northern \epscha/LCC members}
\label{table:lcc}
\begin{tabular}{rlccccccccccccccc}
\hline
& & \multicolumn{3}{c}{\bf Observed} & \multicolumn{4}{c}{\bf \epscha} & \multicolumn{4}{c}{\bf LCC \citep{Chen11}} \\
MML & Star & \multicolumn{2}{c}{$(\mu_{\alpha}\cos\delta,\mu_{\delta})^{\dagger}$} & RV$^{\ddag}$ & \multicolumn{2}{c}{$(\mu_{\alpha}\cos\delta,\mu_{\delta})$} & RV  & $d_{\rm kin}$ & \multicolumn{2}{c}{$(\mu_{\alpha}\cos\delta,\mu_{\delta})$} & RV & $d_{\rm kin}$\\
ID$^{\star}$ & & \multicolumn{2}{c}{[\masyr]} & [\kms]  & \multicolumn{2}{c}{[\masyr]} & [\kms] & [pc] & \multicolumn{2}{c}{[\masyr]} & [\kms] & [pc]\\
\hline
& 2M J1221$-$71\ & $-$39.4 & $-$11.8 & 11.5 & $-$39 & $-$11 & 13.4 & 110 & $-$42 & $-$4 & 14.6 & 89\\
1 & CP$-$68 1388 & $-36.1$ & $+5.7$ & 15.9 & $-$37 & $+$2 & 15.8 & 112 & $-$36 & $+$8 & 16.7 & 91 \\
7 & HD 105923 & $-38.9$ & $-8.3$ & 14.2 & $-$39 & $-$9 & 13.7 & 112 & $-$40 & $-$3 & 14.9  & 92 \\
27 & CD$-$69 1055 & $-43.1$ & $-18.7$ & 12.8 & $-$43 & $-$19 & 12.2 & 99 & $-$46 & $-$12 & 13.6 & 82 \\
34 & MP Mus & $-40.8$ & $-23.0$ & 11.6 & $-$41 & $-$23 & 11.4 & 101 & $-$45 & $-$17 & 12.9 & 84 \\
\hline
\end{tabular}\\
($\star$): Identifier from \cite*{Mamajek02}\\
($\dagger$): Proper motions from Tycho-2 and SPM4 (2MASS J12210499-7116493)\\
($\ddag$): Radial velocities from \citet{Torres06} and \citet{Kiss11} (2MASS J12210499-7116493)\\
\end{minipage}
\end{table*}

2MASS J12210499$-$7116493  \citep{Kiss11} and four stars proposed by \citet{Torres08} lie north of the majority of \epscha\ members within the southern reaches of LCC. \citet*{Mamajek02} attributed the four stars to LCC in their survey of solar-type Sco-Cen members and they also have isochronal ages in Table~\ref{table:members} older than the majority of \epscha\ members (although there may be systematic errors in the ages of solar-type stars, see \S\ref{sec:ages}). Table~\ref{table:lcc} compares their observed and expected kinematics. With the exception of CP$-$68 1388, which is a marginally better match to LCC ($\Delta K=1$~\kms), their proper motions and radial velocities agree with the space motion of \epscha\ at 100--110~pc. The 80--90~pc distances implied by membership in LCC are well below the 100--120~pc typically attributed to the subgroup.  Furthermore, the K7 2MASS J12210499$-$7116493 has an Li\,\textsc{i} $\lambda$6708 EW much greater than similar LCC members recently proposed by \citet{Song12} (or the roughly coeval $\beta$ Pic). At the late-G and early-K spectral types of the four \citeauthor{Torres08} members there is little difference in lithium depletion at ages $\lesssim$20~Myr. Pending better knowledge of the low-mass population of southern LCC, we retain these stars as members of \epscha\ as defined in this study (but see below).

However, given their similar ages, kinematics, distances and the observed trends in these parameters, a useful demarcation between the southern extent of LCC and \epscha\ may simply not exist. In this scenario it is possible the four \citet{Torres08} stars in Table~\ref{table:lcc} could be comoving with \epscha\ and have similar ages but were not born near the majority of its members. \citet{Song12} have identified several young, low-mass stars immediately north of \epscha\ (Fig.\,\ref{fig:etaeps}) and the provisional candidate \rxjtwelvetwo\ may be an outlying LCC member juxtaposed on the central region of \epscha.  A similar problem exists at the northern boundary of LCC and TW Hya, where young stars at 70--150~pc have ambiguous memberships \citep{Lawson05,Mamajek05b}.  Clarifying this messy picture of star formation will require a larger sample of young stars with reliable distances and ages. The soon-to-be-launched \emph{Gaia} mission will help immensely in this regard.

In the interim we can only speculate how $\eta$ and \epscha\ formed within the greater Sco-Cen region.  A schematic picture was provided by \cite{Feigelson96}, who proposed young stars born in different parts of a dynamically unbound giant molecular cloud (GMC) are dispersed by internal turbulent velocities. In this picture some parts of the GMC may collapse quickly to form rich clusters while other regions remain stable against collapse as they disperse, forming stellar aggregates of a range of ages and sizes over wide areas. \citet{Preibisch08} proposed that the bulk of star formation in Sco-Cen probably proceeded in this way as a series of small clusters and filaments, each containing tens to hundreds of stars. The local virial balance of the nascent cloudlets meant some resulting groups are compact (\echa\ and the core of \epscha), while others appear unbound and widely dispersed (e.g. TW Hya, $\beta$ Pic and the outer members of \epscha) \citep{Feigelson03}. 

\section{Summary}\label{sec:epschafinal}

We have critically re-examined membership of the young \epscha\ association using the best-available spectroscopic and kinematic information. The main results of this study are:

-- Of the 52 candidates proposed in the literature, we confirm 35 stars as members, with spectral types of B9 to mid-M. Six candidates are classified as provisional members requiring better kinematics, trigonometric parallaxes or binary information.

-- \epscha\ lies at a mean distance of $110\pm7$~pc and has a mean space motion of $(U,V,W)=(-10.9\pm0.8,-20.4\pm1.3,-9.9\pm1.4)$~\kms. Comparison of its HR diagram to theoretical evolutionary models yields a median age of 3--5~Myr, distinguishing it as the youngest moving group in the solar neighbourhood.

-- Fifteen members have infrared SEDs attributable to circumstellar discs, including 11 stars whose strong H$\alpha$ emission indicates they are accreting. As expected of a rapidly-evolving intermediate-age population, these stars show a variety of SED morphologies, from optically-thick accretion discs, to weak-excess debris discs. Both the disc and accretion fractions ($29^{+8}_{-6}$ and $32^{+9}_{-7}$ per cent, respectively) are consistent with a typical 3--5 Myr-old population.

-- A comparative age analysis shows that \epscha\ is approximately 1--3 Myr younger than the nearby open cluster \echa. Contrary to previous studies which assumed $\eta$ and $\epsilon$ Cha are coeval and formed in the same location, we find the groups were separated by $\sim$30~pc when $\eta$ Cha was born, followed soon after by \epscha. 

-- The physical properties, locations and kinematics of $\eta$ and $\epsilon$ Cha are consistent with them being formed in the turbulent ISM surrounding the Scorpius-Centaurus OB association. They are likely products of the last burst of star formation in southern Sco-Cen, which earlier formed the sparse TW Hya and $\beta$ Pic groups and several thousand older (now field) stars. 

-- After considering all the available observations we rejected several proposed members. They instead belong to the background Cha I and II cloud populations and other nearby young groups. In the absence of parallaxes and precise stellar ages we emphasise the importance of a holistic and conservative approach to assigning stars to kinematic groups, many of which have only subtly different properties and ill-defined memberships.

-- In this study we considered only those stars proposed as members in the literature. There are likely many new members of \epscha\ awaiting discovery in contemporary proper motions catalogues (e.g. SPM4, PPMXL). These candidates can be tested against the definition of \epscha\ presented here to extend its membership to lower masses and wider areas. 

\section*{Acknowledgments}

We thank Pavel Kroupa, Eric Mamajek and Eric Feigelson for their considered comments on the thesis of SJM from which this work is based, and the anonymous referee for their thorough review of the manuscript. We  also thank the ANU TAC for their generous allocation of telescope time. This research has made extensive use of the VizieR and SIMBAD services provided by CDS,  Strasbourg and the \textsc{topcat} software package developed by Mark Taylor (U.\,Bristol). Many catalogues used in this work are hosted by the German Astrophysical Virtual Observatory. \emph{WISE} is a joint project of the University of California, Los Angeles, and the Jet Propulsion Laboratory/California Institute of Technology, funded by NASA. 2MASS is a joint project of the University of Massachusetts and the Infrared Processing and Analysis Centre/California Institute of Technology, funded by NASA and the NSF. 

\footnotesize{

}

\appendix

\section{Notes on individual candidates}\label{sec:candidates}

\subsection{Confirmed members}

\emph{CXOU J120152.8$-$781840 (\epscha\ 9):}\quad \citet{Fang13} questioned  membership of this star based on its PPMXL proper motion. It was not recovered in UCAC4 but using higher-precision SPM4 astrometry we find it is a kinematic member at 121~pc.

\emph{RX J1149.8$-$7850 (\epscha\ 18):}\quad \citet{Malo13} assigned the star to $\beta$ Pic. Their Bayesian analysis (which did not include \epscha) gave a distance of 71~pc, which would make it one of the most distant members. Given its large Li\,\textsc{i} $\lambda$6708 equivalent width it is likely younger than $\beta$ Pic and we find an excellent match to \epscha\ at a distance of 110~pc.

\emph{\rxjelevenfiftynine:}\quad Using the \citet{Terranegra99} proper motion (\S\ref{sec:discordpm}) we find \rxjelevenfiftynine\ to be a kinematic member at 108~pc. The star has a close companion \citep{Kohler01b} and we caution that the WiFeS velocity may not be systemic.

\emph{\rxjelevenfiftyeight B (\epscha\ 21+20):}\quad   The 104~pc kinematic distance adopted for \rxjelevenfiftyeight\ just agrees with its \emph{Hipparcos} parallax ($90.4^{+14}_{-11}$~pc). \citet{Kohler01b} detected a 0.07 arcsec companion and the system is only 16~arcsec from \epscha\ 20 (GSC 9415-2676, $d_{\rm kin}=119$~pc).  Perhaps as well as being responsible for the velocity variation (Table~\ref{table:epschavels}), the close companion distorted the parallax over the short-baseline \emph{Hipparcos} observations. The triple system is only 5~arcmin from HD~104036, which has a \emph{Hipparcos} distance of $108\pm4$~pc.

\emph{HD 104237A (\epscha\ 5):}\quad A core member of \epscha, the Herbig Ae star has a space motion only 1.3~\kms\ from the mean and a kinematic distance which agrees with its 114~pc \emph{Hipparcos} parallax. HD 104237A was rejected in the convergent analysis because of its position above the isochrone. This is likely due to a combination of binarity \citep{Bohm04}, infrared excess and an uncertain spectral type/reddening \citep{Lyo08}.

\emph{HD~104237B--E (\epscha\ 3--7):}\quad All four late-type components of the HD~104237 system show X-ray emission and are almost certainly associated with their primary. \citet{Fang13} estimated the mass of HD~104237C to be 13--15~$M_{\rm Jup}$ from near-infrared photometry, making it the lowest-mass member of \epscha. Components D and E  have strong lithium absorption. Both stars were rejected from the convergent solution by their discrepant kinematics. Their proper motions (and the photometry of HD~104237D) are likely influenced by HD~104237A. 

\emph{2MASS J12014343$-$7835472 (\epscha\ 11):}\quad We find an excellent proper motion match at 100~pc with a predicted radial velocity 6 \kms\ lower than measured. This may be due to unresolved binarity or strong accretion activity (see \S\ref{epschactts}).  Three WiFeS velocities showed no trend over $\sim$1~yr.  \citet{Luhman04} attributed its under-luminosity to obscuration by an edge-on circumstellar disc. This was confirmed by \citet{Fang13}. We confirm membership in \epscha\ based on a congruent proper motion, Li\,\textsc{i}  equivalent width and location in the core of the association.

\subsection{Provisional members}

\emph{CXOU J115908.2$-$781232 (\epscha\ 1):}\quad The smaller proper motion yields a kinematic distance of 165~pc ($K=5.2$~\kms). The star is unlikely to have been ejected from the core as its predicted radial velocity is only $\sim$1~\kms\ from observed.  \epscha\ 1 is not found in UCAC4 but its PPMXL proper motion ($\mu_{\alpha\cos\delta}, \mu_{\delta} = -36\pm14, -6\pm14$~\masyr) is consistent with membership at 120~pc ($K=1.5$~\kms), notwithstanding the large errors. Given its location near the centre of \epscha\ we assign it provisional membership pending better kinematics.

\emph{\rxjtwelvetwo:}\quad In the absence of accelerated lithium depletion \citep[e.g.][]{Baraffe10} the star's 300~m\AA\ Li\,\textsc{i} $\lambda$6708 equivalent width suggests an age closer to the TW Hya and $\beta$ Pic associations (8--12 Myr). The only nearby population with a similar age is the Lower Centaurus Crux (LCC) subgroup of Sco-Cen, whose southern extent may be as young as $\sim$10~Myr (see \S\ref{sec:lcc}).  \rxjtwelvetwo\ is an equally good match to the LCC space motion \citep{Chen11} at 110~pc. Given its location close to the core of \epscha\  we retain it as a provisional member, though it may be an older, dispersed member of Sco-Cen. 

\emph{CM Cha:}\quad Previously assigned to Cha II, the star was tentatively reclassified as a member of \epscha\ by \citet{Lopez-Marti13} by its UCAC3 proper motion. With SPM4 we find a somewhat poor ($K=5.2$~\kms) kinematic match to \epscha\ at 133~pc. A marginally better match to Cha II \citep{Lopez-Marti13} is found at $\sim$120~pc, though this is considerably closer  than recent distance estimates to the cloud \citep[180--210~pc;][]{Knude10}.  Given its location near Cha II and discordant proper motions (see \S\ref{sec:discordpm}) we assign provisional membership in \epscha\ and await improved kinematics or a parallax.

\subsection{Non-members}

\begin{figure}
   \centering
   \includegraphics[width=0.9\linewidth]{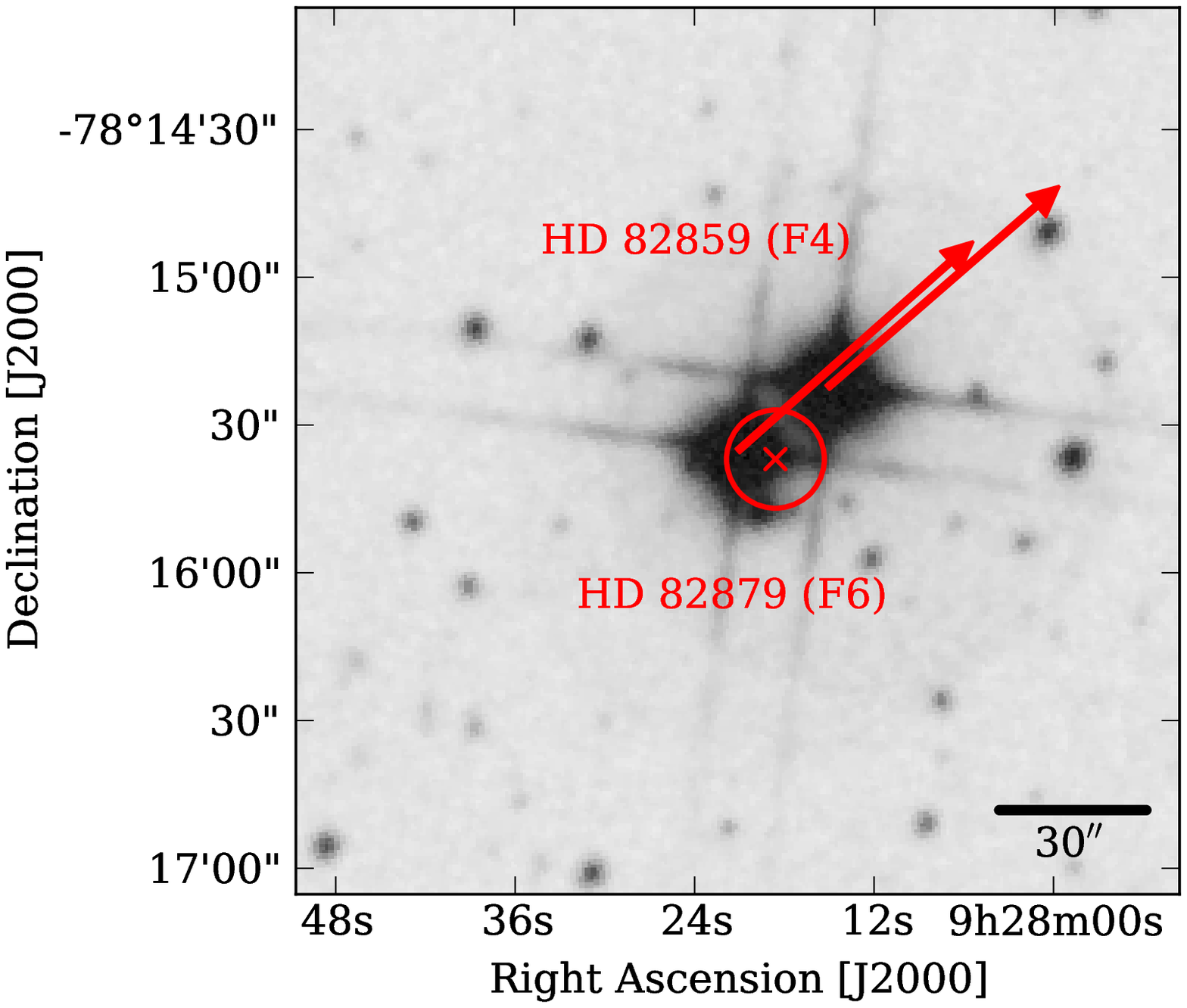} 
   \caption{POSS2-Red 3$\times$3~arcmin image centred on HD 82879. Arrows show the Tycho-2 proper motions over 2000~yr. The cross and circle give the position of the \emph{ROSAT} source (RX J0928.5$-$7815) and its uncertainty.}
   \label{fig:hd82879}
\end{figure}

\emph{HD 82879:}\quad The F6 star is only 24~arcsec from and comoving with HD 82859  (F4, Fig.\,\ref{fig:hd82879}).  It is associated with the \emph{ROSAT} X-ray source RX J0928.5$-$7815 and was proposed as an \epscha\ member by \citet{Torres08}. \cite{Lopez-Marti13} also listed it as a member but neither study noted its binarity. HD 82879 is a good isochronal and kinematic match to \epscha. However, given its large total mass ($\sim$2.8~\msun) and isolation it is unlikely the system could have been dispersed from $\epsilon$ or \echa\ within their lifetimes.  Neither star was observed by \emph{Hipparcos} but HD~82859 has an ill-constrained Tycho-2 parallax of 110$^{+330}_{-50}$~pc. \citet{da-Silva09} suggested HD 82879 as a member of the AB Doradus association, however we find a poor kinematic match at any distance.

\emph{HIP 55746:}\quad This star has the largest offset from the mean \epscha\ space motion ($K=6.7$~\kms) and is the closest proposed member ($89\pm5$~pc, \emph{Hipparcos}). However, it is only 3~\kms\ from the space motion of AB Dor \citep{Torres08} with a distance  similar to other members. We follow \citet{Torres08} and assign the star to AB Dor.

\emph{RX J1137.4$-$7648:}\quad Despite the excellent kinematic match ($K$=1.7~\kms), the 4~mag under-luminosity, outlying distance (67~pc) and lack of Li\,\textsc{i} $\lambda$6708 absorption rule RX J1137.4$-$7648 out as a member of \epscha. The 3-arcsec visual binary is likely an nearby, older field system.

\emph{2MASS J11334926$-$7618399 (\epscha\ 15):}\quad Located between Cha I and the core of \epscha, the star was proposed by \citet{Luhman08b} as having a proper motion (not reported in that work) more similar to \epscha\ than the cloud sources. Although it lies on the mean isochrone only 2.6~\kms\ from the \epscha\ space motion, its 197~pc distance is much greater than other members. The star is not in the USNO-B1 or UCAC catalogues and its SuperCosmos proper motion is erroneous.  Its PPMXL and SPM4 kinematics and modest reddening are consistent with an outlying Cha I member at a distance of $\sim$160~pc. 

\emph{\rxjtwelvefortythree:}\quad The close binary \citep{Kohler01b} lies a few degrees north of Cha II and its space motion gives a reasonable agreement ($K=3.7$~\kms) with the cloud sources at $\sim$260~pc. This is considerably larger than recent distance estimates to the cloud \citep[180--210~pc;][]{Knude10}. If we use the \citet{Terranegra99} proper motion a better match is found at 200~pc. The predicted radial velocity agrees with the mean WiFeS value, implying we observed the star close to its systemic velocity.  We classify \rxjtwelvefortythree\ as an outlying member of Cha II, but caution that its proper motion may be affected by the close companion. 

\emph{VW Cha:}\quad The star is unlikely to be a member of \epscha\ based on its large implied distance (217~pc) and position well above the isochrone. It was classified as a Cha I member by \citet{Luhman07} and again by \citet{Lopez-Marti13}. Its proper motion, large extinction and location inside the southern cloud core support this assignment. We find the best match to the Cha I space motion at $\sim$160~pc, the distance to the cloud.

\emph{\rxjelevenfiftyfour\ (\epscha\ 19):}\quad \citet{Torres08} rejected membership in \epscha\ on the basis of a $-3.3\pm1.0$~\kms\ velocity from \citet{Guenther07}, but this was almost certainly not systemic, nor is the mean WiFeS value. At its 106~pc kinematic distance \rxjelevenfiftyfour\ lies 1.4~mag below the isochrone, more if its unresolved companion contributes significantly to the $I$-band flux. Despite strong lithium absorption and a congruent proper motion we rule out membership in \epscha. A systemic velocity would be useful. \citet{Wahhaj10} classified the star as T Tauri based on a broad H$\alpha$ line, but this is likely the result of binarity. 

\emph{2MASS J12074597$-$7816064 (\epscha\ 12):}\quad A proper motion outlier, \epscha\ 12 also shows evidence of spectroscopic binarity. We find a good match to the \epscha\ space motion at 85 pc, but the star is under-luminous at this distance. It is located in the core of the association and its strong lithium absorption implies an age comparable to other members. The companion, if it exists, may have influenced the star's proper motions but even at 110~pc \epscha\ 12 remains significantly under-luminous.  We refrain from assigning it to \epscha\ pending better kinematics and confirmation of binarity.

\emph{RX J1123.2$-$7924 (\epscha\ 14):}\quad  \citet{Lopez-Marti13} reaffirmed membership in \epscha\ based on a marginally-consistent proper motion. However, the star's low Li\,\textsc{i} equivalent width (130~m\AA) and position below the isochrone mean this is unlikely. The proper motion may have been influenced by the spectroscopic companion, but we did find an good kinematic match to the $\sim$20~Myr-old Octans association \citep{Torres08} at 110~pc. The lithium depletion, CMD placement and heliocentric position of RX J1123.2$-$7924 are consistent with this assignment.   

\subsection{Candidates without radial velocities} 

Five candidates lack any radial velocity information. To calculate kinematic distances for these stars we compared their proper motions to the \epscha\ solution or used parallaxes if  available. TYC 9414-191-1, HD 105234 and HIP 59243 may be members based on their kinematic match and CMD placement. Radial velocities and a Li\,\textsc{i} $\lambda$6708 measurement for TYC 9414-191-1 are necessary to confirm membership. The two remaining Tycho sources (TYC 9420-676-1, TYC 9238-612-1) have positions below the empirical isochrone and are unlikely to be members.

\section{Candidate measurements}

The astrometric and spectroscopic measurements adopted for each candidate are listed in Table~\ref{table:bigtable} with appropriate references. Near-infrared photometry from DENIS, 2MASS and \emph{WISE} is given in Table~\ref{table:phot}. 

\begin{table*}
\caption{Adopted astrometry and spectroscopic measurements of $\epsilon$ Cha  candidates. Spectral types, velocities and equivalent width measurements are derived from our WiFeS spectroscopy unless specified.}
\label{table:bigtable}
\begin{minipage}{\textwidth}
($\ddag$) References: (T06) \citet{Torres06}, (T08) \citet{Torres08}, (L04) \citet{Luhman04}, (LL04) \citet{Luhman04b}, (ZS04) \citet{Zuckerman04a}, (K12) \citet{Kastner12}, (M00) \citet{Mamajek00}, (R06) \citet{Riaz06}, (C97) \citet{Covino97}, (G07) \citet{Guenther07}, (N04) \citet{Nordstrom04}, (TCha) Same velocity as T Cha, (J06) \citet{James06}, (G99) \citet{Grenier99}, (BB00) \citet{Barbier-Brossat00}, (G04) \citet{Grady04}, (K11) \citet{Kiss11}, (dS09) \citet{da-Silva09}, (M11) \citet{Manoj11}, (S09) \citet{Schisano09}, (F03) \citet{Feigelson03}, (S08) \citet{Spezzi08}.\\
($\dag$) Proper motion references: (TYC) Tycho-2/\citet{Hog00}, (HIP) \emph{Hipparcos}/\citet{van-Leeuwen07}, (UCAC4) \citet{Zacharias13}, (SPM4) \citet{Girard11}, (PPMXL) \citet{Roeser10}.
\end{minipage}
\end{table*}

\begin{table*}
\centering
\begin{minipage}{0.83\textwidth}
\caption{Adopted photometry for \epscha\ candidates. Unless specified, $I$-band data is from the DENIS survey \citep{Epchtein99} with $JHK_{s}$ from 2MASS \citep{Skrutskie06} and $W1,W2,W3,W4$  from \emph{WISE} \citep{Wright10}.}
\label{table:phot}
\begin{tabular}{@{}rlclccclcccc@{}}
\hline
ID & Name & $I$ & Ref.$^{\ddag}$ & $J$ & $H$ & $K_{s}$ & Ref.$^{\ddag}$ & $W1$ & $W2$ & $W3$ & $W4$\\
& & [mag] & & [mag] & [mag] & [mag] & & [mag] & [mag] & [mag] & [mag] \\
\hline
 & HD 82879 & 8.95 &  & 8.12 & 7.93 & 7.83 & & 7.78 & 7.80 & 7.82 & 7.92 \\
 & CP$-$68 1388 & 9.28 & T06 & 8.48 & 8.01 & 7.79 & & 7.72 & 7.74 & 7.68 & 7.59 \\
 & VW Cha & 11.03 & B01 & 8.70 & 7.64 & 6.96 & & 6.15 & 5.40 & 4.12 & 1.85 \\
 & TYC 9414-191-1 & 9.59 &  & 8.41 & 7.76 & 7.53 & & 7.43 & 7.55 & 7.45 & 7.43 \\
13 & 2MASS J11183572$-$7935548 & 12.22 &  & 10.49 & 9.89 & 9.62 & & 9.42 & 9.14 & 7.67 & 4.47 \\
14 & RX J1123.2$-$7924 & 11.62 &  & 10.52 & 9.84 & 9.67 & & 9.52 & 9.46 & 9.31 & 8.61 \\
 & HIP 55746 & 7.15 & $^{\dag}$ & 6.73 & 6.57 & 6.49 & & 6.40 & 6.44 & 6.46 & 6.41 \\
15 & 2MASS J11334926$-$7618399 & 14.08 &  & 12.15 & 11.50 & 11.18 & & 11.08 & 10.86 & 10.72 & 8.90 \\
 & RX J1137.4$-$7648 & 12.20 & $^{\#}$ & 11.85 & 10.48 & 10.14 & & 9.80 & 9.70 & 9.59 & 8.94 \\
16 & 2MASS J11404967$-$7459394 & 15.02 &  & 12.68 & 12.15 & 11.77 & & 11.58 & 11.31 & 11.04 & 9.31 \\
 & TYC 9238-612-1 & 9.98 &  & 9.39 & 9.02 & 8.86 & & 8.78 & 8.80 & 8.74 & 8.49 \\
17 & 2MASS J11432669$-$7804454 & 13.51 &  & 11.62 & 10.97 & 10.60 & & 10.23 & 9.84 & 8.71 & 7.12 \\
 & RX J1147.7$-$7842 & 10.92 &  & 9.52 & 8.86 & 8.59 & & 8.47 & 8.33 & 8.22 & 8.42 \\
18 & RX J1149.8$-$7850 & 11.01 &  & 9.45 & 8.72 & 8.49 & & 8.20 & 7.67 & 4.54 & 1.82 \\
 19 & RX J1150.4$-$7704 & 10.54 &  & 9.71 & 9.13 & 8.97 & & 8.86 & 8.87 & 8.79 & 8.71 \\
 & RX J1150.9$-$7411 & 12.06 & $^{\dag}$ & 10.60 & 9.78 & 9.48 & & 9.31 & 9.14 & 8.98 & 8.64 \\
 & 2MASS J11550485$-$7919108 & 13.26 &  & 11.22 & 10.46 & 10.08 & & 9.87 & 9.65 & 9.27 & 6.84 \\
 & T Cha & 10.28 &  & 8.96 & 7.86 & 6.95 & & 5.84 & 5.01 & 4.66 & 2.56 \\
20 & RX J1158.5$-$7754B & 11.81 &  & 10.34 & 9.71 & 9.44 & & 9.35 & 9.20 & 9.08 & 8.75 \\
21 & RX J1158.5$-$7754A & 9.76 & $^{\dag}$ & 8.63 & 7.56 & 7.40 & & 7.27 & 7.28 & 7.19 & 7.08 \\
 & HD 104036 & 6.49 & HIP & 6.29 & 6.22 & 6.11 & & 6.10 & 6.08 & 6.13 & 6.08 \\
1 & CXOU J115908.2$-$781232 & 13.83 & F03 & 12.01 & 11.45 & 11.17 & & 10.97 & 10.74 & 10.19 & 8.32 \\
2 & \epscha\ A & 5.39 & F03$^{\star}$ & 5.52 & 5.04 & 4.98 & & 5.25 & 4.97 & 5.11 & 4.84 \\
 & RX J1159.7$-$7601 & 10.18 &  & 9.14 & 8.47 & 8.30 & & 8.16 & 8.19 & 8.08 & 7.85 \\
3 & HD 104237C & \dots & \dots & 15.28 & 14.85 & 14.48 & G04 & \dots & \dots & \dots & \dots \\
4 & HD 104237B & \dots & \dots & 11.43 & 10.27 & 9.52 & G04 & \dots & \dots & \dots & \dots \\
5 & HD 104237A & 6.31 & HIP & 5.81 & 5.25 & 4.58 & & 3.89 & 2.47 & 0.93 & -0.91 \\
6 & HD 104237D & 11.62 & F03 & 10.53 & 9.73 & 9.67 & G04 & \dots  &   \dots & \dots &  \dots \\
7 & HD 104237E & 10.28 & F03 & 9.10 & 8.05 & 7.70 & G04 & \dots & \dots & \dots & \dots \\
10 & 2MASS J12005517$-$7820296 & 14.00 &  & 11.96 & 11.40 & 11.01 & & 10.62 & 10.16 & 8.52 & 6.61 \\
 & HD 104467 & 7.81 & T06 & 7.26 & 6.97 & 6.85 & & 6.81 & 6.80 & 6.78 & 6.82 \\
11 & 2MASS J12014343$-$7835472 & 15.96 &  & 14.36 & 13.38 & 12.81 & & 12.36 & 11.60 & 7.59 & 5.24 \\
8 & USNO-B 120144.7$-$781926 & 13.72 & F03 & 11.68 & 11.12 & 10.78 & & 10.16 & 9.74 & 8.35 & 6.70 \\
9 & CXOU J120152.8$-$781840 & 13.52 & F03 & 11.63 & 11.04 & 10.77 & & 10.57 & 10.35 & 10.05 & 9.32 \\
 & RX J1202.1$-$7853 & 10.49 &  & 9.21 & 8.46 & 8.31 & & 8.10 & 8.04 & 7.92 & 8.13 \\
 & RX J1202.8$-$7718 & 11.90 &  & 10.51 & 9.82 & 9.59 & & 9.53 & 9.40 & 9.20 & 8.78 \\
 & RX J1204.6$-$7731 & 11.25 &  & 9.77 & 9.12 & 8.88 & & 8.74 & 8.59 & 8.50 & 8.81 \\
 & TYC 9420-676-1 & 9.73 &  & 9.24 & 9.09 & 8.94 & & 8.86 & 8.88 & 8.86 & 9.15 \\
 & HD 105234 & 7.17 & HIP & 6.87 & 6.76 & 6.68 & & 6.55 & 6.48 & 5.43 & 4.57 \\
12 & 2MASS J12074597$-$7816064 & 13.11 &  & 11.55 & 10.98 & 10.67 & & 10.50 & 10.33 & 10.19 & 8.88 \\
 & RX J1207.7$-$7953 & 12.06 &  & 10.43 & 9.76 & 9.57 & & 9.45 & 9.31 & 9.24 & 8.62 \\
 & HIP 59243 & 6.56 & HIP & 6.34 & 6.23 & 6.17 & & 6.11 & 6.12 & 6.16 & 6.05 \\
 & HD 105923 & 8.31 & T06 & 7.67 & 7.31 & 7.17 & & 7.07 & 7.08 & 7.04 & 6.96 \\
 & RX J1216.8$-$7753 & 11.65 &  & 10.09 & 9.46 & 9.24 & & 9.14 & 9.03 & 8.91 & 8.76 \\
 & RX J1219.7$-$7403 & 10.95 &  & 9.75 & 9.05 & 8.86 & & 8.75 & 8.67 & 8.54 & 8.28 \\
 & RX J1220.4$-$7407 & 10.80 & $^{\dag}$ & 9.43 & 8.61 & 8.37 & & 8.26 & 8.15 & 8.04 & 8.11 \\
 & 2MASS J12210499$-$7116493 & 10.21 &  & 9.09 & 8.42 & 8.24 & & 8.16 & 8.16 & 8.08 & 7.99 \\
 & RX J1239.4$-$7502 & 9.21 & T06 & 8.43 & 7.95 & 7.78 & & 7.72 & 7.75 & 7.71 & 7.45 \\
 & RX J1243.1$-$7458 & 12.72 & $^{\dag}$ & 11.27 & 10.15 & 9.81 & & 9.54 & 9.37 & 9.34 & 9.48 \\
 & CD$-$69 1055 & 8.89 & T06 & 8.18 & 7.70 & 7.54 & & 7.43 & 7.47 & 7.42 & 7.44 \\
 & CM Cha & 11.80 & T06 & 10.02 & 9.16 & 8.52 & & 7.62 & 7.16 & 5.06 & 3.06 \\
 & MP Mus & 9.18 & T06 & 8.28 & 7.64 & 7.29 & & 6.58 & 6.18 & 4.06 & 1.59 \\
\hline
 \end{tabular}\\
 ($\ddag$) Photometry references: (T06) SACY/\citet{Torres06}, (HIP) \emph{Hipparcos} catalogue \citep{Perryman97}, (F03) \citet{Feigelson03}, (G04) \citet{Grady04}, (B01) \citet{Brandeker01}.\\
 ($\dag$) DENIS $I$ and 2MASS $J$ photometry corrected for multiplicity using the $K$-band flux ratios of \cite{Kohler01b} and main-sequence colours of \citet{Kraus07c}.\\
 ($\star$) F03 $I$ and 2MASS $J$ photometry corrected for multiplicity using the $V$-band magnitudes in \citet{Feigelson03} and main-sequence colours of \citet{Kraus07c}.\\
 (\#) DENIS $I$ and 2MASS $J$ photometry corrected assuming an equal-mass system.
 \end{minipage}
\end{table*}
\end{document}